\documentclass[twocolumn,secnumarabic,amssymb, nobibnotes, aps, prd]{revtex4-2}

\newcommand{\env}[1]{\texttt{#1}}
\usepackage{subfig}
\usepackage{graphicx}
\usepackage{amsmath}
\usepackage{array} 
\newcolumntype{L}[1]{>{\raggedright\let\newline\\\arraybackslash\hspace{0pt}}m{#1}}
\newcolumntype{C}[1]{>{\centering\let\newline\\\arraybackslash\hspace{0pt}}m{#1}}
\newcolumntype{R}[1]{>{\raggedleft\let\newline\\\arraybackslash\hspace{0pt}}m{#1}}
\newcolumntype{N}{@{}m{0pt}@{}}
\usepackage{multirow}
\usepackage{fixltx2e}
\usepackage{color}
\usepackage{amsmath}
\usepackage{mathtools}

\begin{document}

\title{Partially-Averaged Navier-Stokes Closure Modeling for Variable-Density Turbulent Flow}%

\author{F.S. Pereira\textsuperscript{1}}\email[F.S. Pereira: ]{fspereira@lanl.gov}
\author{F.F. Grinstein\textsuperscript{2}}
\author{D.M. Israel\textsuperscript{2}}
\author{R. Rauenzahn\textsuperscript{2}}
\author{S.S. Girimaji\textsuperscript{3}}
\affiliation{\textsuperscript{1}Los Alamos National Laboratory, T Division, Los Alamos, New Mexico, USA}
\affiliation{\textsuperscript{2}Los Alamos National Laboratory, X Division, Los Alamos, New Mexico, USA}
\affiliation{\textsuperscript{3}Texas A\&M University, Ocean Engineering, College Station, Texas, USA}


\begin{abstract}
This work extends the framework of the partially-averaged Navier-Stokes (PANS) equations to variable-density flow, \text{i.e.}, multi-material and/or compressible mixing problems with density variations and production of turbulence kinetic energy by both shear and buoyancy mechanisms. The proposed methodology is utilized to derive the PANS BHR-LEVM closure. This includes \textit{a-priori} testing to analyze and develop guidelines toward the efficient selection of the parameters controlling the physical resolution and, consequently, the range of resolved scales of PANS. Two archetypal test-cases involving transient turbulence, hydrodynamic instabilities, and coherent structures are used to illustrate the accuracy and potential of the method: the Taylor-Green vortex (TGV) at Reynolds number $\mathrm{Re}=3000$, and the Rayleigh-Taylor (RT) flow at Atwood number $0.5$ and $(\mathrm{Re})_{\max}\approx 500$. These representative problems, for which turbulence is generated by shear and buoyancy processes, constitute the initial validation space of the new model, and their results are comprehensively discussed in two subsequent studies. The computations indicate that PANS can accurately predict the selected flow problems, resolving only a fraction of the scales of large eddy simulation and direct numerical simulation strategies. The results also reiterate that the physical resolution of the PANS model must guarantee that the key instabilities and coherent structures of the flow are resolved. The remaining scales can be modeled through an adequate turbulence scale-dependent closure.
\end{abstract}

\maketitle
%
\section{Introduction}
\label{sec:1}

The numerical prediction of variable-density (multi-material and/or compressible) flows is crucial to numerous applications of fundamental and applied fluid mechanics -- e.g., scramjets, mixing problems, oceanography, supernova explosions, and inertial confinement fusion. In addition to the shear production mechanism of constant density turbulence, variable-density flows include baroclinic production due to local mean density and pressure gradients.  These flows are inherently transient and also include complex physics involving dilatation effects, characteristic hydrodynamic instabilities and coherent structures \cite{HELMHOLTZ_PMJS_1868,KELVIN_PMJS_1871,RAYLEIGHT_PLMS_1882,TAYLOR_PRSA_1950,RICHTMEYER_CPAM_1960,MESHKOV_SFD_1969}, and interactions between material and velocity fields. All these aspects make modeling and simulation of variable-density flows rife with challenges.

Direct numerical simulation (DNS) is the ideal option for prediction of any continuum fluid flow problem because it resolves all scales of motion.  Large-eddy simulation (LES) \cite{SMAGORINSKY_MWR_1963}, since it still resolves most of the turbulence spectrum, is expected to lead to accurate representations of the flow dynamics. Yet, such high-fidelity scale-resolving simulations (SRS) come at with computational expense that may be too prohibitive for practical applications. Also, selecting proper initial and boundary conditions for DNS and LES of variable-density problems is difficult.

The Reynolds-averaged Navier-Stokes (RANS) equations \cite{REYNOLDS_PTRSL_1985,FAVRE_CRAS_1958,FAVRE_JM_1965,FAVRE_JM2_1965,FAVRE_CRASP_1971} are an alternative formulation to simulate practical flows of variable-density. In contrast to DNS and LES,
RANS relies on a fully statistical description of turbulence, in which all turbulence scales are modeled through a constitutive relationship named turbulence closure \cite{BESNARD_TREP_1992,CHASSAING_FTC_2001,CHASSAING_BOOK_2002,WILCOX_BOOK_2010}. This modeling strategy entails computation of the mean and coherent fields \cite{HUSSAIN_JFM_1970,SCHIESTEL_PF_1987} and, consequently, significantly reduces the computations' cost. However, RANS closures are usually inaccurate predicting many problems of interest in this study.

The caveats of the former modeling strategies drove the emergence of a new paradigm of practical SRS methods specially designed to predict complex flows efficiently. This class of models named bridging methods was proposed by \citeauthor{GERMANO_JFM_1992} \cite{GERMANO_JFM_1992,GERMANO_DLES_1999} and \citeauthor{SPEZIALLI_ICNMFD_1996} \cite{SPEZIALLI_ICNMFD_1996} to seamlessly operate between DNS and RANS, and only resolve the flow scales not amenable to modeling. The remaining scales can be accurately represented through an adequate closure \cite{PEREIRA_JCP_2018}. This strategy is responsible for the potential efficiency (accuracy vs. cost) of bridging methods, and introduces the idea of accuracy-on-demand. Very Large-Eddy Simulation (VLES) \cite{SPEZIALLI_ICNMFD_1996}, Limited Numerical Scales (LNS) \cite{BATTEN_AIAA38_2000}, Flow Simulation Methodology (FSM) \cite{FASEL_JFE_2002}, Partially-Integrated Transport Model (PITM) \cite{SCHIESTEL_TCFD_2005,CHAOUAT_PF_2005}, and Partially-Averaged Navier-Stokes (PANS) equations \cite{GIRIMAJI_JAM_2005} are examples of bridging formulations.

Despite being widely used in many scientific areas, bridging models are still not common in the variable-density flow community. There are three main factors contributing to this outcome:
\begin{itemize}
\item[$i)$] \textbf{complexity - }bridging closures are typically based on one-point RANS models, which are calibrated for total turbulent quantities. However, since bridging methods can operate at any range of resolved scales, the inability to reliably estimate RANS variables may lead to calibration deficiencies of the closure \cite{PEREIRA_JFE_2019}. This is expected to be particularly relevant for low physical resolutions, transient and transitional flows, and second-moment closures \cite{BESNARD_TREP_1992,CHASSAING_FTC_2001,CHASSAING_BOOK_2002,ZARLING_TREP_2011,SCHWARZKOPF_FTC_2016}. Further, the RANS closures for variable-density flow possess terms to account for density variations that can be difficult to extend to the bridging framework.
\item[$ii)$] \textbf{physical resolution - } bridging models normally require a physical resolution parameter which determines the range of resolved scales and, consequently, the computational efficiency. This parameter defines the fraction of the dependent quantities of the turbulence closure being modeled, e.g., Reynolds-stress tensor, turbulence kinetic energy, dissipation, etc. Whereas excessively large physical resolutions increase the simulations range of resolved scales and cost unnecessarily, low values of such a parameter can lead to inaccurate computations. This makes the selection of the physical resolution crucial to the simulations accuracy. 

A closure for variable-density flow will rely on multiple model evolution equations \cite{CHASSAING_FTC_2001,CHASSAING_BOOK_2002}. This raises the question, how do different modeled turbulence quantities behave with the physical resolution.  I.e., what is the fraction of each dependent turbulence quantity being modeled for a given range of resolved scales. VLES, LNS, and FSM formulations use a pragmatic approach where the magnitude of the Reynolds-stress tensor is scaled by a given factor. Despite being simple, this strategy does not yield the correct fixed point behavior for the closure system \cite{GIRIMAJI_JAM_2006}. PANS and PITM, on the other hand, do not experience these issues since they use parameters to define the modeled-to-total ratio of each turbulence dependent quantity. Yet, these ratios need to be defined in a physically consistent manner.
\item[$iii)$] \textbf{commutation errors - }SRS formulations are based upon the scale-invariance property of the Navier-Stokes equations \cite{GERMANO_JFM_1992}. This property is responsible for the formal similarity between the filtered Navier-Stokes and RANS equations, and requires the implicit model filter operator to commute with temporal and spatial differentiation \cite{GERMANO_JFM_1992}. If this condition does not hold, the filtered Navier-Stokes equations have additional terms which are difficult to model and are often neglected. This creates the so-called commutation error \cite{HAMBA_PF_2011,GIRIMAJI_JT_2013}.

Most SRS computations are conducted with a spatially varying physical resolution to optimize the use of the grid. Even for conventional LES models, this raises potential modeling issues (as shown in \citeauthor{HAMBA_PF_2011} \cite{HAMBA_PF_2011}).  For bridging models, such as PANS, spatial variation of the resolution control parameters lead to additional commutation terms (\citeauthor{GIRIMAJI_JT_2013} \cite{GIRIMAJI_JT_2013}), that can reach the magnitude of the convective terms of the filtered Navier-Stokes equations. This is expected to be relevant for variable-density turbulent flows due to their transient and transitional nature.
\end{itemize}

This work proposes a PANS bridging model specifically designed for variable-density turbulent flow. Thus, the PANS framework of \citeauthor{GIRIMAJI_JAM_2005} \cite{GIRIMAJI_JAM_2005} and \citeauthor{SUMAN_FTC_2010} \cite{SUMAN_FTC_2010} is extended to variable-density flow, and such a methodology is utilized to derive the PANS version of the six-equation BHR-LEVM (linear eddy viscosity model) closure \cite{BANERJEE_PRE_2010,ZARLING_TREP_2011}. In addition, we perform \textit{a-priori} testing to analyze and develop guidelines toward the efficient selection of the parameters controlling the physical resolution of the PANS model. The accuracy and potential of the model is evaluated through the prediction of two benchmark problems: the Taylor-Green vortex (TGV) \cite{TAYLOR_RSA_1937} at Reynolds number (Re) 3000 and initial Mach number (Ma) $0.28$, and the Rayleigh-Taylor flow \cite{RAYLEIGH_PLMS_1882,TAYLOR_PRSA_1950} at Atwood number (At) $0.50$, $(\mathrm{Re})_{\max}\approx 500$, and $\mathrm{Ma}<0.10$. The first validation test-case assesses the ability of PANS BHR-LEVM predicting the onset and development of turbulence in a transient problem where turbulence is produced by shear processes. The second flow also includes multi-material mixing and turbulence produced by buoyancy mechanisms. These two canonical problems constitute the initial validation space \cite{TRUCANO_SANDIA_2002,OBERKAMPF_BOOK_2010} of the PANS BHR-LEVM model,  and are comprehensively analyzed in two subsequent studies \cite{PEREIRA_PRF_2021,PEREIRA_POF_2021}. All simulations are conducted at multiple physical resolutions to evaluate the effect of this parameter on the simulations accuracy. Also, the physical resolution is set constant in space and time to prevent commutation errors. 

The remainder of this paper is structured as follows. Section \ref{sec:2} presents the derivation of the governing equations of PANS BHR-LEVM. A consistent framework and nomenclature is defined. Next, Section \ref{sec:3} analyses the evolution of PANS turbulence dependent quantities with the physical resolution, and proposes guidelines toward the efficient selection of the parameters controlling the physical resolution of the model. Section \ref{sec:4} describes the selected test cases, while Section \ref{sec:5} discusses the main results. Section \ref{sec:6} concludes this paper with a summary of the major findings.
%
%
\section{Governing Equations}
\label{sec:2}

The partially-averaged Navier-Stokes (PANS) equations are based upon the scale-invariance property of the Navier-Stokes equations. This property has been demonstrated by \citeauthor{GERMANO_JFM_1992} \cite{GERMANO_JFM_1992} for incompressible flow, and extended to compressible flow by \citeauthor{SUMAN_FTC_2010} \cite{SUMAN_FTC_2010}. To derive the PANS equations for variable-density flow (multi-material and/or compressible), let us start by considering a general linear and constant preserving filtering operator $\langle\ \cdot \ \rangle$,
\begin{equation}
\label{eq:2_1}
\langle \Phi_1 + \Phi_2 \rangle =  \langle \Phi_1 \rangle + \langle \Phi_2 \rangle \; , 
\end{equation}
\begin{equation}
\label{eq:2_2}
\langle \alpha \Phi \rangle = \alpha \langle  \Phi \rangle \; ,
\end{equation}
where $\Phi$ is a generic variable, and $\alpha$ is a constant. This filter commutes with spatial and temporal differentiation so that
\begin{equation}
\label{eq:2_3}
\left\langle \frac{\partial  \Phi }{\partial x_i} \right\rangle =  \frac{\partial \langle \Phi  \rangle}{\partial x_i} \; , \hspace{1cm}
\left\langle \frac{\partial  \Phi }{\partial t} \right\rangle =  \frac{\partial \langle \Phi  \rangle}{\partial t} \; ,
\end{equation}
and decomposes any instantaneous flow quantity $\Phi$ into a filtered (resolved), $\langle \Phi \rangle$, and modeled (unresolved), $\phi$, component,
\begin{equation}
\label{eq:2_4}
\Phi \equiv \langle \Phi \rangle + \phi \; .
\end{equation}
This decomposition can be extended to variable-density flow through the concept of Favre-averaging \cite{FAVRE_CRAS_1958,FAVRE_JM_1965,FAVRE_JM2_1965,FAVRE_CRASP_1971},
\begin{equation}
\label{eq:2_5}
\Phi \equiv \{ \Phi \} + \phi^* \; ,
\end{equation}
where $\{ \Phi \}$ and $\phi^*$ are the density-weighted filtered, $\{ \Phi \} \equiv \langle \rho \Phi \rangle /\langle \rho \rangle$, and modeled fluctuating, $\phi^*=\Phi - \{\Phi\}$, components of $\Phi$. In the limit of all turbulence scales being modeled, the former decompositions are equivalent to Reynolds- \cite{REYNOLDS_PTRSL_1985} and Favre-averaging \cite{FAVRE_CRAS_1958,FAVRE_JM_1965,FAVRE_JM2_1965,FAVRE_CRASP_1971}, 
\begin{equation}
\label{eq:2_6}
\Phi=\overline{\Phi}+\phi' \; ,
\end{equation}
\begin{equation}
\label{eq:2_7}
\Phi=\tilde{\Phi}+\phi'' \; ,
\end{equation}
$\overline{\Phi}$ and $\phi'$ being the (time, ensemble or spatial) averaged and turbulent components of $\Phi$, whereas $\tilde{\Phi}$ and $\phi''$ the density-weighted averaged and fluctuating counterparts of $\Phi$.

The application of such filtering operators to the conservation equations for mass, momentum, total energy, and fluid species \cite{WILLIAMS_BOOK_1965,COOK_POF_2009} leads to the filtered or partially-averaged form of the Navier-Stokes equations for variable-density flow \cite{GERMANO_JFM_1992,GIRIMAJI_JAM_2005,SUMAN_FTC_2010,PEREIRA_PRF_2021},
\begin{equation}
\label{eq:2_8}
\frac{ \partial \langle \rho \rangle }{\partial t}+\frac{\partial \left(\langle \rho \rangle \{ V_i \} \right)}{\partial x_i}=0 \; ,
\end{equation}
\begin{equation}
\label{eq:2_9}
\begin{split}
\frac{ \partial \left( \langle \rho \rangle \{V_i \} \right) }{\partial t} &+ \frac{\partial \left( \langle \rho \rangle \{V_j \} \{V_i\} \right)}{\partial x_j}=  - \frac{\partial \langle P \rangle}{\partial x_i} + \frac{\partial \langle \sigma_{ij}\rangle}{\partial x_j}\\
&+ \frac{\partial \left(\langle \rho \rangle \tau^1 \left(V_i,V_j\right) \right)}{\partial x_j} + \langle \rho \rangle g_i
\end{split}
\; ,
\end{equation}
\begin{equation}
\label{eq:2_10}
\begin{split}
\frac{ \partial \left( \langle \rho \rangle \{E\} \right) }{\partial t} &+  \frac{\partial \left( \langle \rho \rangle \{E\} \{V_j\} \right) }{\partial x_j}= - \frac{\partial \left( \langle \rho \rangle \tau^1({V_j,E}) \right)}{\partial x_j} \\
& - \frac{\partial \left( \{V_j \} \langle P\rangle \right)}{\partial x_j} - \frac{\partial \tau^2(V_j,P)}{\partial x_j}\\
& + \frac{\partial \left( \{V_i \} \langle \sigma_{ij}\rangle \right)}{\partial x_j} + \frac{\partial \tau^2(V_i,\sigma_{ij})}{\partial x_j}\\
& - \frac{\partial \langle q_j^c\rangle}{\partial x_j}  - \frac{\partial \langle q_j^h\rangle}{\partial x_j} 
\end{split}
\; ,
\end{equation}
\begin{equation}
\label{eq:2_11}
\frac{ \partial \left( \langle \rho \rangle \{c^n \} \right)}{\partial t} +  \frac{\partial \left(  \langle \rho \rangle \{c ^n\} \{V_j  \} \right)  }{\partial x_j}=- \frac{\partial \langle J^n_j \rangle}{\partial x_j}  \; .
\end{equation}
Here, $t$ is the time, $x_i$ are the coordinates of a Cartesian system, $\rho$ is the fluid density, $V_i$ are the Cartesian velocity components, $P$ is the pressure, $\sigma_{ij}$ is the viscous-stress tensor assuming Newtonian fluid,
\begin{equation}
\label{eq:2_12}
\langle \sigma_{ij} \rangle = 2 \mu \left( \{S_{ij}\} - \frac{2}{3} \frac{\partial \{ V_k\}}{\partial x_k} \delta_{ij} \right)  \; ,
\end{equation}
$\{ S_{ij} \}$ is the resolved strain-rate tensor,
\begin{equation}
\label{eq:2_13}
\{S_{ij}\} = \frac{1}{2} \left( \frac{\partial \{ V_i\}}{\partial x_j} + \frac{\partial \{ V_j\}}{\partial x_i} \right)\; ,
\end{equation}
$\mu$ is the fluid's dynamic viscosity, $\delta_{ij}$ is the Kronecker delta, $g_i$ is the gravitational acceleration vector, $E=\frac{1}{2}V_i^2+e$ is the total energy of the fluid, $e$ is the internal energy, $c^n$ is the mass concentration of material $n$, $q^c$ is the conductive heat flux, $q^d$ is the interdiffusional enthalpy flux, and $J^n$ is the mass fraction diffusivity flux of material $n$. Also, $\tau^1(\Phi_i,\Phi_j)$ and $\tau^2(\Phi_i,\Phi_j)$ are generalized central second moments which account for the effect of the modeled turbulence in the resolved flow field. Expressing the PANS equations in terms of generalized central second moments \cite{GERMANO_JFM_1992} guarantees scale-invariance. Such tensors are formally defined as \cite{GERMANO_JFM_1992,SUMAN_FTC_2010}
\begin{equation}
\label{eq:2_14}
\tau^1(\Phi_i,\Phi_j)\equiv \{ \Phi_i \Phi_j \} - \{ \Phi_i \} \{ \Phi_j \} \; ,
\end{equation}
\begin{equation}
\label{eq:2_15}
\tau^2(\Phi_i,\Phi_j)\equiv \langle \Phi_i \Phi_j \rangle - \{ \Phi_i \} \langle \Phi_j \rangle \; .
\end{equation}
In equations \ref{eq:2_8} to \ref{eq:2_11}, the pressure is calculated assuming a thermally perfect gas ($P=\rho R T$). Thus, its resolved component is given by \cite{SUMAN_FTC_2010},
\begin{equation}
\label{eq:2_16o}
\langle P \rangle = \left( \gamma -1 \right) \langle \rho \rangle \left( \{E\} - \frac{\{V_k\}\{V_k\}}{2} - k_u \right)  \; ,
\end{equation}
where $T$ is the temperature, $\gamma$ is the ratio between specific heats, and $k_u$ is the unresolved or modeled specific turbulence kinetic energy.

The generalized central second-moments and fluxes present in the PANS equations need modeling to close the resultant system of equations. In the present work, this is accomplished through the Boussinesq approximation \cite{BOUSSINESQ_MPDSAS_1877},
\begin{equation}
\label{eq:2_16}
\tau^1({V_i,V_j})=2 \nu_u \{S_{ij}\} - \frac{2}{3} k_u \delta_{ij}  \; ,
\end{equation}
and the relationships given in \citeauthor{BESNARD_TREP_1992} \cite{BESNARD_TREP_1992}, \citeauthor{SUMAN_FTC_2010} \cite{SUMAN_FTC_2010}, \citeauthor{ZARLING_TREP_2011} \cite{ZARLING_TREP_2011}, and \citeauthor{SCHWARZKOPF_FTC_2016} \cite{SCHWARZKOPF_FTC_2016}. These lead to the partially-averaged form of the energy and fluid species equations, 
\begin{equation}
\label{eq:2_17}
\begin{split}
\frac{ \partial \left( \langle \rho \rangle \{E\} \right) }{\partial t} &+  \frac{\partial \left( \langle \rho \rangle \{E\} \{V_j\} \right) }{\partial x_j}= - \frac{\partial \left( \{V_j \} \langle P\rangle \right)}{\partial x_j} \\
& + \frac{\partial \left( \{V_i \} \langle \sigma_{ij}\rangle \right)}{\partial x_j} + \frac{\partial \left( \langle \rho \rangle \{V_i \} \tau^1(V_i,V_j) \right)}{\partial x_j} \\
& + \frac{\partial}{\partial x_j} \left[ \left( \mu + \frac{\mu_u}{\sigma_k}\right)\frac{\partial k_u}{\partial x_j}\right] \\
& - \frac{\partial}{\partial x_j} \left[ c_p\left( \frac{ \mu}{\mathrm{Pr}} + \frac{ \mu_u}{\mathrm{Pr}_t}\right)\frac{\partial \langle T \rangle}{\partial x_j}\right] \\
& - \frac{\partial}{\partial x_j} \left[ \sum_{n=1}^{n_t}  h_n \langle J_j^n \rangle \right]
\end{split}
\; ,
\end{equation}
\begin{equation}
\label{eq:2_18}
\begin{split}
\frac{ \partial \left( \langle \rho \rangle \{c^n \} \right)}{\partial t} &+  \frac{\partial \left(  \langle \rho \rangle \{c ^n\} \{V_j  \} \right)  }{\partial x_j}=- \frac{\partial \langle J^n_j \rangle}{\partial x_j}  \\
&=- \frac{\partial }{\partial x_j} \left[ \langle \rho \rangle\left( {\cal{D}} +  \frac{\nu_u}{\sigma_c} \right) \frac{\partial \{ c^n \}}{\partial x_j} \right] 
\end{split}
\; ,
\end{equation}
where $\nu_u=\mu_u/\langle \rho \rangle$ is the kinematic turbulent viscosity of the unresolved scales, $\sigma_k$ and $\sigma_c$ are turbulent diffusion coefficients, $c_v$ is the constant specific heat (ideal gas is assumed), $\mathrm{Pr}$ is the Prandtl number, $\mathrm{Pr_t}$ is the turbulent Prandtl number defined as $\mathrm{Pr_t}=c_v \nu_u/ \kappa$, $\kappa$ is the effective thermal conductivity, and $h^n$ is the enthalpy of material $n$. Throughout this manuscript, all modeled/unresolved PANS and RANS turbulence quantities are denoted by the subscripts \textit{u} and \textit{t}, respectively. 

The relationships above create two additional turbulence quantities, $k_u$ and $\nu_u$, that need modeling. This is accomplished through the BHR-LEVM \cite{BANERJEE_PRE_2010,ZARLING_TREP_2011} closure model which is now derived for PANS.
%
%
%
\subsection{PANS BHR-LEVM closure}
\label{sec:2.1}

\begin{table*}
\centering
\setlength\extrarowheight{3pt}
\caption{Coefficients of BHR-LEVM closure \cite{ZARLING_TREP_2011}.}
\label{tab:2_1}    
\begin{tabular}{C{1.1cm}C{1.1cm}C{1.1cm}C{1.1cm}C{1.1cm}C{1.1cm}C{1.1cm}C{1.1cm}C{1.1cm}C{1.1cm}C{1.1cm}}
\hline
$c_1$ & $c_2$ &  $c_4$ &  $c_{a_1}$ &  $c_{b}$ &   $c_\mu$  & $\sigma_a$ & $\sigma_b$ & $\sigma_c$ & $\sigma_k$ & $\sigma_S$  \\[3pt] \hline
0.06 & 0.42 & 0.45 & 6.00 & 0.45 & 0.28  & 1.00 & 3.00 & 0.60 & 1.00 & 0.10 \\[3pt] \hline
\end{tabular}
\end{table*}

Most PANS closures are based on one-point, linear turbulent viscosity RANS closures. This modeling strategy is chosen to balance sufficient complexity to accurately operate at any degree of physical resolution with scale-aware minor modifications \cite{PEREIRA_ACME_2021}, without the loss of robustness observed for full Reynolds-stress closures. For variable-density flow we choose the BHR model originally proposed by \citeauthor{BESNARD_TREP_1992} \cite{BESNARD_TREP_1992}, in $k_t-S_t$ linear turbulent viscosity form found in \cite{BANERJEE_PRE_2010,ZARLING_TREP_2011}. The model requires transport equations for six turbulent dependent variables: the turbulence kinetic energy, $k_t$, turbulence dissipation length-scale, $S_t$, velocity mass flux, $a_{i_t}$,
\begin{equation}
\label{eq:2.1_01}
 a_{i_t}= \frac{\overline{\rho' v_i'}}{\overline{\rho}} \; ,
\end{equation}
and density-specific volume correlation, $b_t$,
\begin{equation}
\label{eq:2.1_02}
b_t=-\overline{\rho'(1/\rho)'}\; .
\end{equation}
In equations \ref{eq:2.1_01} and \ref{eq:2.1_02}, the primes refer to the fluctuating component over the mean value so that these quantities can be divided into a coherent and turbulent part \cite{PALKIN_FTC_2016}. For the $b_t$ equation, we use the newer formulation of \citeauthor{SCHWARZKOPF_JOT_2011} \cite{SCHWARZKOPF_JOT_2011}.  The subscript $t$ indicates a total turbulence quantity, that is, the quantity predicted by the RANS model that includes the action of all the turbulent scales of motion.  The subscript $u$ will be used for partial-averaged quantities, including only the unresolved portion of the turbulent scales.

The RANS BHR-LEVM model used in this work calculates the total kinematic turbulent viscosity as,
\begin{equation}
\label{eq:2.1_1}
\nu_t= \frac{\mu_t}{\overline{\rho}} =  c_\mu S_t \sqrt{k_t} \; ,
\end{equation}
where $c_\mu$ is a coefficient given in table \ref{tab:2_1}, and $S_t$ is the turbulence dissipation length-scale defined as
\begin{equation}
\label{eq:2.1_2}
S_t=\frac{k_t^{3/2}}{\varepsilon_t} \; ,
\end{equation}
and $\varepsilon_t$ is the specific total turbulence dissipation. The turbulence quantities $k_t$ and $S_t$ are obtained from the following evolution equations,
\begin{equation}
\label{eq:2.1_3}
\frac{\partial {\color{black}\overline{\rho}} k_t}{\partial t} + \frac{\partial {\color{black}\overline{\rho}}k_t\tilde{V}_j}{\partial x_j} = {\cal{P}}_{b_t} + {\cal{P}}_{s_t} - {\color{black}\overline{\rho}}\frac{k_t^{3/2}}{S_t} + \frac{\partial}{\partial x_j}\left( \frac{\overline{\rho} \nu_t}{\sigma_k} \frac{\partial k_t}{\partial x_j}\right)\; ,
\end{equation}
\begin{equation}
\label{eq:2.1_4}
\begin{split}
\frac{\partial {\color{black}\overline{\rho}} S_t}{\partial t} + \frac{\partial {\color{black}\overline{\rho}} S_t\tilde{V_j}}{\partial x_j} &= \frac{S_t}{k_t} \left( c_4 {\cal{P}}_{b_t} + c_1 {\cal{P}}_{s_t} \right) - {\color{black}\overline{\rho}} c_2 \sqrt{k_t} \\
 &+ \frac{\partial}{\partial x_j}\left( \frac{\overline{\rho}\nu_t}{\sigma_S} \frac{\partial S_t}{\partial x_j} \right) 
\end{split}
 \; ,
\end{equation}
\begin{equation}
\label{eq:2.1_5}
\begin{split}
\frac{\partial {\color{black}\overline{\rho}} a_{i_t}}{\partial t} + \frac{\partial {\color{black}\overline{\rho}} a_{i_t} \tilde{V_j}}{\partial x_j} &= b_t \frac{\partial \overline{P}}{\partial x_i}
+ R^1(V_i,V_j) \frac{\partial \overline{\rho} }{\partial x_j} \\
&  + {\color{black}\overline{\rho}}\frac{\partial \left( a_{i_t}a_{j_t} \right)}{\partial x_j} - c_{a_1}{\color{black}\overline{\rho}}a_{i_t} \frac{\sqrt{k_t}}{S_t}\\
& - {\color{black}\overline{\rho}}a_{j_t} \frac{\partial \overline{V}_i}{\partial x_j}  + \frac{\partial}{\partial x_j}\left( \frac{{\color{black}\overline{\rho}}\nu_t}{\sigma_a} \frac{\partial a_{i_t}}{\partial x_j} \right) 
\end{split}
 \; ,
\end{equation}
\begin{equation}
\label{eq:2.1_6}
\begin{split}
\frac{\partial {\color{black}\overline{\rho}} b_t}{\partial t} + \frac{\partial{\color{black}\overline{\rho}} b_t\tilde{V_j}}{\partial x_j} &=
 2{\color{black}\overline{\rho}}a_{j_t} \frac{\partial b_t}{\partial x_j}
  - 2a_{j_t}\left(b_t+1 \right) \frac{\partial \overline{\rho} }{\partial x_j} \\
 &- c_{b}{\color{black}\overline{\rho}}b_t \frac{\sqrt{k_t}}{S_t}
  + \overline{\rho}{\color{black}^2}\frac{\partial}{\partial x_j}\left( \frac{\nu_t}{ \overline{\rho} \sigma_b} \frac{\partial b_t}{\partial x_j} \right) 
\end{split}
 \; ,
\end{equation}
where ${\cal{P}}_{b_t}$ and ${\cal{P}}_{s_t}$ are the specific total production of turbulence kinetic energy by buoyancy and shear mechanisms,
\begin{equation}
\label{eq:2.1_7}
{\cal{P}}_{b_t} =  a_{j_t} \frac{\partial \overline{P}}{\partial x_j} \; ,
\end{equation}
\begin{equation}
\label{eq:2.1_8}
{\cal{P}}_{s_t} = - {\color{black}\overline{\rho}}{R^1(V_i,V_j)} \frac{\partial \tilde{V}_i}{\partial x_j}  \; ,
\end{equation}
$R^1(V_i,V_j)$ is the Reynolds stress tensor ($R^1(V_i,V_j)$ and $\tau^1(V_i,V_j)$ are equivalent when all turbulence scales are modeled), and $c_1$, $c_2$, $c_4$, $c_{a_1}$, $c_{b}$, $\sigma_k$, $\sigma_S$, $\sigma_a$, and $\sigma_b$ are coefficients of the original RANS BHR-LEVM model, whose values are given in table \ref{tab:2_1}.

Equations \ref{eq:2.1_3} to \ref{eq:2.1_6} have been designed to operate exclusively with RANS variables: Reynolds averaged, $\overline{\Phi}$, density-weighted averaged, $\tilde{\Phi}$, and total turbulence, $\Phi_t$, quantities. We now derive their PANS counterpart by extending the framework proposed by \citeauthor{GIRIMAJI_JAM_2005} \cite{GIRIMAJI_JAM_2005} to variable-density flow. To this end, the parameters $f_\phi$ defining the ratios of modeled-to-total specific turbulence kinetic energy, $f_k$, dissipation length-scale, $f_S$, mass flux velocity, $f_{a_i}$, and density-specific volume correlation, $f_b$,
\begin{equation}
\label{eq:2.1_9}
f_k\equiv \frac{k_u}{k_t}\;, \hspace{0.6cm}  f_S \equiv \frac{S_u}{S_t}\;, \hspace{0.6cm} f_{a_i} \equiv \frac{{a_{i_u}}}{a_{i_t}}\;,  \hspace{0.6cm} f_{b} \equiv \frac{{b_{u}}}{b_t}  \; ,
\end{equation}
need to be included in equations \ref{eq:2.1_3} to \ref{eq:2.1_6}. These define the physical resolution and, as such, the fraction of the dependent quantities of the turbulent closure being modeled. Also, they enable the closure to operate at any range of resolved scales, i.e., from RANS ($f_\phi = 1$), where turbulence is fully represented by the closure so that $\tau^1(V_i,V_j)=R^1(V_i,V_j)$, to DNS ($f_\phi = 0$, no closure), where turbulence is fully resolved so that $\tau^1(V_i,V_j)=0$. $f_S$ can also be calculated as a function of $f_k$ and $f_\varepsilon$,
\begin{equation}
\label{eq:2.1_10}
f_S \equiv \frac{S_u}{S_t} = \left(\frac{k_u^{3/2}}{\varepsilon_u}\right) \left( \frac{\varepsilon_t}{k_t^{3/2}}\right)=\frac{f_k^{3/2}}{f_\varepsilon}  \; ,
\end{equation}
where $f_\varepsilon$ is the ratio of modeled-to-total of specific turbulence dissipation. Since $f_\varepsilon$ is physically more intuitive than $f_S$, the evolution equations for $k_u$, $S_u$, $a_{i_u}$, and $b_u$ are derived in terms of $f_k$, $f_\varepsilon$, $f_{a_i}$, and $f_b$.
%
%
%
\subsubsection{$k_u$ evolution equation}
\label{sec:2.1.1}

It has been demonstrated by \citeauthor{GIRIMAJI_JAM_2005} \cite{GIRIMAJI_JAM_2005} and \citeauthor{SUMAN_FTC_2010} \cite{SUMAN_FTC_2010} that the scale-invariant form of $k_u$ equation can be written as,
\begin{equation}
\label{eq:2.1.1_1}
\frac{\partial{\color{black}\langle \rho \rangle} k_u}{\partial t} +   \frac{\partial {\color{black}\langle \rho \rangle} k_u\{V_j \}}{\partial x_j} = {\cal{P}}_{b_u} + {\cal{P}}_{s_u} - {\color{black}{\cal{E}}_u}+   {\cal{T}}_u\; ,
\end{equation}
where ${\cal{P}}_{b_u}$ and ${\cal{P}}_{s_u}$ are the production of $k_u$ by buoyancy and shear mechanisms,
\begin{equation}
\label{eq:2.1.1_2}
{\cal{P}}_{b_u} =  a_{j_u} \frac{\partial \langle P \rangle}{\partial x_j} \; ,
\end{equation}
\begin{equation}
\label{eq:2.1.1_3}
{\cal{P}}_{s_u} = - {\color{black}\langle \rho \rangle} \tau^1(V_i,V_j) \frac{\partial \{V_i\}}{\partial x_j}  \; ,
\end{equation}
and ${\color{black}{\cal{E}}_u=\langle \rho \rangle \varepsilon_u}$ and ${\cal{T}}_u$ represent the dissipation and transport of modeled turbulence kinetic energy. For constant $f_k$, differentiation commutes in time and space and so it possible to establish a relationship between the equations for $k_u$ (PANS) and $k_t$ (RANS),
\begin{equation}
\label{eq:2.1.1_4}
\frac{\partial {\color{black}\overline{\rho}} k_u}{\partial t} + \frac{\partial {\color{black}\overline{\rho} } k_u\tilde{V_j}}{\partial x_j} =  f_k \left[ \frac{\partial {\color{black}\overline{\rho}} k_t}{\partial t} + \frac{\partial {\color{black}\overline{\rho} } k_t\tilde{V_j}}{\partial x_j}  \right] \; .
\end{equation}
Since PANS calculates filtered or partially-averaged dependent variables, the former relationship can be rewritten as follows,
\begin{equation}
\label{eq:2.1.1_5}
\begin{split}
\frac{\partial {\color{black}\overline{ \rho }}k_u}{\partial t} + \frac{\partial {\color{black}\overline{\rho}}k_u\{V_j\}}{\partial x_j} &=  f_k \left[ \frac{\partial {\color{black}\overline{\rho}} k_t}{\partial t} + \frac{\partial {\color{black}\overline{ \rho }} k_t\tilde{V_j}}{\partial x_j}  \right] \\
&+\frac{\partial }{\partial x_j}\left( k_u  {\color{black}\overline{\rho}}\left( \{ V_j \} - \tilde{V}_j\right) \right)
\end{split}
 \; .
\end{equation}
Next, replacing the term between brackets by the right-hand side of equation \ref{eq:2.1_3} {\color{black}and introducing the parameter $f_\rho=\langle \rho \rangle/\overline{\rho}$} leads to
\begin{equation}
\label{eq:2.1.1_6}
\begin{split}
\frac{\partial {\color{black}\langle \rho \rangle }k_u}{\partial t} + \frac{\partial {\color{black}\langle{\rho}\rangle}k_u\{V_j\}}{\partial x_j}  &= {\color{black}f_\rho} f_k \left[ {\cal{P}}_{b_t} +  {\cal{P}}_{s_t} - {\color{black}{\cal{E}}_t} + {\cal{T}}_{t}\right] \\
&+ \frac{\partial }{\partial x_j}\left( k_u{\color{black}\langle \rho \rangle}\left( \{ V_j \} - \tilde{V}_j\right) \right)
\end{split}
\; ,
\end{equation}
and applying self-similarity considerations to the left-hand side, the following relation is obtained,
\begin{equation}
\label{eq:2.1.1_7}
\begin{split}
{\cal{P}}_{b_u} + {\cal{P}}_{s_u} - {\color{black}{\cal{E}}_u} + {\cal{T}}_u  & =  {\color{black}f_\rho}f_k \left[ {\cal{P}}_{b_t} +  {\cal{P}}_{s_t }- {\color{black}{\cal{E}}_t} + {\cal{T}}_{t} \right] \\
&+ \frac{\partial }{\partial x_j}\left( k_u{\color{black}\langle \rho \rangle}\left( \{ V_j \} - \tilde{V}_j\right) \right)
\end{split}
 \; .
\end{equation}
This equation shows the formal similarity between PANS (left-hand side) and RANS (right-hand side) production, dissipation and transport terms. Hence, it is possible to relate the source and sink (local processes), and transport terms as follows,

\begin{subequations}
\begin{equation}
\label{eq:2.1.1_8a}
{\cal{P}}_{b_u}  + {\cal{P}}_{s_u} - {\color{black}{\cal{E}}_u} = {\color{black}f_\rho} f_k \left[ {\cal{P}}_{b_t} +  {\cal{P}}_{s_t} - {\color{black}{\cal{E}}_t} \right]  \; ,
\end{equation}
\begin{equation}
\label{eq:2.1.1_8b}
{\cal{T}}_u ={\color{black}f_\rho}f_k {\cal{T}}_t + \frac{\partial }{\partial x_j}\left( k_u{\color{black}\langle \rho \rangle}\left( \{ V_j \} - \tilde{V}_j\right) \right) \; .
\end{equation}
\end{subequations}
Now, we define the weighting functions $\omega_s$ and $\omega_b$,
\begin{equation}
\label{eq:2.1.1_9}
\omega_s =\frac{{\cal{P}}_{s_u}}{{\cal{P}}_{b_u}+{\cal{P}}_{s_u}}  \; ,
\end{equation}
\begin{equation}
\label{eq:2.1.1_10}
\omega_b =\frac{{\cal{P}}_{b_u}}{{\cal{P}}_{b_u}+{\cal{P}}_{s_u}}  \; ,
\end{equation}
which define the relative weight of the shear, $\omega_s$, and buoyancy, $\omega_b$, mechanisms to the total production of specific turbulence kinetic energy. Thus, their sum is equal to unity, $\omega_s +\omega_b =1$. Using $f_k$, $f_\varepsilon$ and these weighting functions, we can rewrite equation \ref{eq:2.1.1_8a},
\begin{equation}
\label{eq:2.1.1_11}
{\cal{P}}_{b_u}  + {\cal{P}}_{s_u} - {\color{black}{\cal{E}}_u} \omega_s - {\color{black}{\cal{E}}_u} \omega_b =  f_k \left[ {\color{black}{f_\rho}}({\cal{P}}_{b_t} +  {\cal{P}}_{s_t}) - \omega_s \frac{{\color{black}{\cal{E}}_u}}{f_\varepsilon} - \omega_b\frac{{\color{black}{\cal{E}}_u}}{f_\varepsilon}\right]  \; ,
\end{equation}
and obtain the relationships,
\begin{equation}
\label{eq:2.1.1_12}
{\color{black}{f_\rho}}{\cal{P}}_{b_t} = \frac{{\cal{P}}_{b_u}}{f_k} - \omega_b {\color{black}{\cal{E}}_u} \left( \frac{1}{f_k} - \frac{1}{f_\varepsilon} \right)    \; ,
\end{equation}
\begin{equation}
\label{eq:2.1.1_13}
{\color{black}{f_\rho}}{\cal{P}}_{s_t} = \frac{{\cal{P}}_{s_u}}{f_k} - \omega_s {\color{black}{\cal{E}}_u} \left( \frac{1}{f_k} - \frac{1}{f_\varepsilon} \right)    \; .
\end{equation}
The former relations are used to derive the evolution equation for $S_u$. On the other hand, the transport terms of the $k_u$ and $k_t$ equations can be related as follows,
\begin{equation}
\label{eq:2.1.1_14}
\begin{split}
{\cal{T}}_u&={\color{black}f_\rho}f_k\left[ {\cal{T}}_t\right] + \frac{\partial }{\partial x_j}\left( k_u{\color{black}\langle \rho \rangle}\left( \{ V_j \} - \tilde{V}_j\right) \right)  \\
& ={\color{black}f_\rho} f_k\left[  \frac{\partial}{\partial x_j}\left( \frac{\overline{\rho}\nu_t}{\sigma_k} \frac{\partial k_t}{\partial x_j}\right) \right] + \frac{\partial }{\partial x_j}\left( k_u{\color{black}\langle \rho \rangle}\left( \{ V_j \} - \tilde{V}_j\right) \right) \\
&= \frac{\partial}{\partial x_j}\left( \frac{\langle \rho \rangle \nu_u}{\sigma_k}\frac{f_\varepsilon}{f_k^2} \frac{\partial k_u}{\partial x_j}\right) + \frac{\partial }{\partial x_j}\left( k_u{\color{black}\langle \rho \rangle}\left( \{ V_j \} - \tilde{V}_j\right) \right)
\end{split}
\; ,
\end{equation}
where $\nu_u=c_\mu S_u \sqrt{k_u}$. Using scaling arguments, \citeauthor{GIRIMAJI_JAM_2005} \cite{GIRIMAJI_JAM_2005} showed that
\begin{equation}
\label{eq:2.1.1_15}
\frac{\partial }{\partial x_j}\left( k_u{\color{black}\langle \rho \rangle}\left( \{ V_j \} - \tilde{V}_j\right) \right) \approx 0 \; ,
\end{equation}
leading to the so-called zero-transport model (ZTM). The accuracy of this model has been confirmed in the recent study of \citeauthor{TAZRAEI_PRF_2019} \cite{TAZRAEI_PRF_2019}. Also, it is important to highlight that the velocity difference term tends to zero in the limit of $f_k=0.00$ and $1.00$ since
\begin{equation}
\label{eq:2.1.1_16}
 k_u=0\ \mathrm{at} \ f_k=0.0\; , \hspace{5mm} \{ V_j \} - \tilde{V}_j = 0 \ \mathrm{at} \ f_k=1.0\; .
\end{equation}
The derivation of the evolution equation for $k_u$ concludes by combining equations \ref{eq:2.1.1_6}, \ref{eq:2.1.1_14}, and \ref{eq:2.1.1_15}, this leading to its final form,
\begin{equation}
\label{eq:2.1.1_17}
\frac{\partial{\color{black}\langle \rho \rangle} k_u}{\partial t} +   \frac{\partial {\color{black}\langle \rho \rangle} k_u\{V_j \}}{\partial x_j}  =  {\cal{P}}_{b_u} +  {\cal{P}}_{s_u} - {\color{black}{\cal{E}}_u} +\frac{\partial}{\partial x_j}\left( \frac{\langle \rho \rangle \nu_u}{\sigma_k} \frac{f_\varepsilon}{f_k^2} \frac{\partial k_u}{\partial x_j}\right)   \; ,
\end{equation}
where,
\begin{equation}
\label{eq:2.1.1_19}
{\color{black}{\cal{E}}_u =\langle \rho \rangle} \varepsilon_{u} ={\color{black}\langle \rho \rangle } \frac{k_u^{3/2}}{S_u}  \; .
\end{equation}
We recall that the derivation of equation \ref{eq:2.1.1_17} assumes that $f_k$ and $f_\varepsilon$ are constant. If this property does not hold, the model's derivation needs to consider additional terms \cite{GIRIMAJI_JT_2013} and the modeled-to-total ratio of density, $f_\rho$. Despite being commonly neglected, this requirement holds for any bridging and hybrid formulation.
%
%
%
\subsubsection{$S_u$ evolution equation}
\label{sec:2.1.2}

The derivation of the evolution equation for $S_u$ is similar to that for $k_u$. From $f_S$, it is possible to establish the following relationship between the evolution equations for $S_u$ (PANS) and $S_t$ (RANS),
\begin{equation}
\label{eq:2.1.2_1}
\frac{\partial {\color{black}\overline{\rho}} S_u}{\partial t} + \frac{\partial {\color{black}\overline{\rho}} S_u \tilde{V_j}}{\partial x_j} =  f_S \left[ \frac{\partial{\color{black}\overline{\rho}} S_t}{\partial t} + \frac{\partial{\color{black}\overline{\rho}} S_t\tilde{V_j}}{\partial x_j}  \right] \; ,
\end{equation}
which can be rewritten as 
\begin{equation}
\label{eq:2.1.2_2}
\frac{\partial {\color{black}\langle \rho \rangle} S_u}{\partial t} + \frac{\partial {\color{black}\langle \rho \rangle} S_u \{V_j\}}{\partial x_j} \approx  {\color{black}f_\rho}f_S \left[ \frac{\partial {\color{black}\overline{\rho}}S_t}{\partial t} + \frac{\partial {\color{black}\overline{\rho}} S_t \tilde{V_j}}{\partial x_j}  \right] \; ,
\end{equation}
using the zero transport model \cite{GIRIMAJI_JAM_2005}. Now, we replace the material derivative of $S_t$ by the right-hand side of equation \ref{eq:2.1_4},

\begin{equation}
\label{eq:2.1.2_3}
\begin{split}
\frac{\partial{\color{black}\langle \rho \rangle}  {S_u}}{\partial t} + \frac{\partial{\color{black}\langle \rho \rangle}  {S_u}\{V_j\}}{\partial x_j} &= {\color{black}f_\rho}f_S\frac{S_t}{k_t} \left( c_4 {\cal{P}}_{b_t} + c_1 {\cal{P}}_{s_t} \right) - {\color{black}f_\rho}f_S c_2 {\color{black}\overline{\rho}} \sqrt{k_t} \\
 &+ {\color{black}f_\rho}f_S \frac{\partial}{\partial x_j}\left( \frac{\overline{\rho}\nu_t}{\sigma_S} \frac{\partial S_t}{\partial x_j} \right) 
\end{split}
 \; .
\end{equation}
Using the parameters $f_k$ and $f_\varepsilon$, the definition of $f_S$, and relationships \ref{eq:2.1.1_12} and \ref{eq:2.1.1_13}, we get,
\begin{equation}
\label{eq:2.1.2_4}
\begin{split}
\frac{\partial {\color{black}\langle \rho \rangle} S_u}{\partial t} + \frac{\partial {\color{black}\langle \rho \rangle} S_u \{V_j\}}{\partial x_j} &= \frac{\partial}{\partial x_j}\left( \frac{\langle \rho \rangle \nu_u}{\sigma_S} \frac{f_\varepsilon}{f_k^2} \frac{\partial S_u}{\partial x_j} \right)     \\
& - c_2 \frac{f_k}{f_\varepsilon} {\color{black}\langle \rho \rangle}\sqrt{k_u}  \\
& +  \frac{S_u}{k_u} c_4 f_k  \left( \frac{{\cal{P}}_{b_u}}{f_k} - \omega_b {\color{black}\langle \rho \rangle}\frac{k_u^{3/2}}{S_u}\left[\frac{1}{f_k}-\frac{1}{f_\varepsilon}\right] \right)  \\
& +  \frac{S_u}{k_u} c_1 f_k  \left( \frac{{\cal{P}}_{s_u}}{f_k} - \omega_s {\color{black}\langle \rho \rangle}\frac{k_u^{3/2}}{S_u}\left[\frac{1}{f_k}-\frac{1}{f_\varepsilon}\right] \right)  
\end{split}
\; .
\end{equation}
This equation can be rearranged by introducing the coefficient $c_2^*$,
\begin{equation}
\label{eq:2.1.2_5}
\begin{split}
\frac{\partial {\color{black}\langle \rho \rangle} S_u}{\partial t} + \frac{\partial {\color{black}\langle \rho \rangle} S_u \{V_j\}}{\partial x_j} &= \frac{\partial}{\partial x_j}\left( \frac{\langle \rho \rangle \nu_u}{\sigma_S} \frac{f_\varepsilon}{f_k^2} \frac{\partial S_u}{\partial x_j} \right) \\
&  - c_2^*{\color{black}\langle \rho \rangle}\sqrt{k_u} + \frac{S_u}{k_u} \left( c_4 {\cal{P}}_{b_u} + c_1 {\cal{P}}_{s_u} \right)     
\end{split}
\; ,
\end{equation}
\begin{equation}
\label{eq:2.1.2_6}
\begin{split}
c_2^*=c_2 \frac{f_k}{f_\varepsilon} + \left( c_4  \omega_b + c_1  \omega_s \right) \left( 1-\frac{f_k}{f_\varepsilon}\right)  \; .
\end{split}
\end{equation}
%
%
%
\subsubsection{$a_{u_i}$ evolution equation}
\label{sec:2.1.3}

The production terms of $k_u$ and $S_u$ in PANS BHR-LEVM closure require the calculation of the velocity mass flux, $a_{i}$, which is obtained from an additional evolution equation. The derivation of the PANS equation for $a_{i}$ starts by establishing the following relationship,
\begin{equation}
\label{eq:2.1.3_1}
\frac{\partial {\color{black}\overline{\rho}}a_{i_u}}{\partial t} + \frac{\partial {\color{black}\overline{\rho}}a_{i_u}\tilde{V_j}}{\partial x_j} =  f_{a_i} \left[ \frac{\partial {\color{black}\overline{\rho}}a_{i_t}}{\partial t} + \frac{\partial {\color{black}\overline{\rho}}a_{i_t}\tilde{V_j}}{\partial x_j}  \right] \; .
\end{equation}
Following the approach used for $k_u$ and $S_u$ equations, the left-hand side of equation \ref{eq:2.1.3_1} can be approximated as,
\begin{equation}
\label{eq:2.1.3_2}
\frac{\partial {\color{black}\langle \rho \rangle} a_{i_u}}{\partial t} + \frac{\partial {\color{black}\langle \rho \rangle} a_{i_u}\{V_j\}}{\partial x_j} \approx  {\color{black}f_\rho} f_{a_i} \left[ \frac{\partial {\color{black}\overline{\rho}}a_{i_t}}{\partial t} + \frac{\partial {\color{black}\overline{\rho}} a_{i_t}\tilde{V_j}}{\partial x_j}  \right]  \; ,
\end{equation}
using the zero transport model \cite{GIRIMAJI_JAM_2005}. Next, we replace the right-hand side of this equation by that of equation \ref{eq:2.1_5},
\begin{equation}
\label{eq:2.1.3_3}
\begin{split}
\frac{\partial {\color{black}\langle \rho \rangle} a_{i_u}}{\partial t} + \frac{\partial {\color{black}\langle \rho \rangle} a_{i_u}\{V_j\}}{\partial x_j} &=  
{\color{black}f_{\rho}} f_{a_i} \left[ b_t \frac{\partial \overline{P}}{\partial x_i} 
+ R^1(V_i,V_j) \frac{\partial \overline{\rho} }{\partial x_j}\right. \\
&
- {\color{black}\overline{\rho}} a_{j_t} \frac{\partial \overline{V}_i}{\partial x_j} 
+ {\color{black}\overline{\rho}}\frac{\partial \left( a_{i_t}a_{j_t} \right)}{\partial x_j}\\
&
-  c_{a_1}{\color{black}\overline{\rho}}a_{i_t} \frac{\sqrt{k_t}}{S_t} \\
&
\left.
+  \frac{\partial}{\partial x_j}\left( \frac{{\color{black}\overline{\rho}} \nu_t}{ \sigma_a} \frac{\partial a_{i_t}}{\partial x_j} \right) \right]
\end{split}
 \; .
\end{equation}
The final step to derive the $a_{u_i}$ equation is to express the right-hand side of equation \ref{eq:2.1.3_3} in terms of filtered and unresolved quantities. This can be accomplished through the parameters $f_k$, $f_\varepsilon$, $f_{a_i}$, $f_b$, and key closure simplifications,
\begin{equation}
\label{eq:2.1.3_4}
\begin{split}
{\color{black}f_{\rho}} f_{a_i} b_t\frac{\partial \overline{P}}{\partial x_i} &\approx   f_{a_i} \frac{b_u}{f_b  } \frac{\partial \langle P \rangle}{\partial x_i}
\end{split}
\; ,
\end{equation}
\begin{equation}
\label{eq:2.1.3_5}
{\color{black}f_{\rho}}  f_{a_i}  R^1(V_i,V_j) \frac{\partial \overline{\rho} }{\partial x_j}  \approx  \frac{ f_{a_i} }{f_k} \tau^1(V_i,V_j)\frac{\partial \langle \rho \rangle }{\partial x_j} 
\; ,
\end{equation}
\begin{equation}
\label{eq:2.1.3_6}
\begin{split}
{\color{black}f_{\rho}} f_{a_{i}} {\color{black}\overline{\rho}}a_{j_t} \frac{\partial \overline{V}_i }{\partial x_j} &= \frac{f_{a_i}}{f_{a_j}}{\color{black}\langle \rho \rangle} a_{j_u} \frac{\partial \overline{V}_i}{\partial x_j} \\
&= \frac{f_{a_i}}{f_{a_j}} {\color{black}\langle \rho \rangle} a_{j_u} \left[ \frac{\partial \langle V_i \rangle }{\partial x_j} + \left(\frac{\partial \overline{V}_i }{\partial x_j} - \frac{\partial \langle V_i \rangle }{\partial x_j}  \right) \right] \\
& = \frac{f_{a_i}}{f_{a_j}} {\color{black}\langle \rho \rangle}a_{j_u} \frac{\partial \langle V_i \rangle }{\partial x_j} \ \ \ \ \ (\mathrm{ZTM})
\end{split}
\; ,
\end{equation}
\begin{equation}
\label{eq:2.1.3_7}
{\color{black}f_{\rho}} f_{a_i} {\color{black}\overline{\rho}} \frac{\partial \left( a_{i_t}a_{j_t} \right)}{\partial x_j} = \frac{{\color{black}\langle \rho \rangle}}{f_{a_{j}}}  \frac{\partial \left( a_{i_u}a_{j_u} \right)}{\partial x_j}\; ,
\end{equation}
\begin{equation}
\label{eq:2.1.3_8}
{\color{black}f_{\rho}} f_{a_i} c_{a_1}{\color{black}\overline{\rho}} a_{i_t} \frac{\sqrt{k_t}}{S_t} =  c_{a_1} {\color{black}\langle \rho \rangle} a_{i_u} \frac{\sqrt{k_u}}{S_u}\frac{f_k}{f_\varepsilon} \; ,
\end{equation}
\begin{equation}
\label{eq:2.1.3_9}
{\color{black}f_{\rho}} f_{a_i} \frac{\partial}{\partial x_j}\left( \frac{{\color{black}\overline{\rho}} \nu_t}{\sigma_a} \frac{\partial a_{i_t}}{\partial x_j} \right) =   \frac{\partial}{\partial x_j}\left( \frac{{\color{black}\langle \rho \rangle}\nu_u}{\sigma_a} \frac{f_\varepsilon}{f_k^2} \frac{\partial a_{i_u}}{\partial x_j} \right) \; .
\end{equation}
These six terms allow us to rearrange equation \ref{eq:2.1.3_3} and obtain its final form,
\begin{equation}
\label{eq:2.1.3_10}
\begin{split}
\frac{\partial {\color{black}\langle \rho \rangle}a_{i_u}}{\partial t} + \frac{\partial{\color{black}\langle \rho \rangle} a_{i_u}\{V_j\}}{\partial x_j} &= f_{a_i} \frac{b_u}{f_b} \frac{\partial \langle P \rangle}{\partial x_i}  \\
& + \frac{ f_{a_i} }{f_k} \tau^1(V_i,V_j) \frac{\partial \langle \rho \rangle }{\partial x_j} \\ 
& - \frac{f_{a_i}}{f_{a_j}} {\color{black}\langle \rho \rangle}a_{j_u} \frac{\partial \langle V_i \rangle }{\partial x_j} + \frac{{\color{black}\langle \rho \rangle}}{f_{a_{j}}} \frac{\partial \left( a_{i_u}a_{j_u} \right)}{\partial x_j} \\
& - c_{a_1} {\color{black}\langle \rho \rangle}a_{i_u} \frac{\sqrt{k_u}}{S_u}\frac{f_k}{f_\varepsilon} \\
&+ \frac{\partial}{\partial x_j}\left( \frac{{\color{black}\langle \rho \rangle}\nu_u}{\sigma_a} \frac{f_\varepsilon}{f_k^2} \frac{\partial a_{i_u}}{\partial x_j} \right)
\end{split}
\; .
\end{equation}
%
%
%
\subsubsection{$b_u$ evolution equation}
\label{sec:2.1.4}
The derivation of PANS BHR-LEVM closure concludes with the evolution equation for the unresolved density-specific volume correlation, $b_u$. Once again, we start by establishing the following relationship using the parameter $f_b$,
\begin{equation}
\label{eq:2.1.4_1}
\frac{\partial {\color{black}\overline{\rho}} b_u}{\partial t} + \frac{\partial {\color{black}\overline{\rho}} b_u\tilde{V_j}}{\partial x_j} =  f_b \left[ \frac{\partial {\color{black}\overline{\rho}} b_t}{\partial t} + \frac{\partial {\color{black}\overline{\rho}} b_t\tilde{V_j}}{\partial x_j}  \right] \; ,
\end{equation}
which, using the ZTM model \cite{GIRIMAJI_JAM_2005}, can be simplified as follows,
\begin{equation}
\label{eq:2.1.4_2}
\frac{\partial{\color{black}\langle \rho \rangle} b_u}{\partial t} + \frac{\partial {\color{black}\langle \rho \rangle} b_u\{V_j\}}{\partial x_j} \approx  {\color{black}f_\rho} f_b \left[ \frac{\partial {\color{black}\overline{\rho}} b_t}{\partial t} + \frac{\partial {\color{black}\overline{\rho}} b_t\tilde{V_j}}{\partial x_j}  \right] \; .
\end{equation}
Replacing the term between brackets by the right-hand side of equation \ref{eq:2.1_6},
\begin{equation}
\label{eq:2.1.4_3}
\begin{split}
\frac{\partial{\color{black}\langle \rho \rangle} b_u}{\partial t} + \frac{\partial {\color{black}\langle \rho \rangle} b_u\{V_j\}}{\partial x_j} &=   
{\color{black}f_\rho} f_b \left[2 {\color{black}\overline{\rho}}a_{j_t} \frac{\partial b_t}{\partial x_j}  - 2a_{j_t}\left(b_t+1 \right) \frac{\partial \overline{\rho} }{\partial x_j}   \right. \\
&\left. -  c_{b}{\color{black}\overline{\rho}}b_t \frac{\sqrt{k_t}}{S_t} + f_b \overline{\rho}{\color{black}^2}\frac{\partial}{\partial x_j}\left( \frac{\nu_t}{ \overline{\rho} \sigma_b} \frac{\partial b_t}{\partial x_j} \right) \right]
%
\end{split}
\; ,
\end{equation}
and converting total to partial quantities using $f_\phi$, we get the final form of the $b_u$ equation,
\begin{equation}
\label{eq:2.1.4_4}
\begin{split}
\frac{\partial{\color{black}\langle \rho \rangle} b_u}{\partial t} + \frac{\partial {\color{black}\langle \rho \rangle} b_u\{V_j\}}{\partial x_j}&= 2{\color{black}\langle \rho \rangle}\frac{{a_{j_u}}}{f_{a_j}} \frac{\partial b_u}{\partial x_j} - 2\frac{a_{j_u}}{f_{a_j}} \left( b_u + f_b \right) \frac{\partial \langle \rho \rangle }{\partial x_j} \\
& - c_{b}{\color{black}\langle \rho \rangle}b_u \frac{f_k}{f_\varepsilon} \frac{\sqrt{k_u}}{S_u}  \\
&+ \langle \rho \rangle{\color{black}^2}\frac{\partial}{\partial x_j}\left( \frac{\nu_u}{ \langle \rho \rangle \sigma_b}\frac{f_\varepsilon}{f_k^2}  \frac{\partial b_u}{\partial x_j} \right)
\end{split}
\; .
\end{equation}
Thus, the PANS BHR-LEVM closure is composed by equations \ref{eq:2.1.1_17}, \ref{eq:2.1.2_5}, \ref{eq:2.1.3_10}, and \ref{eq:2.1.4_4}. Note that the PANS BHR-LEVM closure recovers its original RANS form when all $f_\phi$ are equal to unity.
%
%
\section{Filter control parameter}
\label{sec:3}

The efficiency of bridging and hybrid formulations is determined by the degree of physical resolution. As the model resolves a wider range of flow scales, both the cost and accuracy of the simulations are expected to grow. Whereas excessive physical resolution reduces the computational efficiency by increasing the cost without commensurate improvement in accuracy, insufficient resolution can compromise accuracy by precluding the model from resolving the scales not amenable to modeling \cite{PEREIRA_JCP_2018,PEREIRA_IJHFF_2019,PEREIRA_OE_2019,PEREIRA_PRF_2021}. Hence, the success of such SRS methods is dictated by the parameters controlling their physical resolution. 

As discussed in \cite{PEREIRA_PRF_2021}, there are three main factors to consider when determining the physical resolution needed for a given model, flow configuration, and quantities of interest:
\begin{itemize}
\item[$i$)] the length- and time-scales that need to be resolved;
\item[$ii)$] the smallest flow scales that the selected spatio-temporal grid resolution and numerical setup can accurately resolve;
\item[$iii)$] the effect of the physical resolution on the dependent variables of the turbulence closure.
\end{itemize}
The authors have recently investigated the first point through the analysis of flows around cylinders and the Taylor-Green vortex \cite{PEREIRA_JCP_2018,PEREIRA_IJHFF_2019,PEREIRA_OE_2019,PEREIRA_PRF_2021}. These studies have shown that the accurate prediction of these complex problems is determined by the mathematical model's ability to resolve the instabilities and coherent structures governing the flow physics. This flow physics is dominated by non-local effects, which most one-point closures cannot represent accurately. Thus, accurate computations of such problems require resolving the Kelvin-Helmholtz rollers observed in flows past cylinders in the sub-critical regime \cite{WILLIAMSON_ARFM_1996,ZDRAVKOVICH_BOOK_1997}, and the vortex-reconnection process of the TGV. The remaining flow scales can be accurately modeled through an adequate turbulence closure model. These studies also illustrate the importance of understanding the flow physics and its main features to select the physical resolution and obtain high-fidelity solutions.

The second aspect is ideally addressed through verification exercises \cite{ROACHE_BOOK_1998,TRUCANO_SANDIA_2002,OBERKAMPF_BOOK_2010}. We have illustrated the crucial role of verification of RANS and SRS predictions in \cite{PEREIRA_IJHFF_2018,PEREIRA_PHD_2018,PEREIRA_OE_2019,PEREIRA_ACME_2021,PEREIRA_PRF_2021}. It is possible to obtain a reasonable \textit{a-priori} estimate of the maximum physical resolution that a given grid can support and, conversely, the dependency of the grid requirements from the physical resolution. Irrespective of the numerical scheme, using Kolmogorov arguments, and assuming high-Re flow ($f_\varepsilon=1.0$), it is possible to obtain an expression to estimate the smallest value of $f_k$ that a spatial grid resolution can accurately resolve \cite{PEREIRA_PRF_2021},
\begin{equation}
\label{eq:3_1}
f_k \ge \left(\frac{1}{c_\mu}\right)^{1/2}\left( \frac{\Delta}{k_t^{1.5}/\varepsilon_t} \right)^{2/3} \; .
\end{equation}
This expression can be rearranged to provide the ratio between the smallest grid size for two values of $f_k$,
\begin{equation}
\label{eq:3_2}
r_\Delta = \frac{\Delta_{f_k}}{\Delta _{(f_k)_\mathrm{ref}}}= \left( \frac{f_k}{(f_k)_\mathrm{ref}}\right)^{3/2} \; ,
\end{equation}
where $\Delta$ is the grid resolution or size, and the subscript ``ref'' denotes a reference $f_k$ ($f_{k_\mathrm{ref}} > 0$). This expression enables the evaluation of the relative evolution of $\Delta$ with the physical resolution. Note that it possible to use different arguments and expressions to perform this simple, \textit{a-priori}, and qualitative assessment of the impact of $f_k$ on the grid requirements of the model.

\begin{figure}[t!]
\centering
\includegraphics[scale=0.105,trim=0 0 0 0,clip]{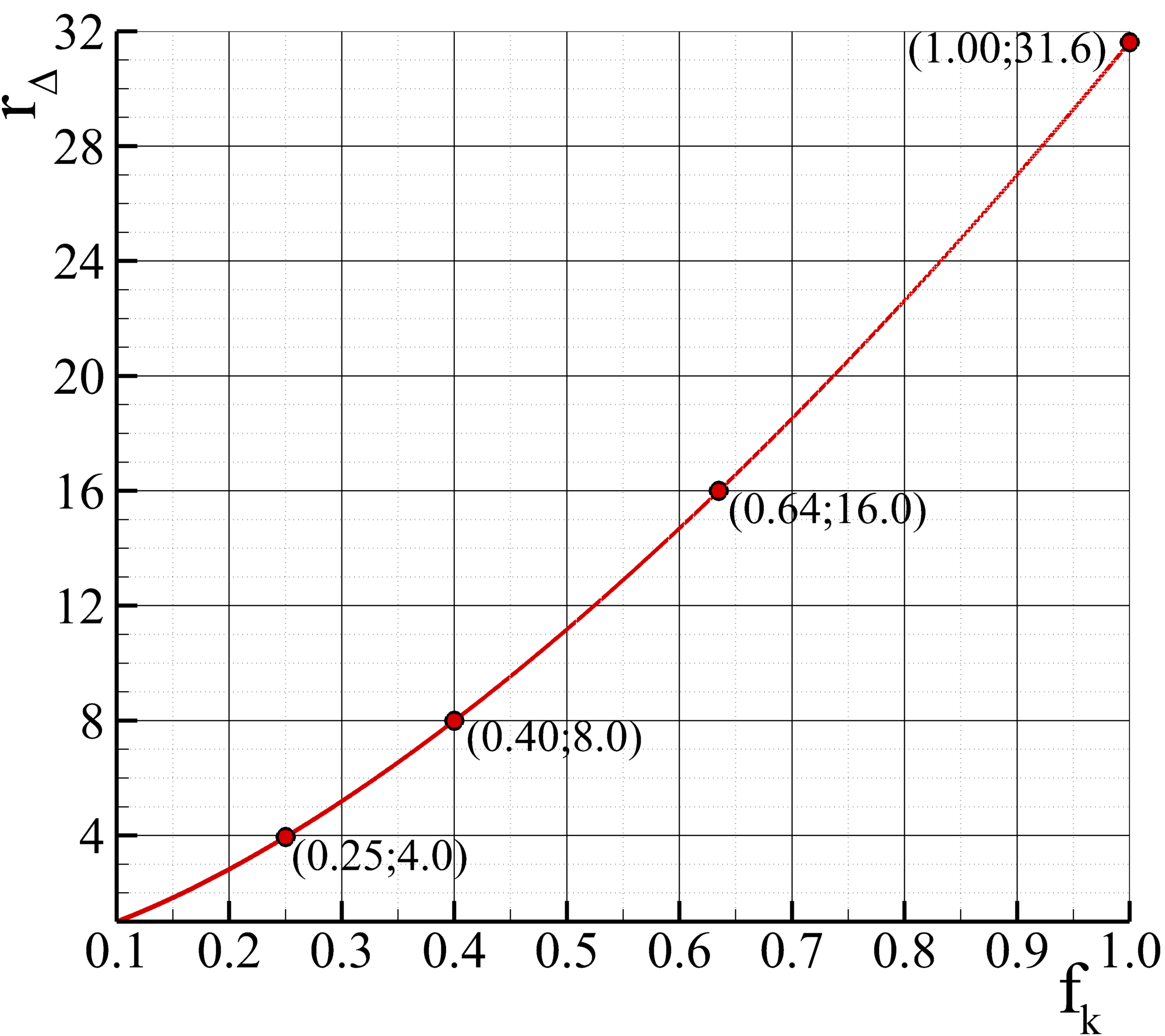}
\caption{Ratio between the minimum spatial grid resolution needed for computations at a given $f_k$ and $(f_k)_\mathrm{ref}=0.10$, $r_\Delta$ \cite{PEREIRA_PRF_2021}.}
\label{fig:3_1}
\end{figure}

Figure \ref{fig:3_1} depicts the evolution of $\Delta$ with $f_k$ relative to the case $(f_k)_\mathrm{ref}=0.10$. The results show the close dependence between the grid and physical ($f_k$) resolutions. As $f_k$ increases, the minimum grid resolution coarsens as $(10 f_k)^{1.5}$. Considering the cases of $f_k=0.25$ and $0.40$, this represents reducing the grid resolution requirements by a factor of four and eight when compared to the case at $f_k=0.10$. Since SRS computations are inherently three dimensional and unsteady, figure \ref{fig:3_1} confirms the potential of bridging methods to predict complex flows efficiently. It also emphasizes the importance of selecting an adequate physical resolution for a given problem and quantities of interest. 

The third factor is caused by the fact that an SRS model's physical resolution does not affect all turbulence dependent quantities equally, i.e., a given range of resolved scales does not lead to equal ratios modeled-to-total for all  dependent variables of the closure model. For instance, it is not expected that the turbulence kinetic energy and dissipation possess similar spectral signatures in a fully-developed turbulent flow \cite{POPE_BOOK_2000,DAVIDSON_BOOK_2006}. This has been recently addressed in \citeauthor{PEREIRA_PRF_2021} \cite{PEREIRA_PRF_2021}. We emphasize that despite most SRS models neglecting this aspect, PANS can consider the spectral signature of each dependent quantity of the closure through $f_\phi$. This benefit comes at the expense of having to determine the other control parameters in such a manner that consistency between the various physical processes can be preserved. 

PANS BHR-LEVM model relies on $f_k$, $f_\varepsilon$, $f_{a_i}$, and $f_b$ to set the physical resolution. The parameter $f_k$ can be either set constant \cite{GIRIMAJI_JAM_2005,PEREIRA_IJHFF_2018,GIRIMAJI_AIAA43_2005,LAKSHMIPATHY_JFE_2010} or dynamically \cite{GIRIMAJI_AIAA43_2005,ELMILIGUI_AIAA22_2004,BASARA_IJHFF_2018,DAVIDSON_JOT_2019} in space and time. Although the second approach may enhance the simulation's efficiency, we choose using constant values of $f_k$ to prevent commutation errors \cite{HAMBA_PF_2011,GIRIMAJI_JT_2013} and enable robust verification and validation exercises where one can evaluate numerical and modeling errors separately to avoid possible error canceling \cite{PEREIRA_JCP_2021}. Regarding $f_\varepsilon$, this parameter is commonly defined constant and equal to one. This modeling assumption stems from the fact that most turbulence dissipation in high-Re flows occurs at the smallest scales \cite{POPE_BOOK_2000,DAVIDSON_BOOK_2006}. For this reason, $f_\varepsilon=1.00$ is often used in practical PANS simulations. The validity of this option has been recently confirmed by the authors \cite{PEREIRA_PRF_2021}, and it is discussed in Section \ref{sec:3.1.1}. The remaining parameters, $f_{a_i}$ and $f_b$, have never been used before and so their definition needs to be investigated, Section \ref{sec:3.1.2}. 

\begin{figure}[t!]
\centering
\subfloat[FHIT - $V_1(\mathbf{x})$]{\label{fig:3_2a}
\includegraphics[scale=0.13,trim=0 0 190 0,clip]{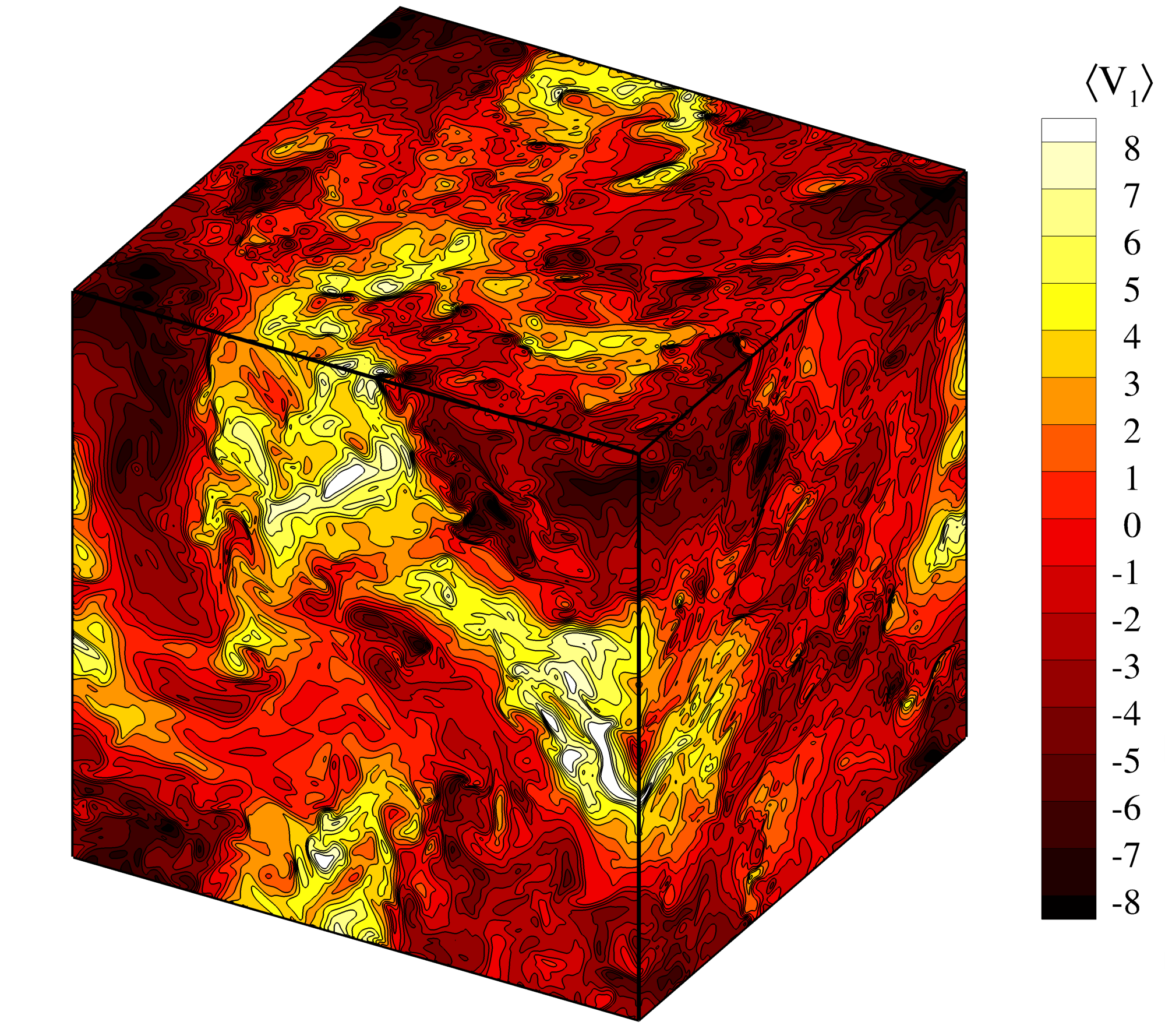}}
\\
\subfloat[HVDT - $\rho(\mathbf{x},t_o)$]{\label{fig:3_2b}
\includegraphics[scale=0.13,trim=0 0 190 0,clip]{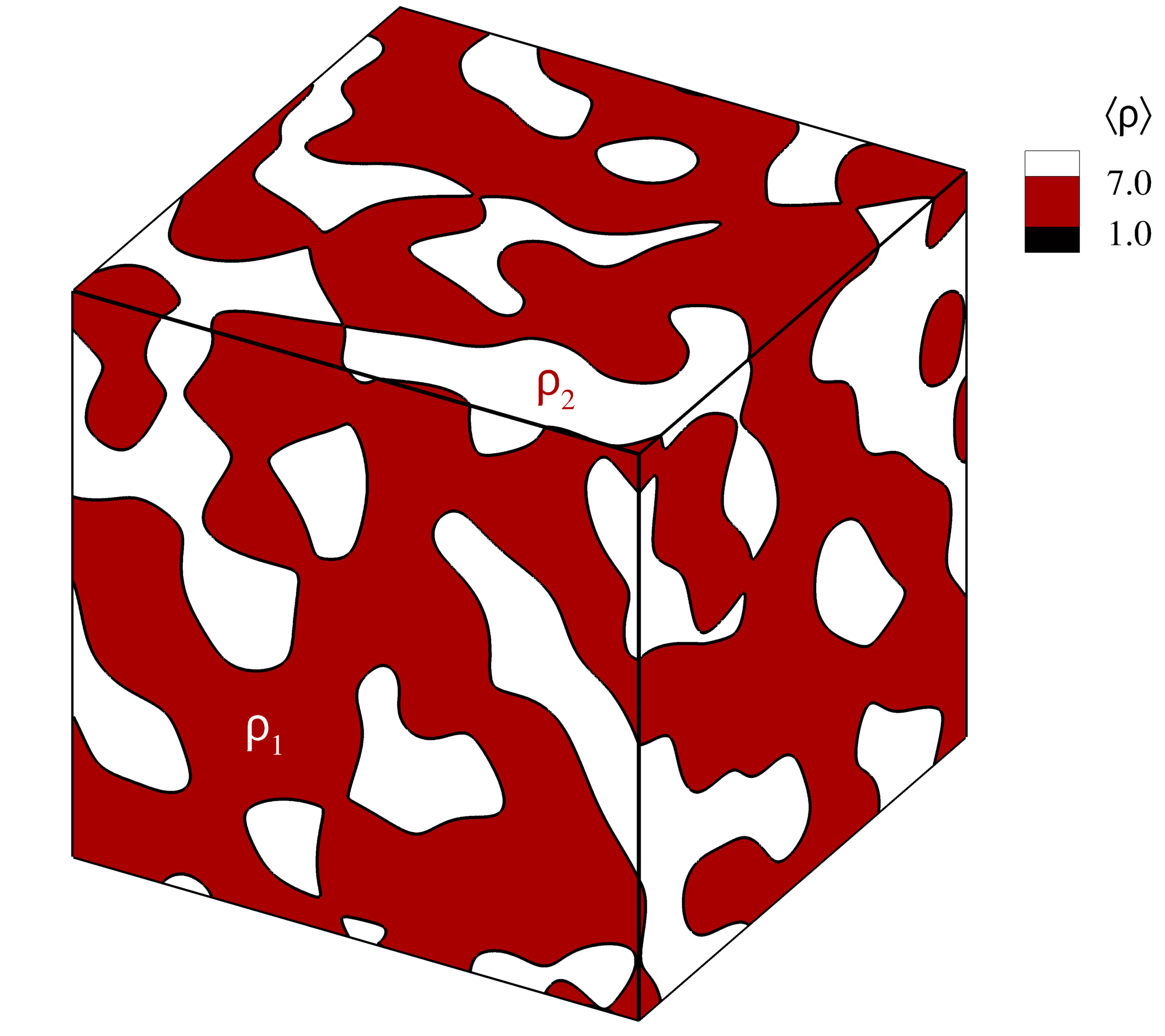}}
\caption{FHIT and HVDT DNS velocity and density ($t=0$) flow fields.}
\label{fig:3_2}
\end{figure}
\begin{figure*}[t]
\centering
\subfloat[$n=1$]{\label{fig:3_3a}
\includegraphics[scale=0.073,trim=0 0 0 0,clip]{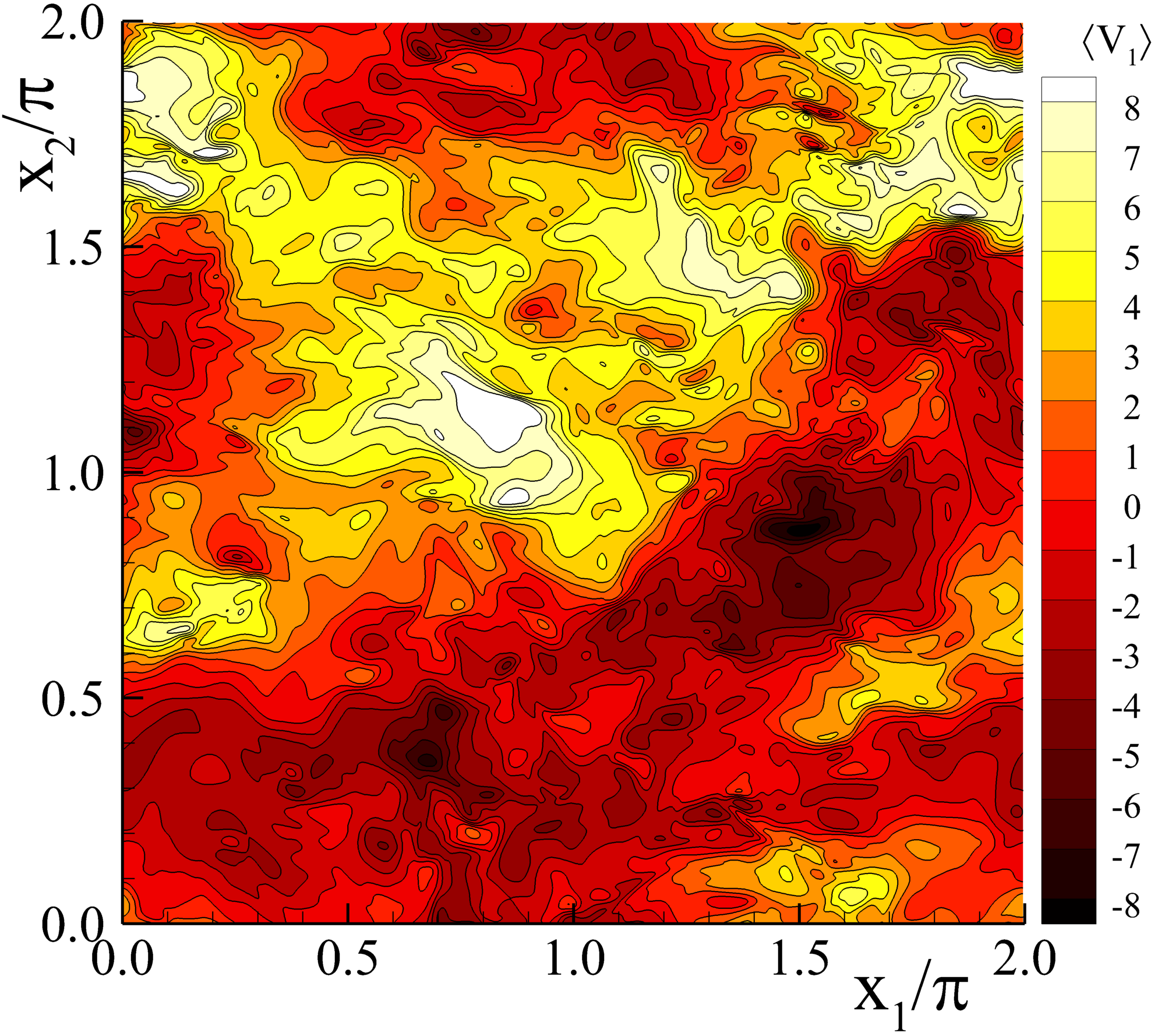}}
~
\subfloat[$n=17$]{\label{fig:3_3b}
\includegraphics[scale=0.073,trim=0 0 0 0,clip]{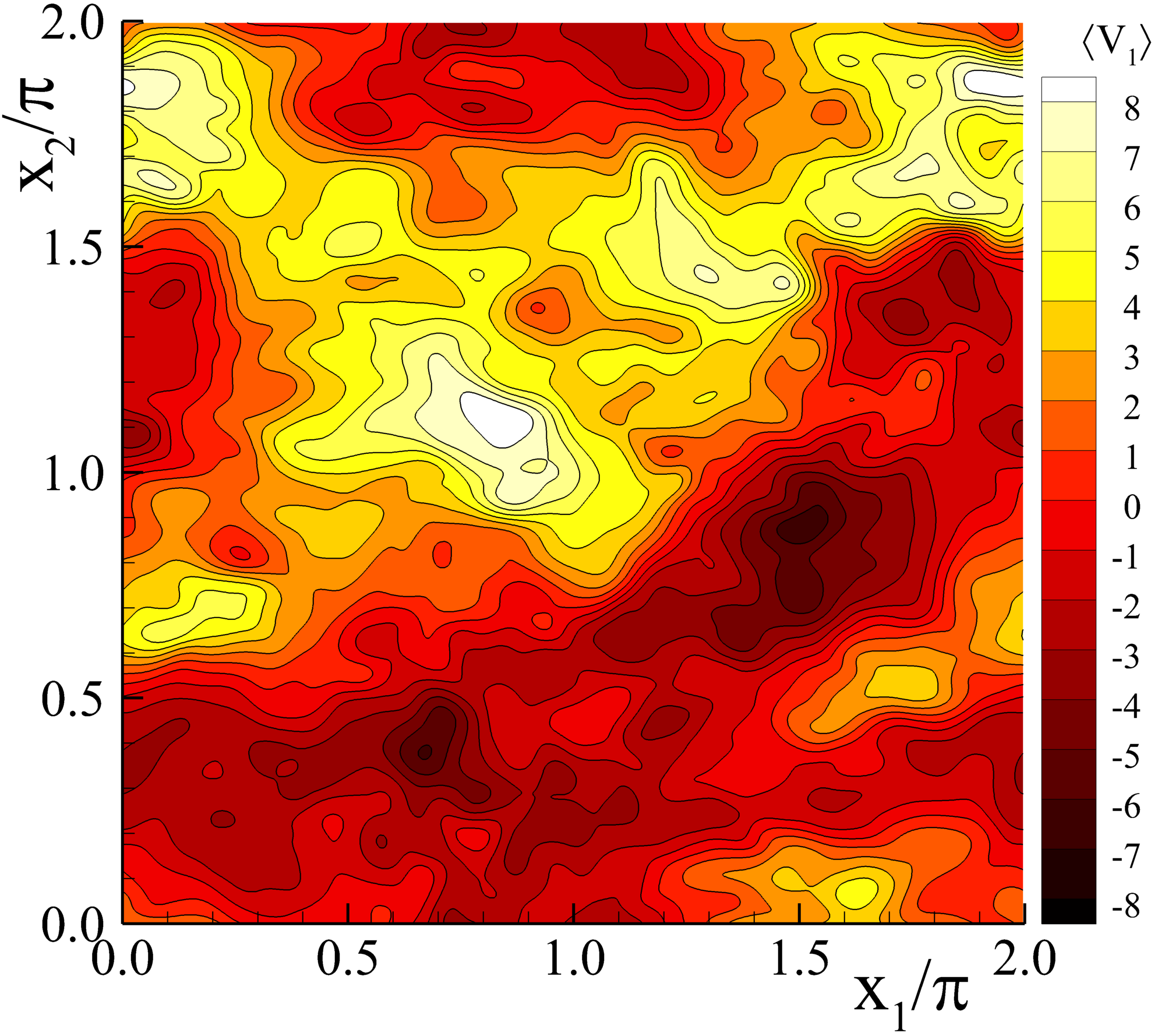}}
~
\subfloat[$n=33$]{\label{fig:3_3c}
\includegraphics[scale=0.073,trim=0 0 0 0,clip]{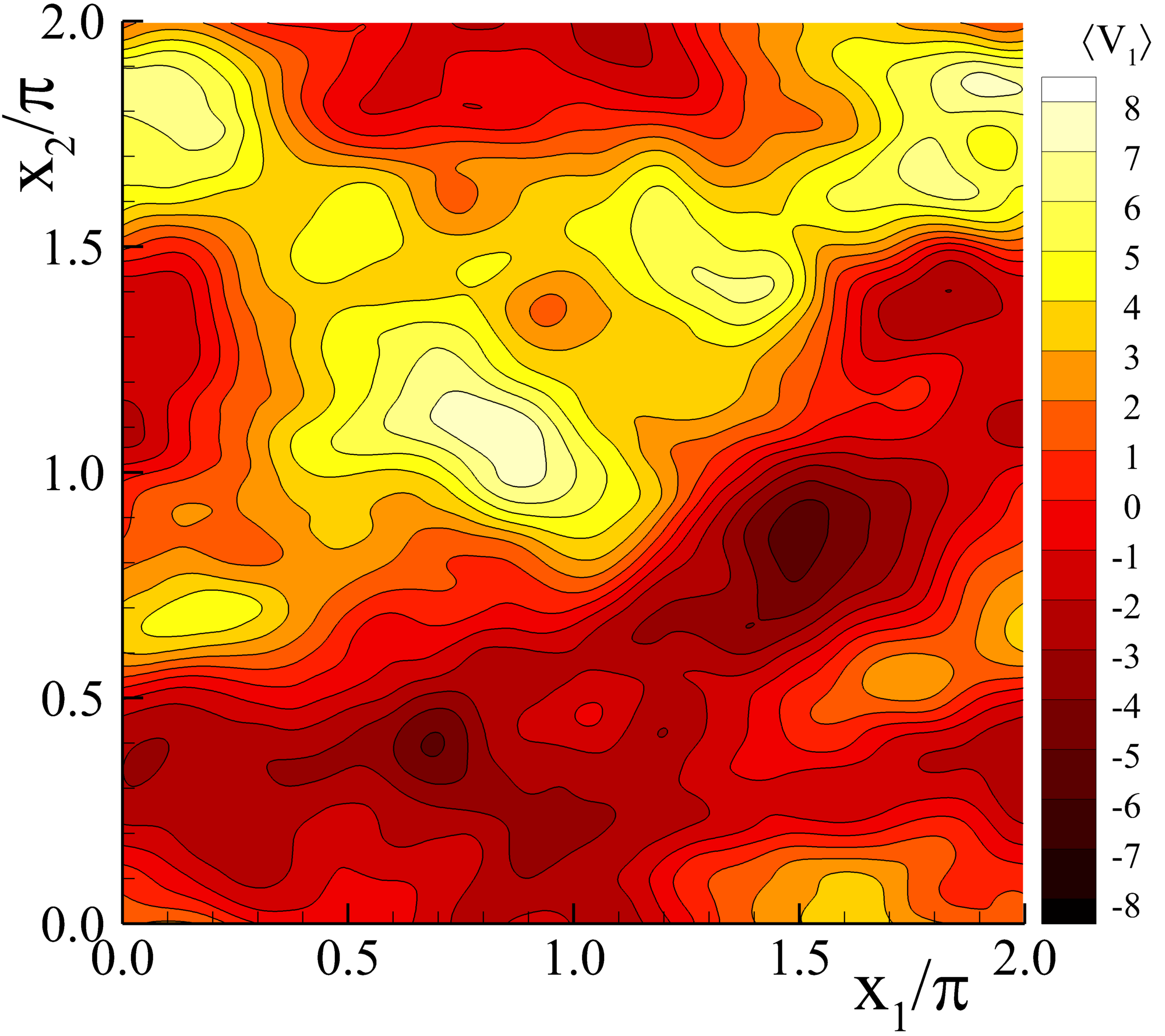}}
\\
\subfloat[$n=69$]{\label{fig:3_3d}
\includegraphics[scale=0.073,trim=0 0 0 0,clip]{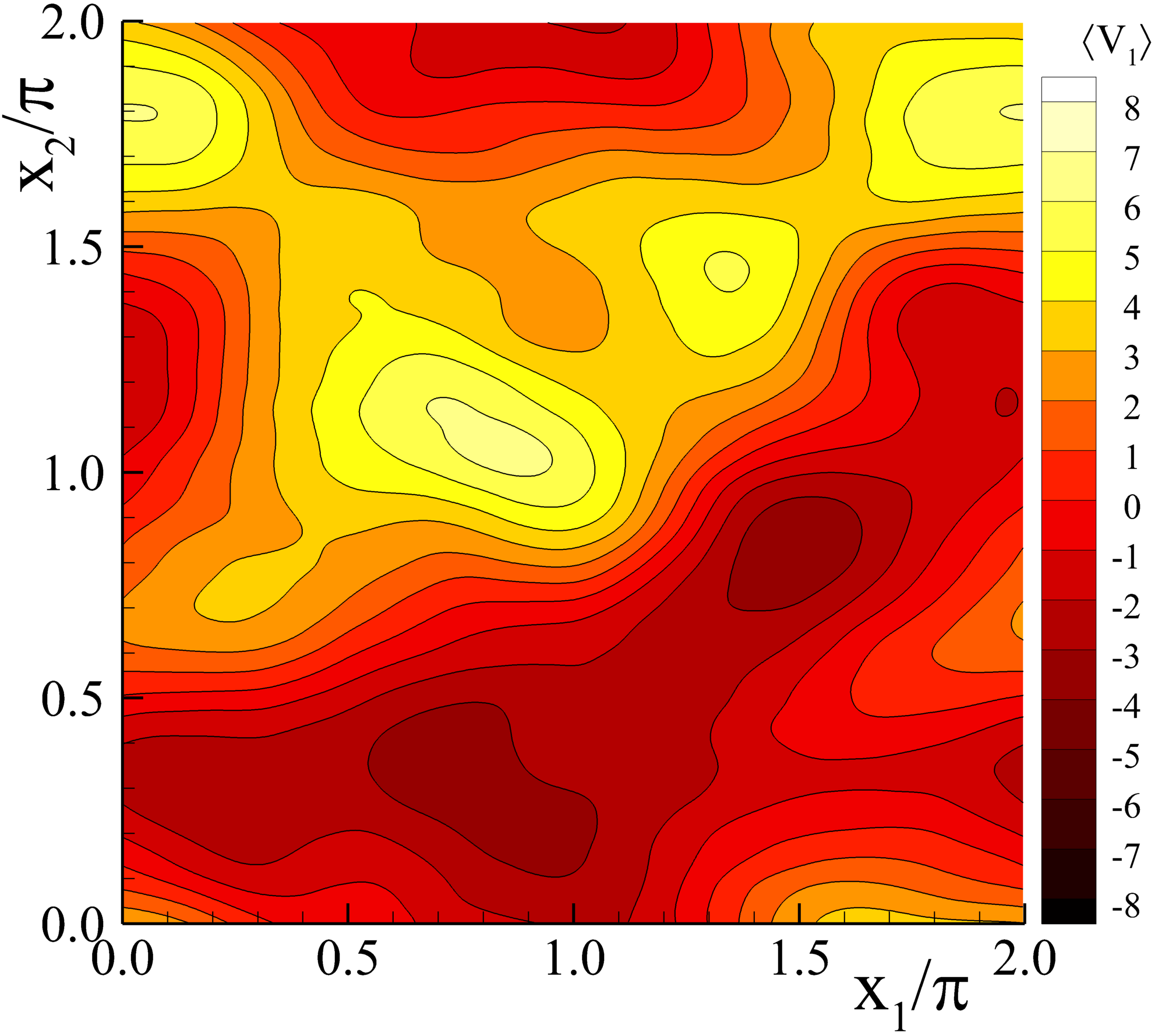}}
~
\subfloat[$n=199$]{\label{fig:3_3e}
\includegraphics[scale=0.073,trim=0 0 0 0,clip]{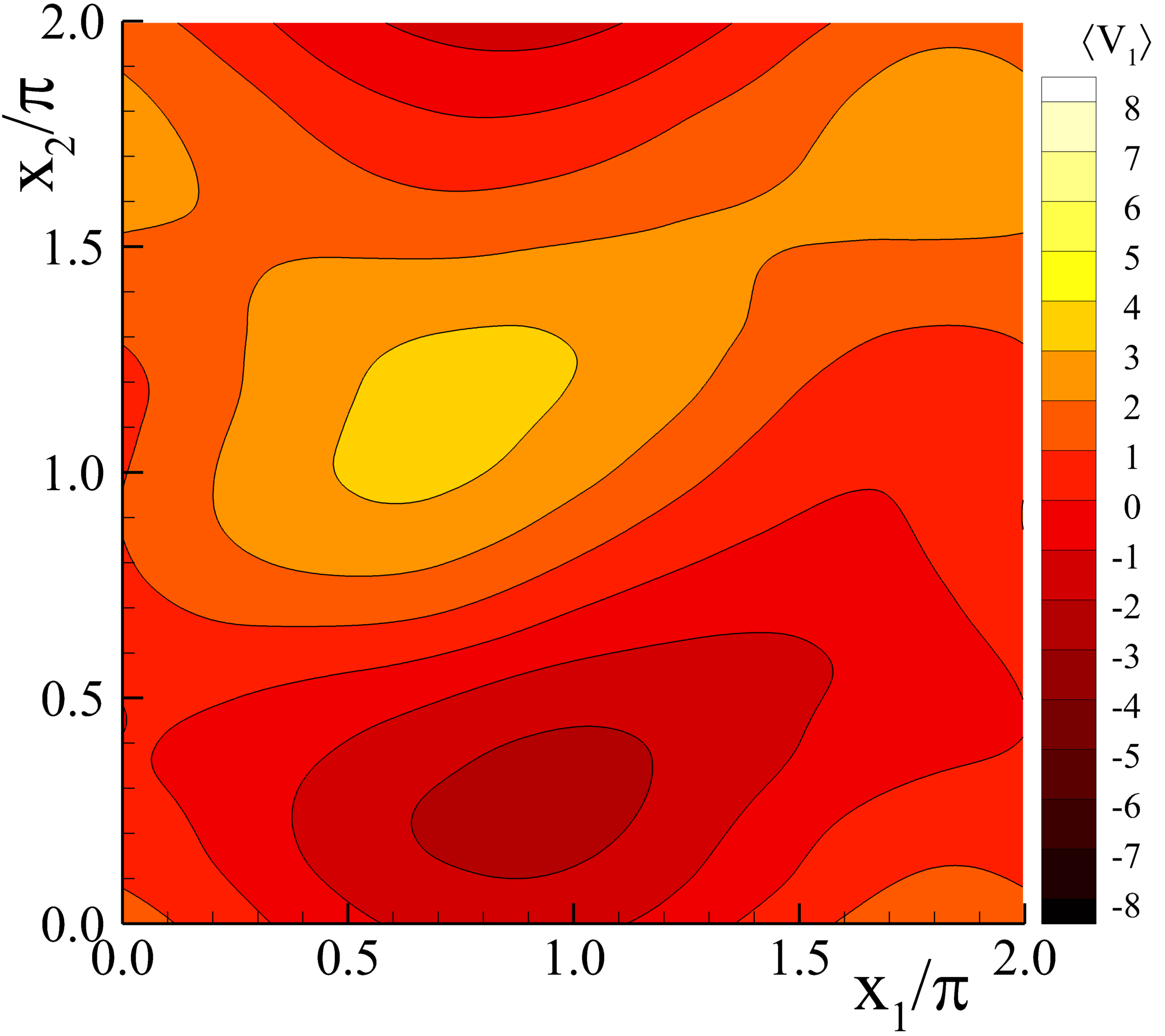}}
~
\subfloat[$n=349$]{\label{fig:3_3f}
\includegraphics[scale=0.073,trim=0 0 0 0,clip]{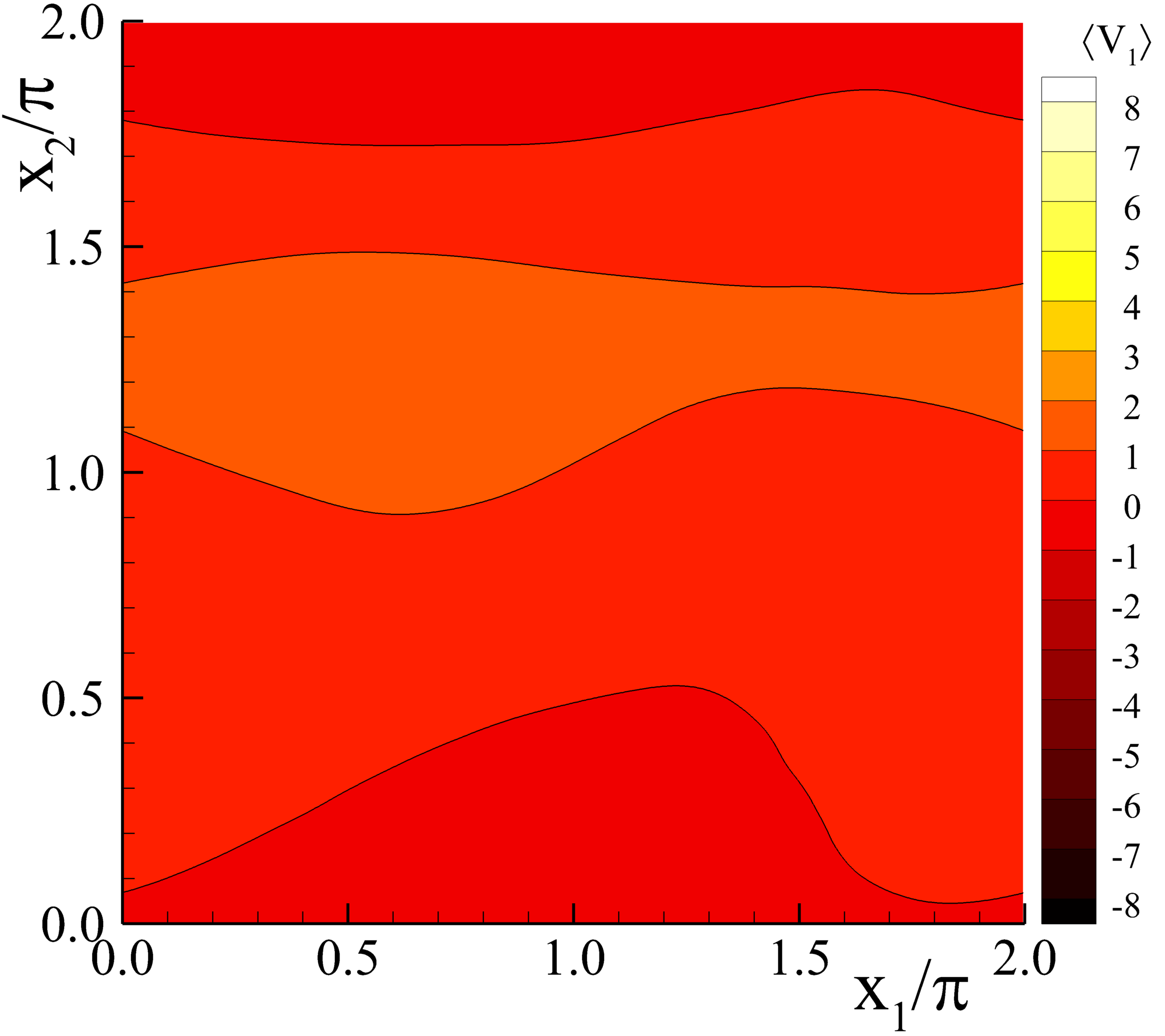}}
\caption{Evolution of $x_1$-component of the FHIT velocity field with the relative filter length size, $n=\Delta/\Delta_\eta$.}
\label{fig:3_3}
\end{figure*}

The remainder of this section addresses the specification of the parameters $f_\phi$ in PANS BHR-LEVM, in a manner that is consistent with the implicit-filter corresponding to $f_k$. Unfortunately, canonical turbulence theories cannot be used for this purpose. Instead, this objective is accomplished through \textit{a-priori} testing, in which the parameters $f_\phi$ are calculated at successively smaller physical resolutions - from $f_k=0.00$ to $f_k=1.00$. The selected canonical flows are the forced homogeneous isotropic turbulence (FHIT) of \citeauthor{SILVA_JFM_2018} \cite{SILVA_JFM_2018} at Taylor Reynolds numbers $\mathrm{Re}_\lambda=140$ and $300$, and the buoyancy driven homogeneous variable-density turbulence (HVDT \cite{BATCHELOR_JFM_1992,SANDOVAL_IUTAM_1997,LIVESCU_JFM_2007,LIVESCU_JFM_2008}) of \citeauthor{ASLANGIL_PD_2020} \cite{ASLANGIL_PD_2020,ASLANGIL_JFM_2020} at At=$0.75$. These flow problems have been simulated by means of DNS in a cubical domain of length $2\pi$, and their streamwise velocity (FHIT) and density (HVDT) fields are depicted in figure \ref{fig:3_2}. All computations were performed in a $1024^3$ mesh, except the FHIT at $\mathrm{Re_\lambda=140}$ which used a $512^3$ grid. The comprehensive description of these data sets is given in \citeauthor{SILVA_JFM_2018} \cite{SILVA_JFM_2018}, \citeauthor{LIVESCU_JFM_2007} \cite{LIVESCU_JFM_2007,LIVESCU_JFM_2008}, and \citeauthor{ASLANGIL_PD_2020} \cite{ASLANGIL_PD_2020,ASLANGIL_JFM_2020}.

The ratios modeled-to-total, $f_\phi$, of the dependent quantities of PANS BHR-LEVM at distinct filter's cut-off are computed as follows. The DNS flow fields are filtered using the operator \cite{SILVA_PHD_2001,SILVA_JT_2008},
\begin{equation}
\label{eq:3_3}
\langle \Phi  \rangle (\mathbf{x})= \int_{-\Delta/2}^{+\Delta/2}\int_{-\Delta/2}^{+\Delta/2}\int_{-\Delta/2}^{+\Delta/2} \Phi(\mathbf{x}) \ G_\Delta(\mathbf{x}-\mathbf{x'})d\mathbf{x'} \; ,
\end{equation}
where bold symbols denote vectors, $\Delta$ is the filter's width, and $G_\Delta$ is the kernel of the filtering operator.  Here, we use a box filter so that,
\begin{equation}
\label{eq:3_4}
G_\Delta(\mathbf{x}-\mathbf{x'})= 
\left\{
\begin{array}{ll}
\Delta^{-1}\; &,\  |\mathbf{x-x'}|< 0.5\Delta \\
0\; &, \  \text{otherwise}
\end{array}\right.
\; .
\end{equation}
Figure \ref{fig:3_3} illustrates the effect of varying this operator filter's width on the velocity field of the FHIT flow. As the filter's width increases, the magnitude of the filtered turbulent velocity field asymptotes to zero, and its gradients get smoother. This is the reasoning for the cost reduction observed from DNS to RANS \cite{PEREIRA_ACME_2021}. It is important to note that the shape of the filter implied by the PANS decomposition is not known, however the utilization of a box filter is not expected to alter the conclusions of this class of studies \cite{SILVA_POF_2007,BORUE_JFM_1998,LIU_JFM_1994} (see Section \ref{sec:3.1.2} for HVDT). This idea is confirmed by comparing our HVDT results with those of \citeauthor{SAENZ_POF_2021} \cite{SAENZ_POF_2021} obtained with a Gaussian filter. Both studies lead to similar qualitative conclusions. In contrast, cut-off filters are not suitable for this exercise since practical SRS computations do not rely on such operators \cite{VREMAN_JFM_1994,SILVA_POF_2007}. It is also generally accepted that the box filtering operator is the closest approach to the implicit filtering of finite-difference and finite-volume discretization schemes utilized in engineering computations \cite{SILVA_JT_2008,SCHUMANN_JCP_1975,ROGALLO_ARFM_1984}. For all these reasons, the use of a box filter operator is not expected to affect the qualitative conclusions of the present analysis.

For any given filter $\{\ \cdot \ \}$ or $\langle \ \cdot \ \rangle$ chosen, the unresolved dependent variables of the BHR-LEVM closure are calculated from relations \ref{eq:2_14} and \ref{eq:2_15} \cite{MOIN_PFA_1991,GERMANO_JFM_1992,SUMAN_FTC_2010}, 
\begin{equation}
\label{eq:3_5}
k_u=0.5 \left(  \{V_iV_i\}-\{V_i\}\{V_i\} \right)   \; ,
\end{equation}
\begin{equation}
\label{eq:3_6}
\varepsilon_u=\nu  \left( \left\{ \frac{\partial V_i }{\partial x_j} \frac{\partial V_i }{\partial x_j} \right\} - \left\{\frac{ \partial  V_i  }{\partial x_j}\right\}  \left\{\frac{ \partial V_i  }{\partial x_j}\right\}\right)  \; ,
\end{equation}
\begin{equation}
\label{eq:3_7}
a_{i_u}= \frac{ \langle \rho' v_i' \rangle - \langle \rho' \rangle \langle  v_i'  \rangle}{\langle \rho \rangle} \; ,
\end{equation}
\begin{equation}
\label{eq:3_8}
b_u = \langle \rho' \rangle \langle (1/\rho)'\rangle - 1\; .
\end{equation}
It is crucial to emphasize that $\rho'$ and $v_i'$ in equations \ref{eq:3_7} and \ref{eq:3_8} consider the fluctuating component of the density and velocity fields, i.e., these quantities can comprise both the coherent and stochastic fields \cite{HUSSAIN_JFM_1970,SCHIESTEL_PF_1987}. Yet, the stochastic field is expected to be the main contributor to $a_{u_i}$ and $b_u$ at late times when the flow exhibits fully-developed and high-intensity turbulence features. The quantities given by relations \ref{eq:3_5} to \ref{eq:3_8} are calculated with $n=\Delta/\Delta_\eta$ up to 349, being $\Delta_\eta$ the grid size used in the DNS simulations. Note that due to the objective and computational cost of these exercises, only the FHIT case at $\mathrm{Re_\lambda=140}$ is filtered until $f_k\approx 1$. This study is performed with the code used in \cite{PEREIRA_PHD_2018,PEREIRA_ACME_2021}.

The outcome of the \textit{a-priori} exercises is now discussed in Section \ref{sec:3.1}. However, before presenting the results, note that the FHIT problem is an archetypal problem widely utilized to investigate the dynamics and modeling of fully-developed incompressible turbulence. For this reason, we use this flow to analyze the dependence of $f_k$ and $f_\varepsilon$ on the range of resolved scales \cite{PEREIRA_PRF_2021}. On the other hand, the HVDT flow is a canonical problem used to study the fundamental physics and modeling of variable-density flow. Hence, we use the HVDT case to investigate the evolution of $f_k$, $f_{a_i}$, and $f_b$ with the physical resolution.
%
%
%
\subsection{\textit{A-priori} testing results}
\label{sec:3.1}
\subsubsection{Forced homogeneous isotropic turbulence}
\label{sec:3.1.1}
\begin{figure}[t!]
\centering
\subfloat[$f_k(n)$]{\label{fig:3.1.1_1a}
\includegraphics[scale=0.095,trim=0 0 0 0,clip]{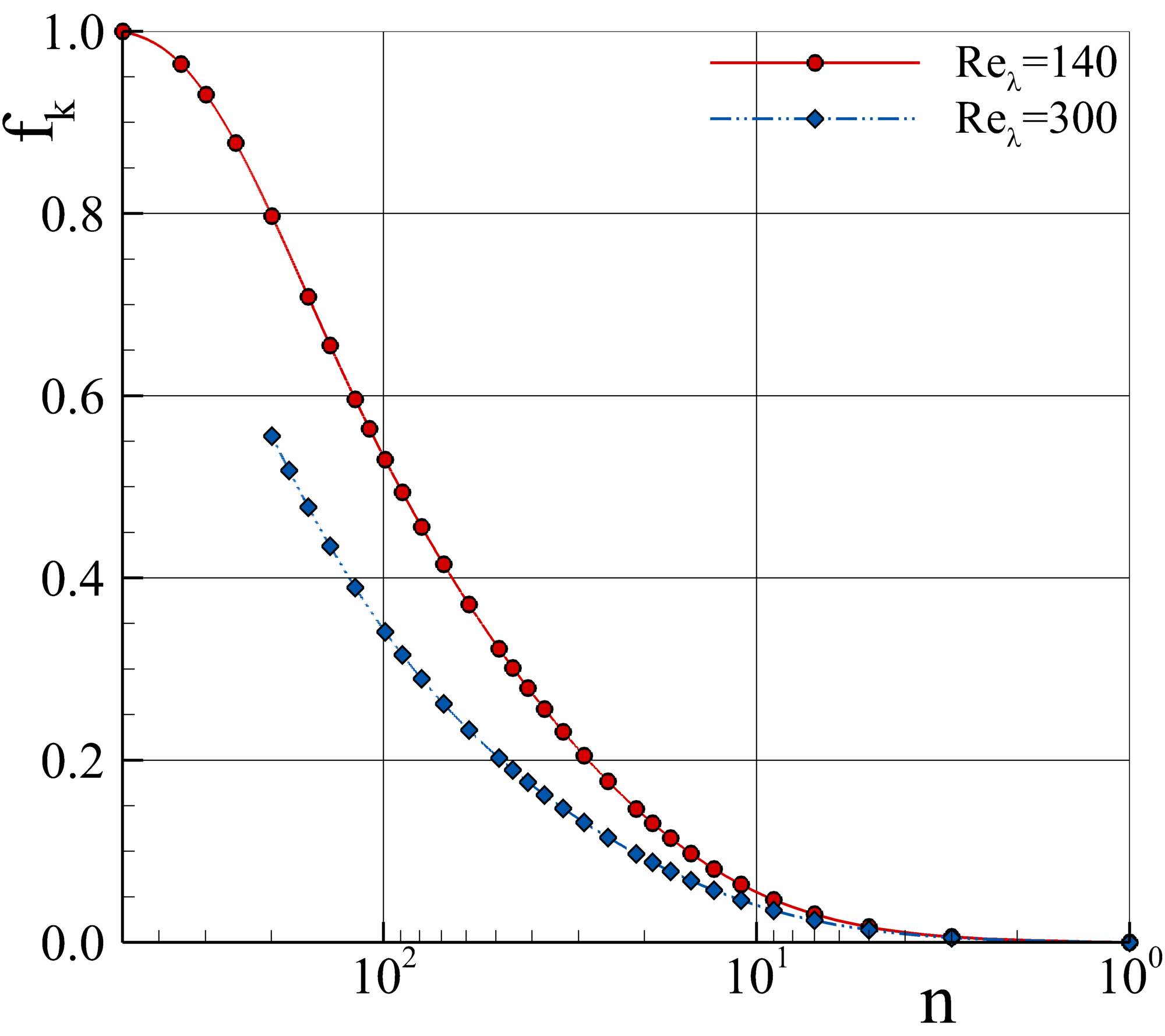}}
\\
\subfloat[$f_\varepsilon(n)$]{\label{fig:3.1.1_1b}
\includegraphics[scale=0.095,trim=0 0 0 0,clip]{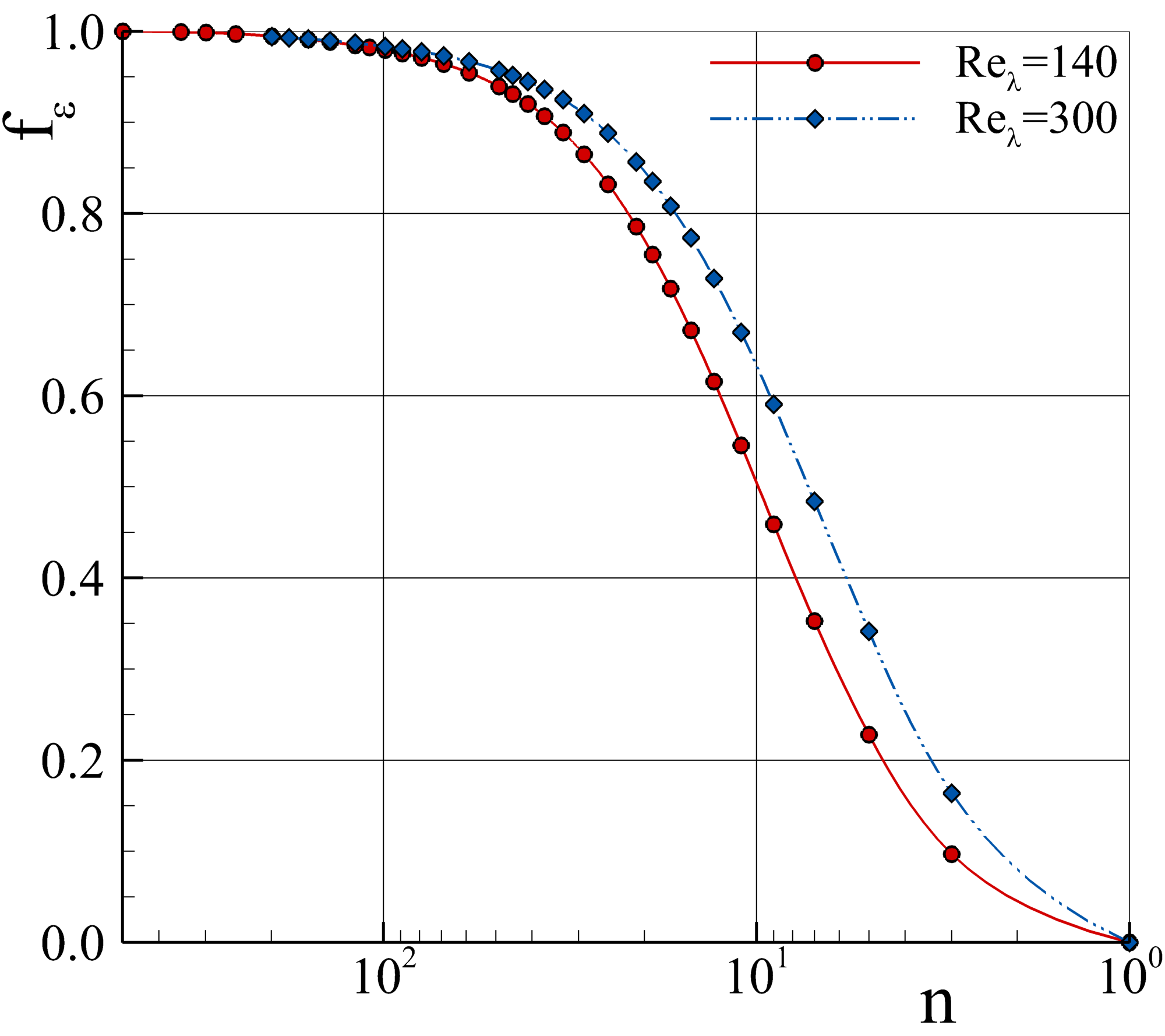}}
\caption{Variation of $f_k$ and $f_\varepsilon$ with the relative filter size, $n$, at $\mathrm{Re_\lambda}=140$ and $300$.}
\label{fig:3.1.1_1}
\end{figure}

Figure \ref{fig:3.1.1_1} presents the variation of $f_k$ and $f_\varepsilon$ with the relative filter size, $n=\Delta/\Delta_\eta$. $n$ indicates how large is the filter size when compared to DNS resolution (so $n=1$ is DNS). As expected, the $f_k(n)$ results indicate that most turbulence kinetic energy is contained at the largest flow scales. This behavior gets more pronounced with increasing $\mathrm{Re}_\lambda$. It is observed that to filter only $20\%$ of the total turbulence kinetic energy we need $n=29$ ($\mathrm{Re}_\lambda=140$) and $49$ ($\mathrm{Re}_\lambda=300$). This clearly illustrates the potential of bridging models to efficiently compute complex flow problems. Also, note that bridging models are usually not used at $f_k<0.20$ (LES range \cite{POPE_BOOK_2000}). Regarding $f_\varepsilon$, the results of figure \ref{fig:3.1.1_1} confirm that most turbulence dissipation occurs at the smallest scales and, as such, $f_\varepsilon$ grows significantly more rapidly than $f_k$ with $n$. The data show $f_\varepsilon=0.22$ ($\mathrm{Re}_\lambda = 140$) and $0.34$ ($\mathrm{Re}_\lambda = 300$) for $n=5$, and $f_\varepsilon>0.99$ for $n=199$ and both Reynolds numbers.

\begin{figure}[t!]
\centering
\includegraphics[scale=0.095,trim=0 0 0 0,clip]{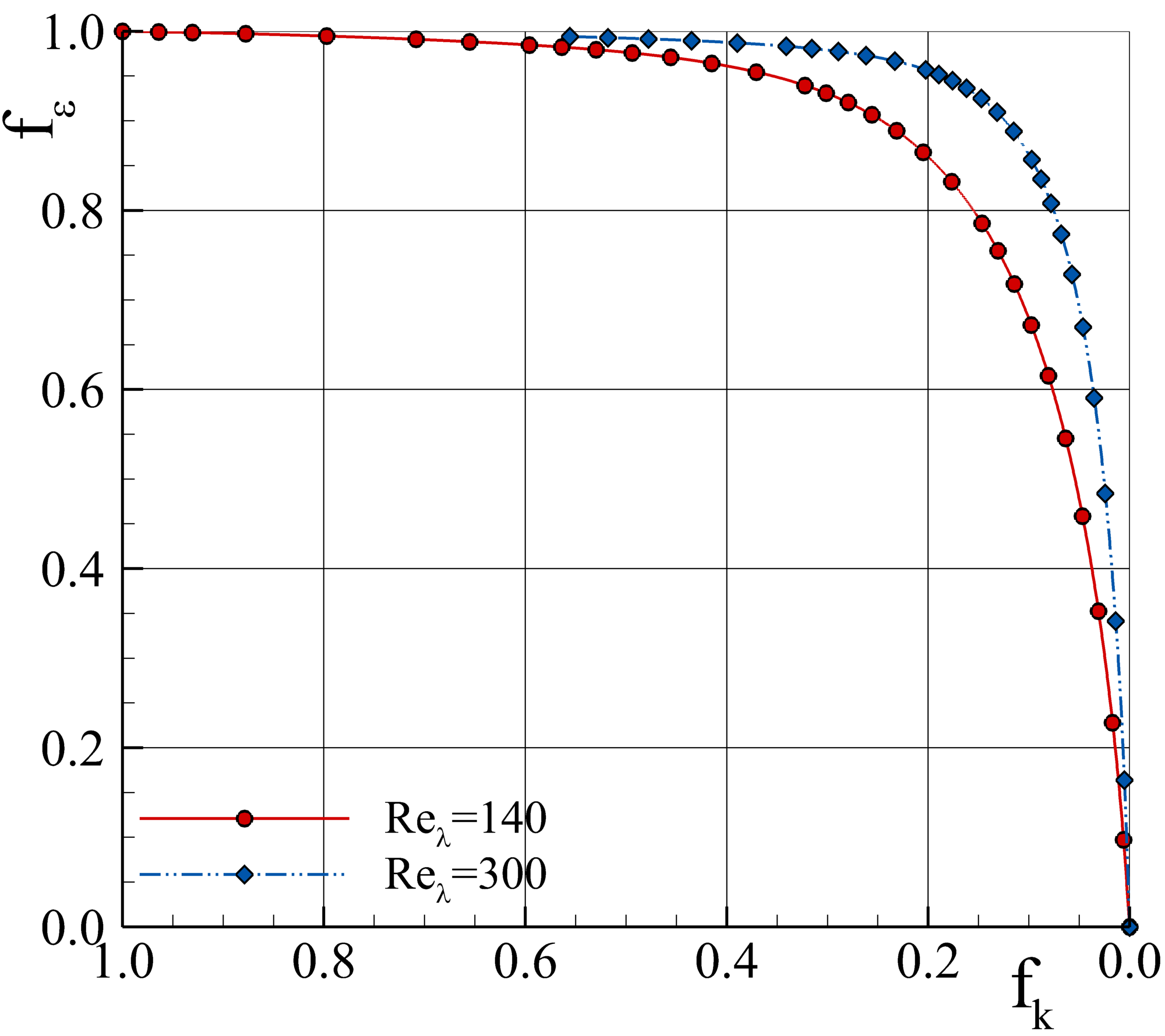}
\caption{$f_\varepsilon$ as a function of $f_k$ at $\mathrm{Re_\lambda}=140$ and $300$.}
\label{fig:3.1.1_2}
\end{figure}

Next, figure \ref{fig:3.1.1_2} presents  $f_\varepsilon$ as a function of $f_k$. The results indicate that $f_\varepsilon$ is only weakly dependent on $f_k$ at coarser $f_k$ values. At $f_k=0.20$, $f_\varepsilon=0.87$ for $\mathrm{Re}_\lambda=140$ and $f_\varepsilon=0.93$ for $\mathrm{Re}_\lambda=300$. Considering that practical simulations of turbulence are expected to operate at $f_k\ge 0.20$ due to the inherent cost and availability of LES formulations, figure \ref{fig:3.1.1_2} confirms that $f_\varepsilon=1.00$ is a good assumption for practical PANS computations.
\subsubsection{Homogeneous variable-density turbulence}
\label{sec:3.1.2}
\begin{figure}[t!]
\centering
\subfloat[$t=t_1$]{\label{fig:3.1.2_1a}
\includegraphics[scale=0.13,trim=0 0 0 0,clip]{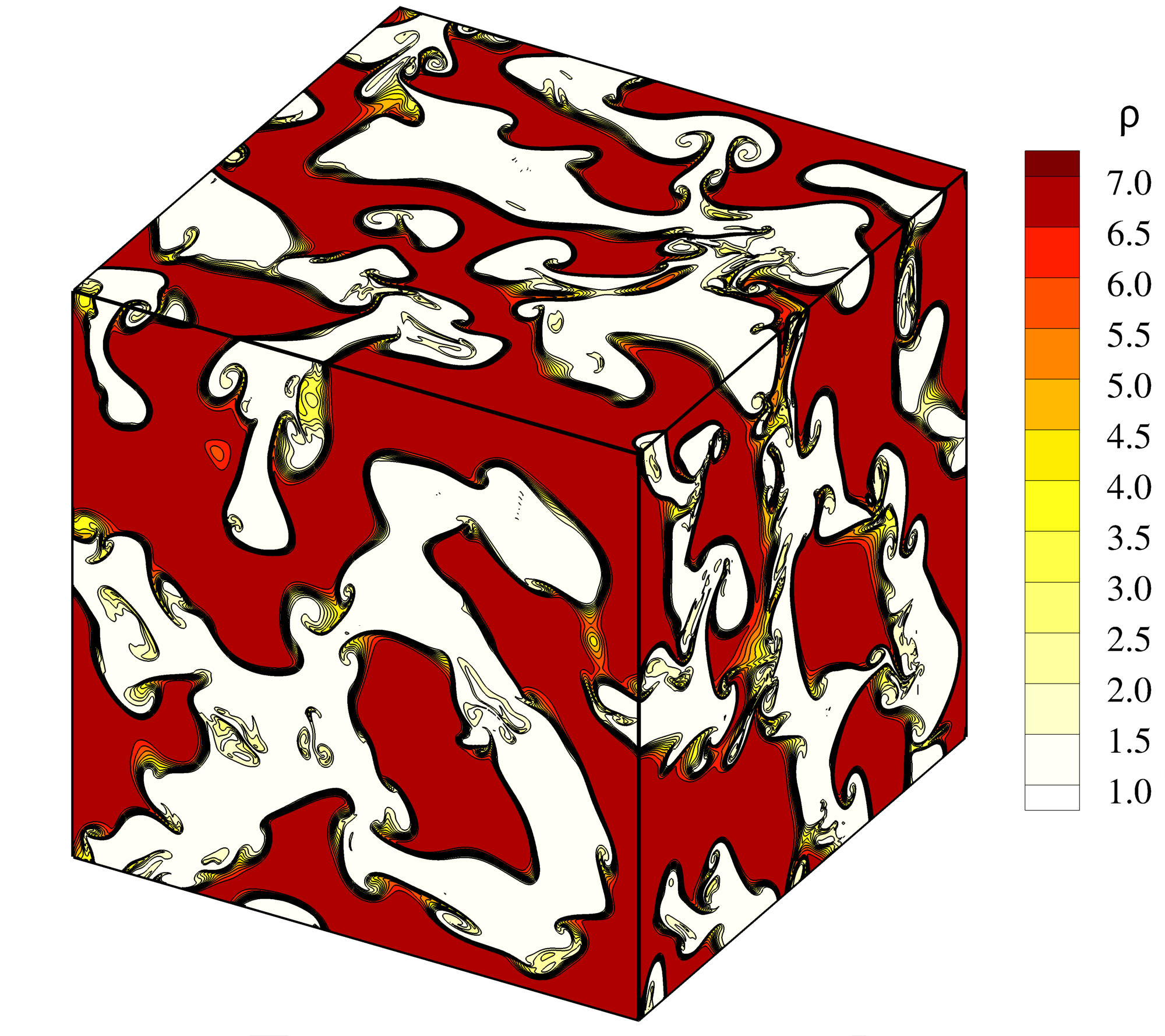}}
\\
\subfloat[$t=t_2$]{\label{fig:3.1.2_1b}
\includegraphics[scale=0.13,trim=0 0 0 0,clip]{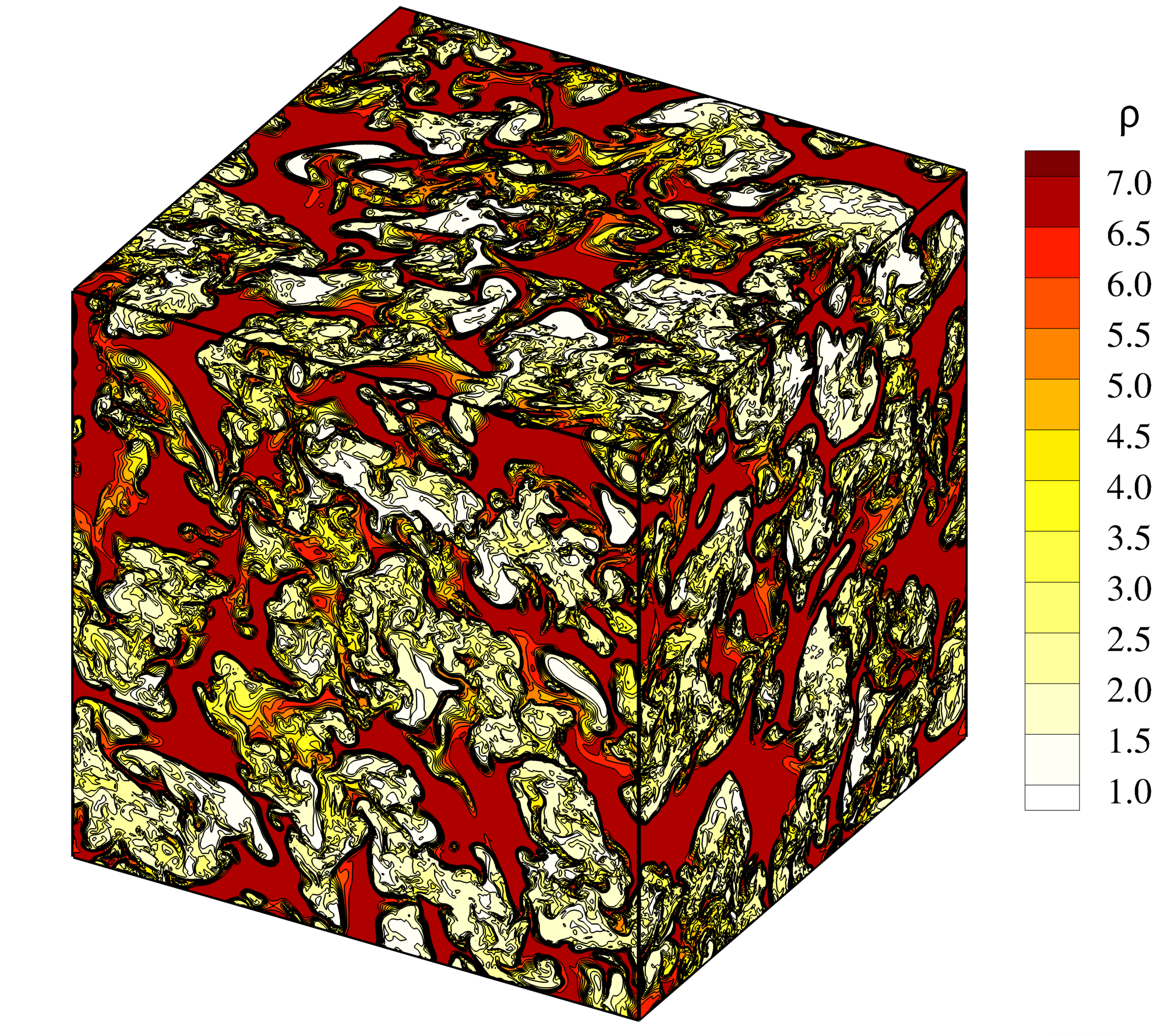}}
\\
\subfloat[$t=t_3$]{\label{fig:3.1.2_1c}
\includegraphics[scale=0.13,trim=0 0 0 0,clip]{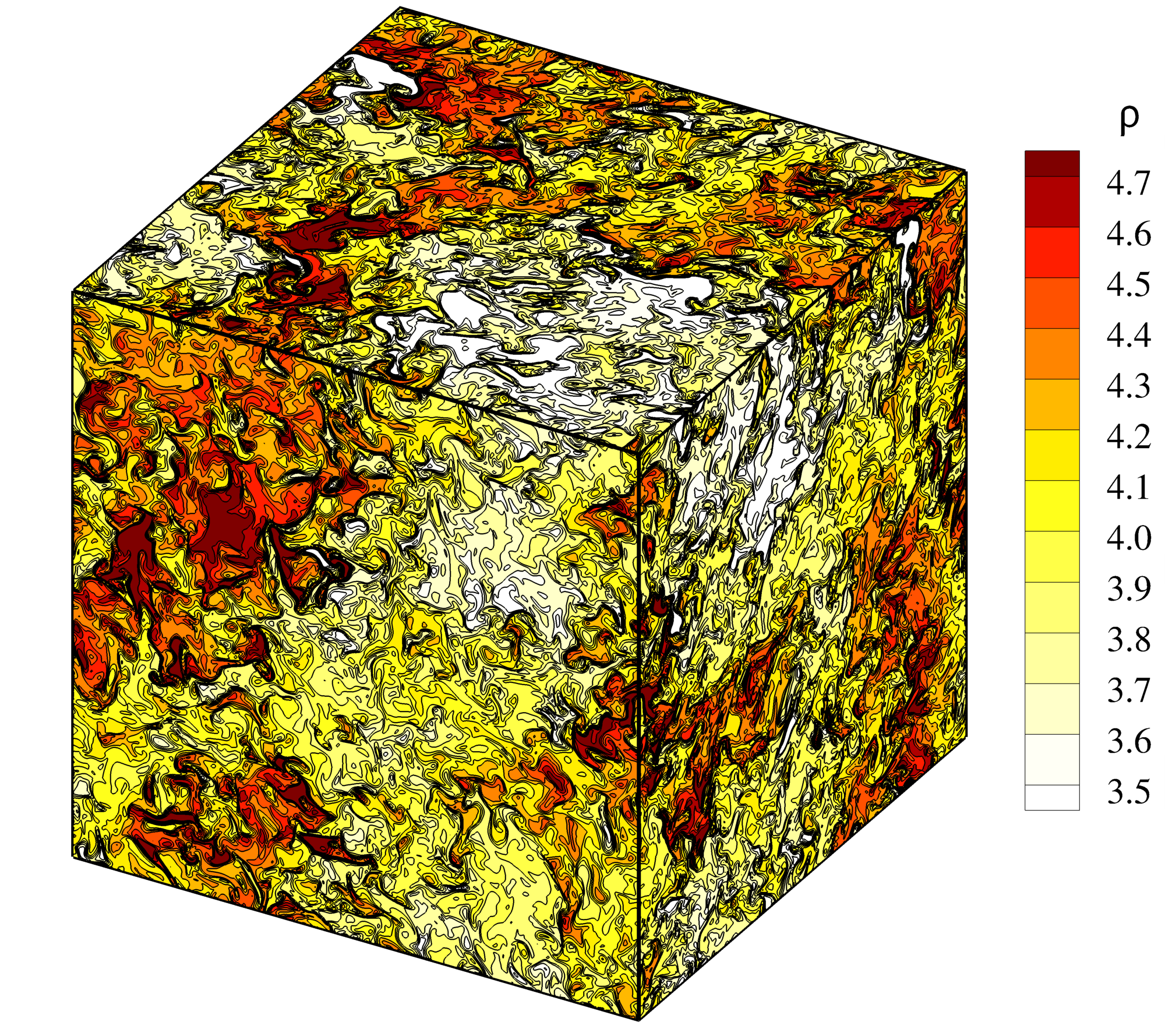}}
\caption{Density field of HVDT flow at distinct times: before ($t_1$), at ($t_2$), and after ($t_3$) the peak of kinetic energy.}
\label{fig:3.1.2_1}
\end{figure}

The HVDT is a transient flow and, as such, the \textit{apriori} tests are conducted at the three distinct and representative times shown in figure \ref{fig:3.1.2_1}: at $t_1=1.8$, the flow is in the so-called explosive growth regime \cite{ASLANGIL_JFM_2020} and the kinetic energy of the system is rapidly increasing through the conversion of potential into kinetic energy.  As shown in figure \ref{fig:3.1.2_1a}, the flow does not exhibit small scale turbulence, and the two fluids ($\rho_1=7.0$ and $\rho_2=1.0$) are mostly unmixed. At $t_2=2.8$, the flow kinetic energy grows, reaching close to its peak. This leads to flow regions characterized by small scale turbulence, where the two fluids mix. Finally, $t_3=4.8$ is just after the fast decay regime where the kinetic energy decays rapidly. Turbulence is the major component of the kinetic energy, and the flow exhibits high-intensity and fully-developed turbulence features. This enhances mixing (compare figures \ref{fig:3.1.2_1b} and \ref{fig:3.1.2_1c}). A comprehensive description of this flow is given in \citeauthor{ASLANGIL_PD_2020} \cite{ASLANGIL_PD_2020,ASLANGIL_JFM_2020}. It is important to emphasize that the transient nature of the HVDT flow and the overlap between coherent and turbulent wavelengths/frequencies hamper \textit{a-priori} exercises of this class of flows. Nonetheless, these studies still provide valuable information about the flow physics and evolution $f_\phi$ with the filter size.

\begin{figure}[t!]
\centering
\subfloat[$f_k(n)$]{\label{fig:3.1.2_2a}
\includegraphics[scale=0.095,trim=0 0 0 0,clip]{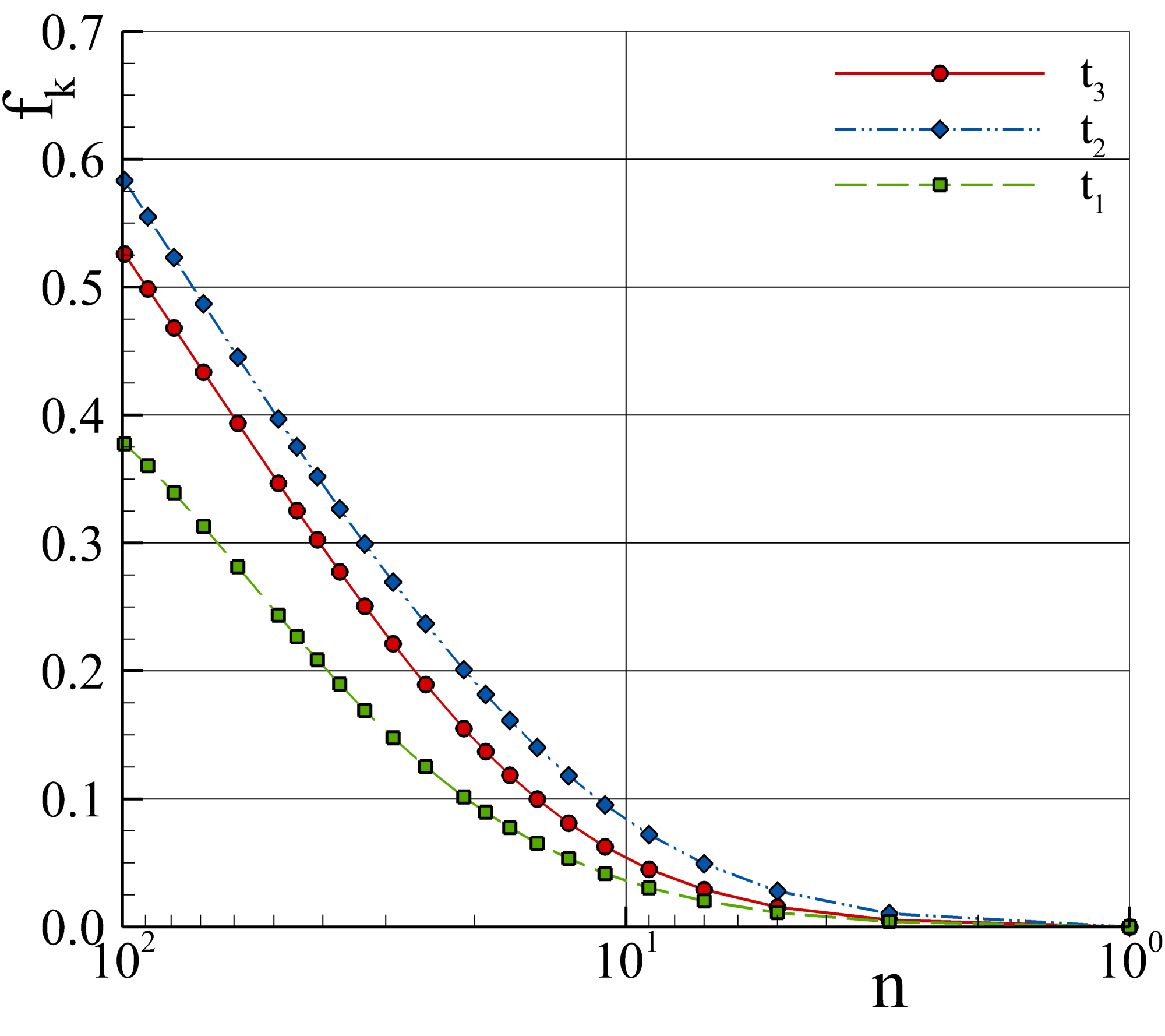}}
\\
\subfloat[$f_a(n)$]{\label{fig:3.1.2_2b}
\includegraphics[scale=0.095,trim=0 0 0 0,clip]{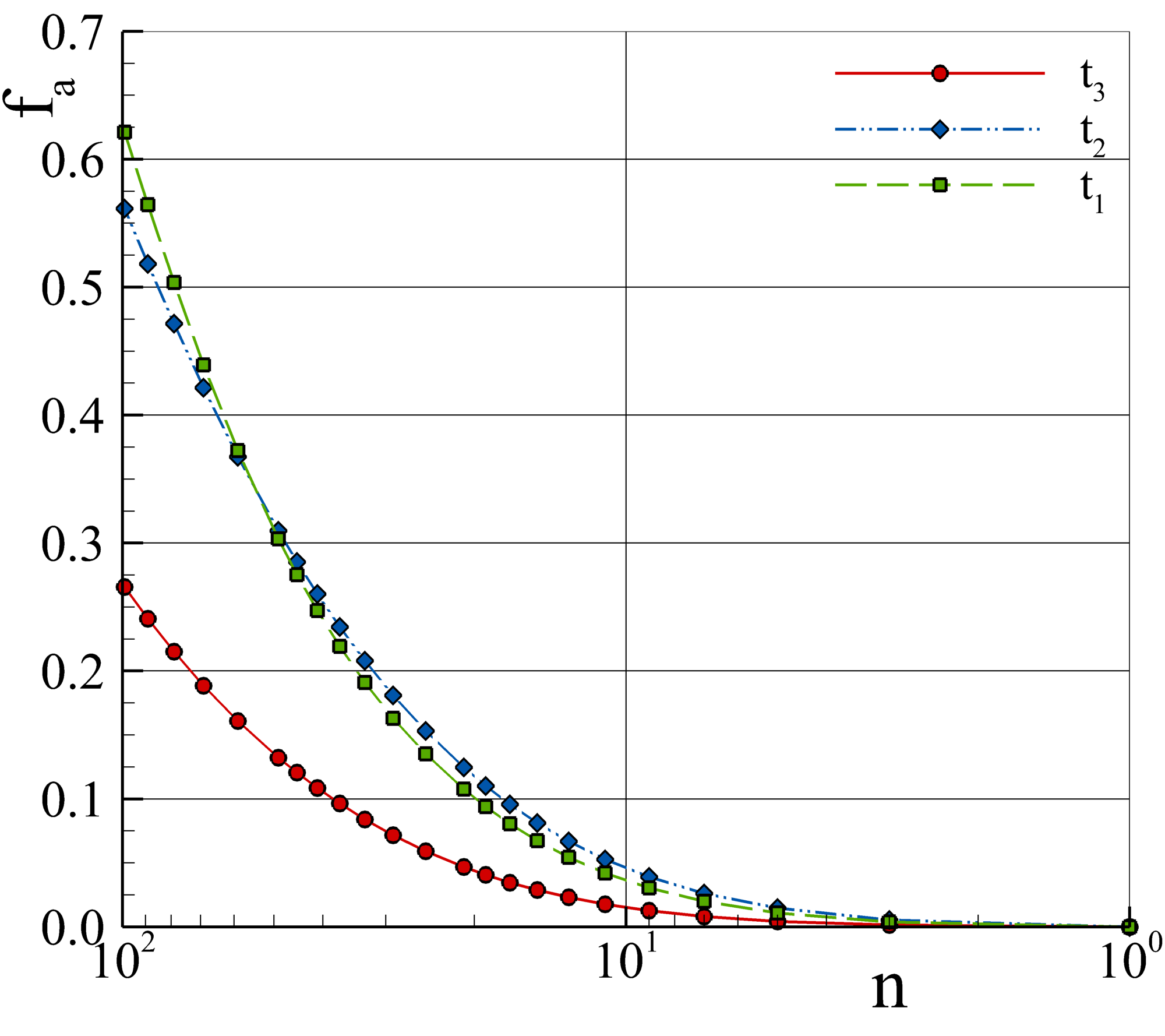}}
\\
\subfloat[$f_b(n)$]{\label{fig:3.1.2_2c}
\includegraphics[scale=0.095,trim=0 0 0 0,clip]{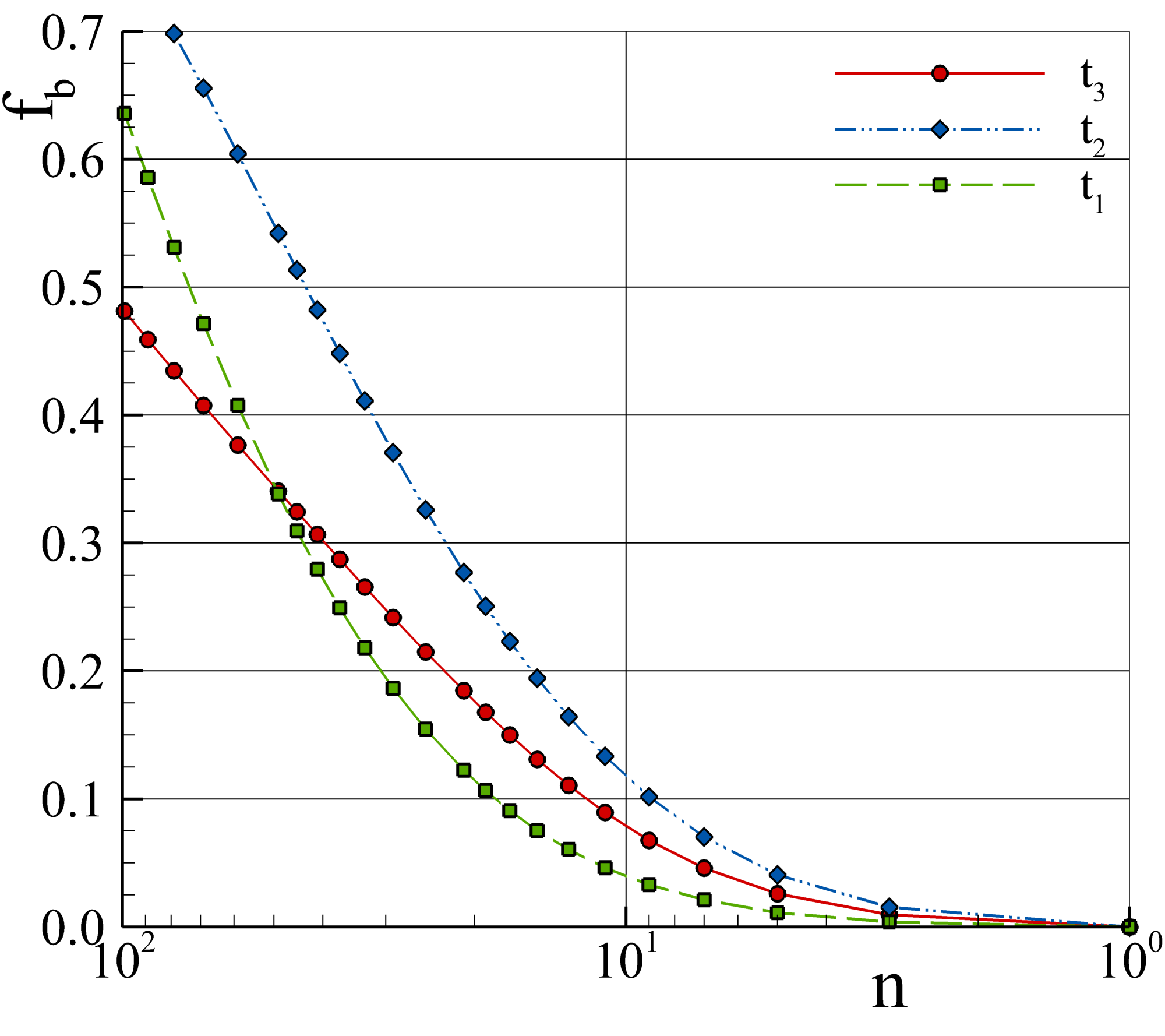}}
\caption{Variation of $f_k$, $f_a$, and $f_b$ with the relative filter size, $n$, at distinct times.}
\label{fig:3.1.2_2}
\end{figure}

\begin{figure}[t!]
\centering
\subfloat[$E(k)$]{\label{fig:3.1.2_3a}
\includegraphics[scale=0.095,trim=0 0 0 0,clip]{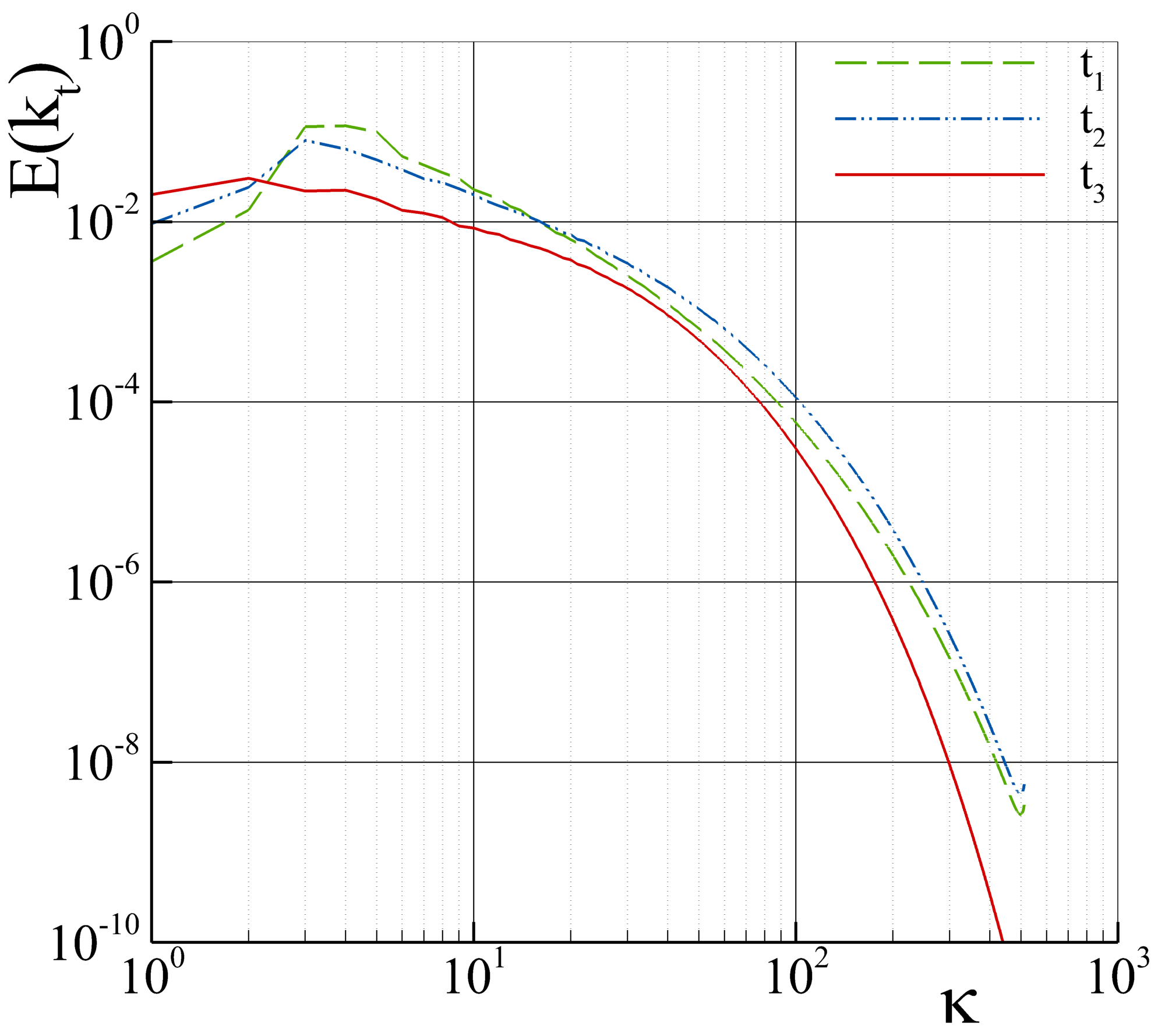}}
\\
\subfloat[$E(a)$]{\label{fig:3.1.2_3b}
\includegraphics[scale=0.095,trim=0 0 0 0,clip]{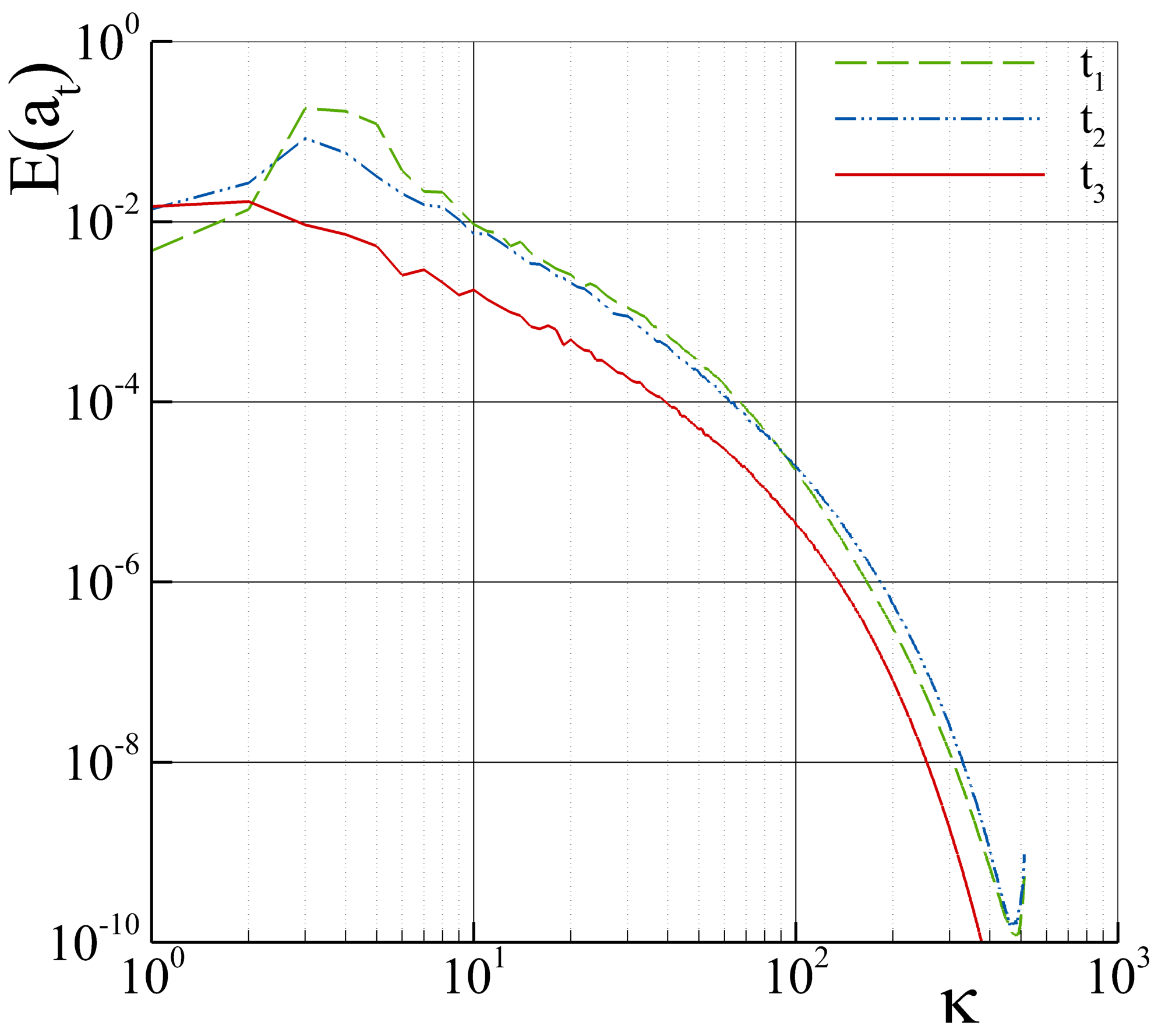}}
\\
\subfloat[$E(b)$]{\label{fig:3.1.2_3c}
\includegraphics[scale=0.095,trim=0 0 0 0,clip]{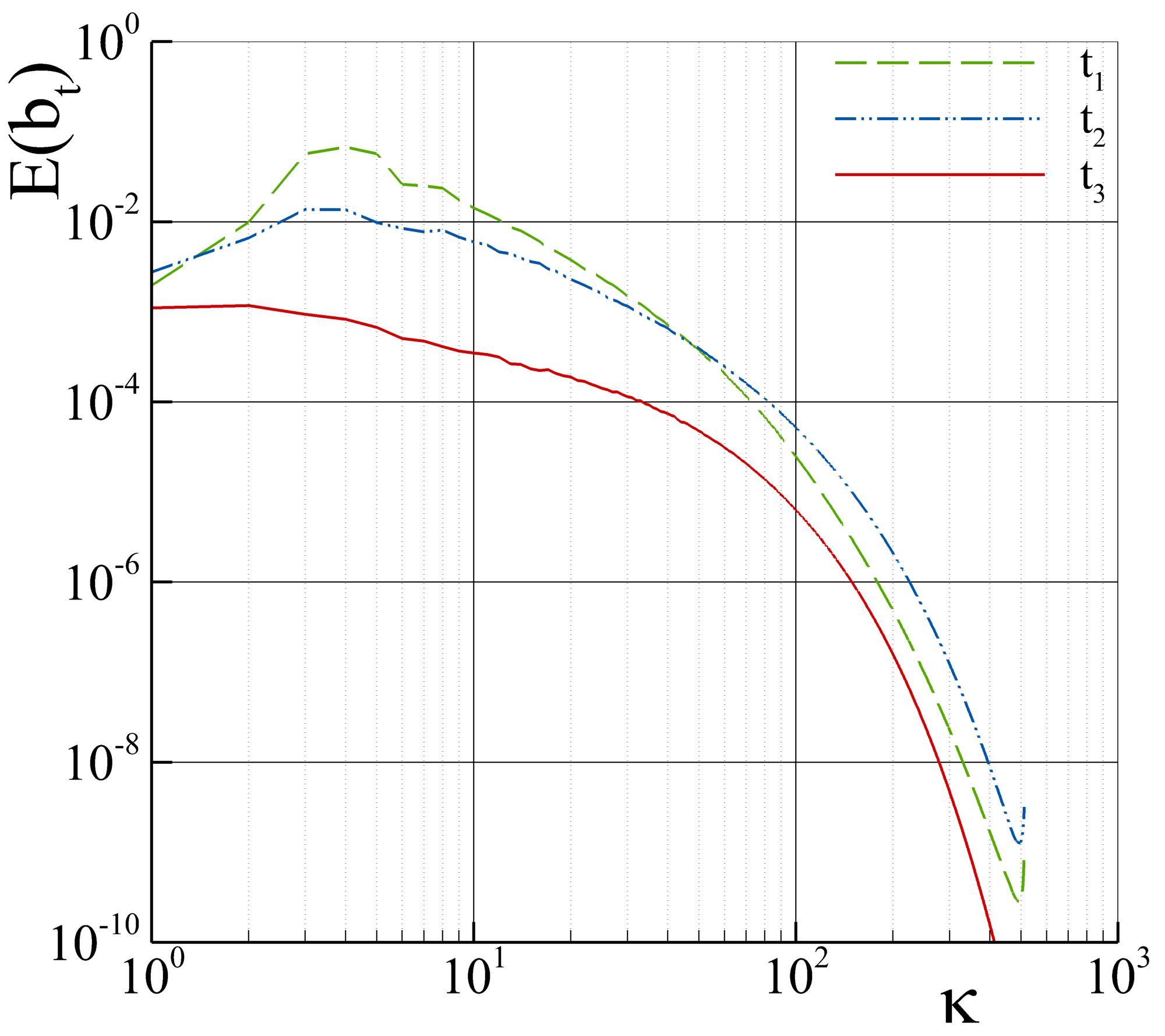}}
\caption{Energy spectra of $k_t$, $a_t$, and $b_t$ at distinct times.}
\label{fig:3.1.2_3}
\end{figure}

Figures \ref{fig:3.1.2_2} and \ref{fig:3.1.2_3} present the variation of $f_k$, $f_a$, and $f_b$ with the relative filter size, and the energy spectra of $k_t$, $a_t$, and $b_t$ for the unfiltered fields. Due to the HVDT flow properties, the quantities $\overline{a}_2$ and $\overline{a}_3$ are equal to zero, so we consider $f_a = f_{a_1}$. The results for $f_k(n)$ indicate that before the peak of $k_t$ ($t=t_1$), most of the kinetic energy is contained in the largest coherent flow scales (blobs of laminar fluid). For this reason, $f_k$ does not exceed $0.38$ when $n=99$. As the flow and turbulence field develop, $t=t_2$, $k_t$ increases to a value close to its maximum, altering the spectral properties of the kinetic energy field. In addition to the conversion of potential into kinetic energy, the energy of the largest scales is transferred to the smallest ones, widening the spectra so that larger fractions of $k_t$ are modeled for the same $n$. At later times, $t=t_3$, when the flow exhibits high-intensity turbulence and mixing, $f_k(n)$ exhibits a slight reduction. This stems from the dissipation of $k_t$ at the smallest scales.

Turning our attention to $f_a(n)$, the results indicate that the evolution of this quantity with the filter width at $t=t_1$ and $t_2$ is nearly identical. This shows that the morphological flow changes occurring at these early instants significantly affect the turbulence kinetic energy but not the velocity fluctuations uncorrelated with the density field (figures \ref{fig:3.1.2_3b} and \ref{fig:3.1.2_3c}). The data of figure \ref{fig:3.1.2_2b} also indicate that approximately $60\%$ of the energy of $a$ at these instants is contained in the smallest flow scales, $n = 99$. At $t=t_3$, the values of $f_a$ are reduced approximately fifty percent. Note that this is when the flow exhibits high-intensity turbulence, a more homogeneous mixture, and diminishing influence of the coherent field (figure \ref{fig:3.1.2_2}).

The variation of $f_b$ with $n$ shows that the magnitude of this quantity increases from $t=t_1$ to $t_2$, reaching values of $0.63$ at $t=t_1$ and $0.77$ at $t_2$. Such result indicate that the density-specific volume correlation is dominated by the smallest wavelengths at these early stages. The observed high-intensity turbulence and enhanced mixing at $t=t_3$ leads to a significant reduction of $f_b$. For $n\leq 99$, $f_b$ does not exceed $0.50$. 

As for the FHIT case, figure \ref{fig:3.1.2_2} illustrates the potential of bridging models to compute complex flow problems efficiently. Considering $f_k$, the data indicate that simulations at $f_k=0.50$ and $0.25$ ($t=t_3$) can run on grid resolutions $89$ and $33$ times coarser (in each direction) than those required by DNS. This constitutes a significant cost reduction. 

\begin{figure}[t!]
\centering
\subfloat[$f_a(f_k)$]{\label{fig:3.1.2_4a}
\includegraphics[scale=0.095,trim=0 0 0 0,clip]{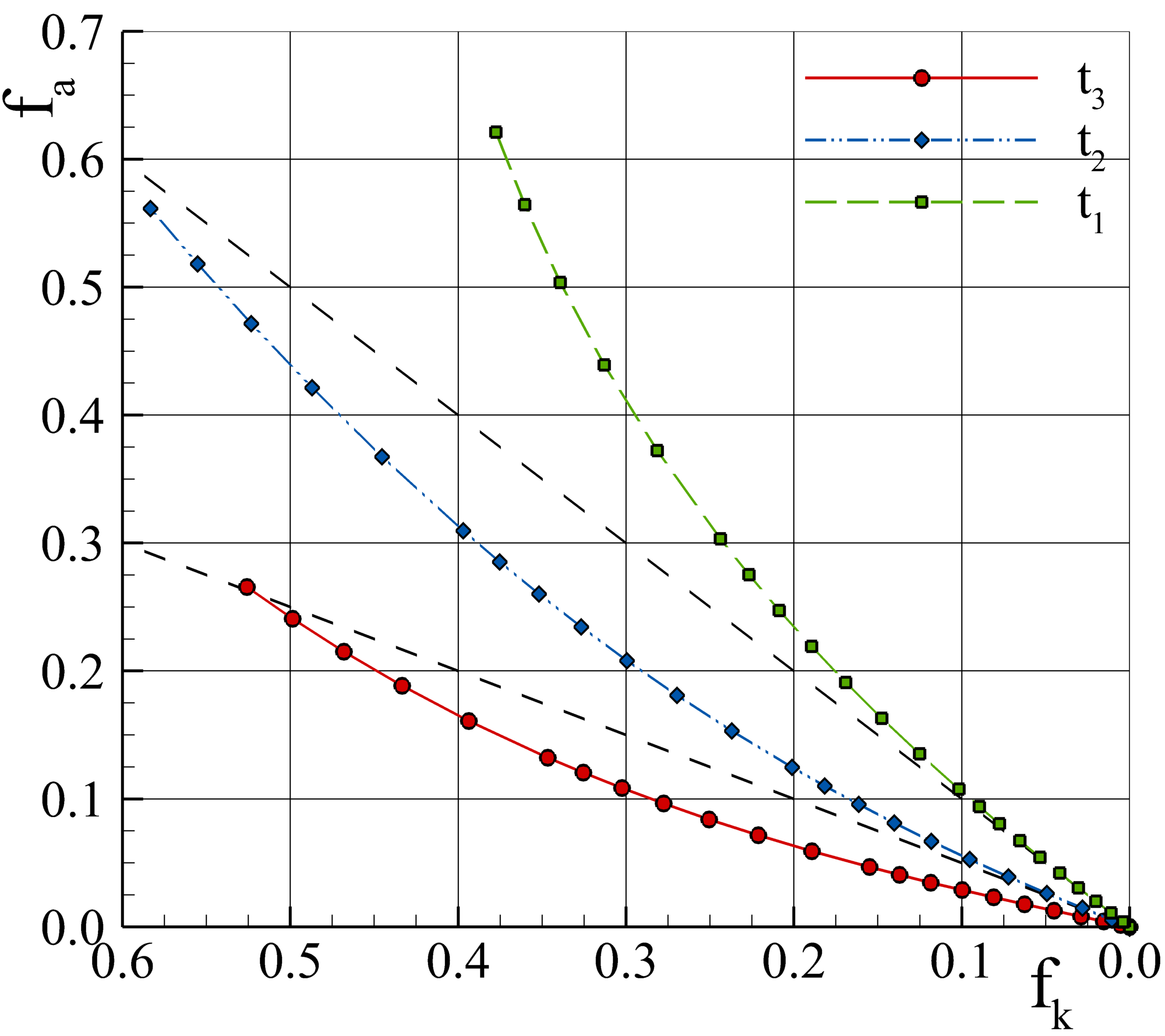}}
\\
\subfloat[$f_b(f_k)$]{\label{fig:3.1.2_4b}
\includegraphics[scale=0.095,trim=0 0 0 0,clip]{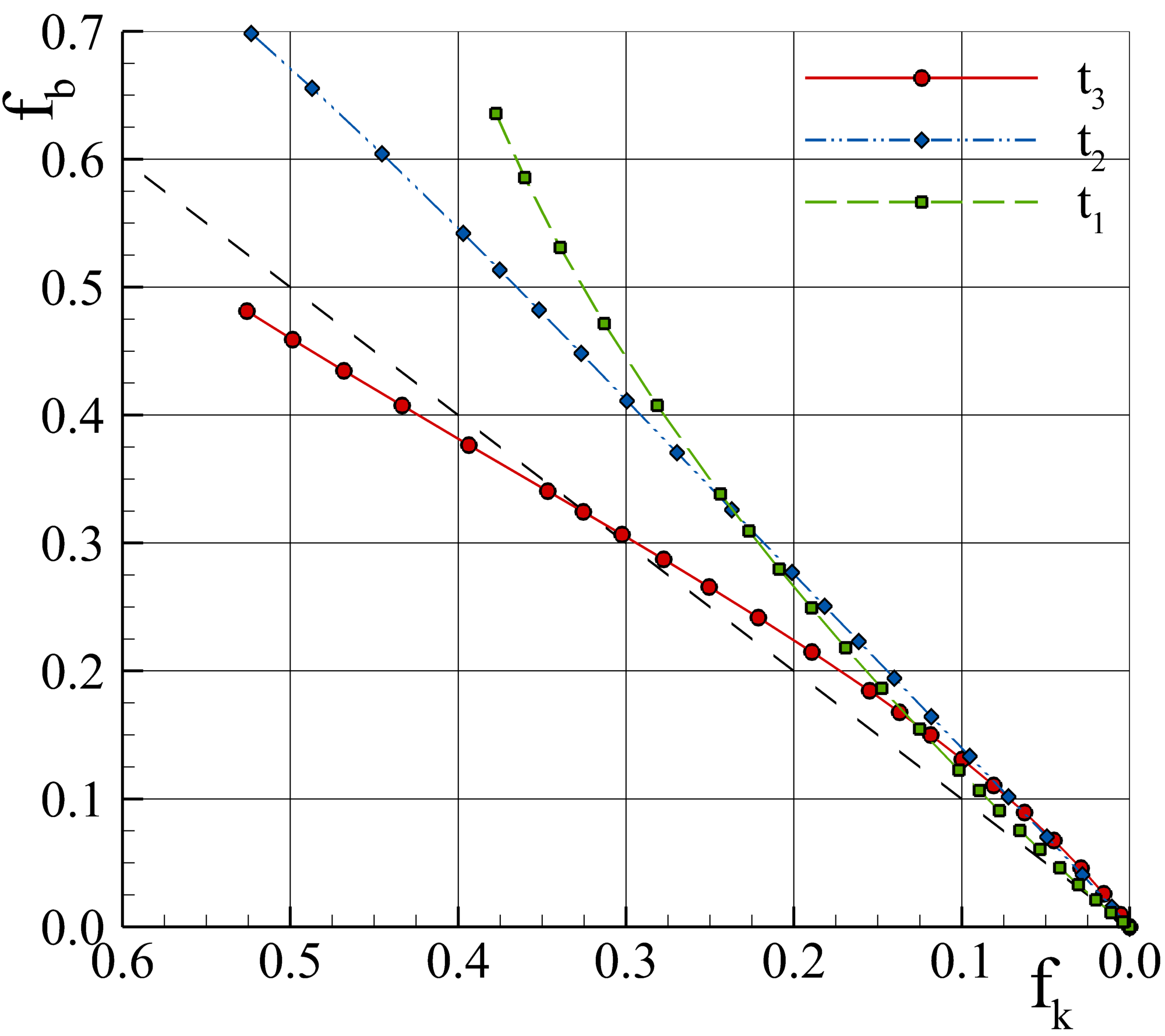}}
\caption{Variation of $f_a$ and $f_b$ with $f_k$ at distinct times.}
\label{fig:3.1.2_4}
\end{figure}

Finally, figure \ref{fig:3.1.2_4} depicts the variation of $f_a$ and $f_b$ as a function of $f_k$. The results show that the ratio $f_a/f_k$ gets smaller in time, and $f_a \approx f_k/2$ at $t=t_3$ (the instant when the flow is characterized by high-intensity turbulence). In contrast, $f_b$ has a small temporal dependence until $t_2$, and $f_b>f_k$. Such a behavior is not observed at $t=t_3$, where $f_b \approx f_k$. We attribute this result to the breakdown into turbulence and dissipation of the coherent field.
\subsubsection{Parameter selection}
\label{sec:3.1.3}
The above \textit{a-priori} tests have been conducted to help us determining the parameters $f_\phi$ of PANS BHR-LEVM, and propose guidelines toward their efficient selection. Nevertheless, we reiterate that the present paper's primary objectives are to extend PANS methodology to variable-density flow and provide the resulting PANS BHR-LEVM model's initial validation space. Closures using different dependent variables may require similar studies to define $f_\phi$ (only $k$ and $\varepsilon$ tend to be used in most closures \cite{CHASSAING_FTC_2001,ZHOU_PR1_2017,ZHOU_PR2_2017}).

The FHIT results have shown that prescribing $f_\varepsilon=1.00$ is a good strategy because most dissipation occurs at the smallest flow scales. These are usually modeled in practical PANS computations ($f_k\ge 0.20$). Although $f_k$ and $f_\varepsilon$ can get closer in transient and/or transitional flows, the results available in the extensive PANS literature have shown that this approach is still good \cite{GIRIMAJI_JAM_2005,BASARA_IJHFF_2018,PEREIRA_IJHFF_2018,PEREIRA_IJHFF_2019,PEREIRA_PRF_2021,TAZRAEI_PRF_2019,GIRIMAJI_AIAA43_2005,LAKSHMIPATHY_JFE_2010,PEREIRA_THMT15_2015,KAMBLE_POF_2020,FOWLER_POF_2020}.
 In these cases, the resulting modeling shortcomings need to be compensated by slighter finer values of $f_k$. Referring to $f_a$ and $f_b$, selecting these parameters is more complex and has never been done before. Our \textit{a-priori} tests suggest setting $f_a=f_k/2$ and $f_b=f_k$ at late times when turbulence is closer to fully-developed, and the coherent field has a diminishing impact on the flow dynamics. However, note that these quantities are inherently time-dependent and contain a meaningful coherent component at early times. Once again, these calibration issues can be overcome through a proper selection of $f_\phi$, and slightly lower values of $f_k$. Although often neglected, it is crucial to emphasize that these issues are common to most SRS models.

Considering the previous points and the results of the \textit{a-priori} tests, the present simulations utilize one of the following strategies to define $f_k$, $f_\varepsilon$, $f_a$ ($f_a=f_{a_i}$), and $f_b$:
\begin{itemize}
\item[$i)$] Upon inspection of the governing equations of the PANS BHR-LEVM closure, it is possible to infer that $f_k$ and $f_\varepsilon$ are the major influence on the production of modeled turbulent kinetic energy by shear and buoyancy effects ($k_u$ and $S_u$ equations), and, consequently, on the modeled turbulent stresses. Thus, we prescribe $f_k$, define $f_\varepsilon=1.00$ based on the outcome of the \textit{a-priori} exercises, and set the remaining parameters equal to one. We expect this approach to be more robust and general so it can be applied to most flows. Yet, it might require slightly smaller values of $f_k$ to compensate for possible calibration deficits.
\item[$ii)$] Disregard the contribution of the density fluctuations to the specification of $f_a$ and $f_b$ (equations \ref{eq:2.1_01}, \ref{eq:2.1_02} and \ref{eq:2.1_9}) by assuming that the coherent field is the main contributor to the magnitude of $a_{i_u}$ and $b_u$. This makes $f_a$ only dependent on the velocity field so that $f_a\approx \sqrt{f_k}$, and $f_b \approx 1.00$. $f_\varepsilon$ is set equal to one based on the \textit{a-priori} results.
\item[$iii)$] Based on the outcome of the \textit{a-priori} exercises of the FHIT and  HVDT, set $f_a \approx f_k/2$, $f_\varepsilon=1.00$, and $f_b\approx f_k$. This strategy is optimized for HVDT type of flows, and best suited for instants characterized by fully-developed turbulence.
\end{itemize}
These approaches are summarized in table \ref{tab:3_1} and tested in Section \ref{sec:5.2}. Note that for $S_3$, the values of $f_a$ are rounded to the closest upper multiple of 0.05.  In the remaining of this paper, we use $f_k$ to refer to the physical resolution of the model. Nonetheless, we stress that each $f_k$ has a corresponding $f_\varepsilon$, $f_a$, and $f_b$ given in table \ref{tab:3_1}.

\begin{table}
\centering
\setlength\extrarowheight{3pt}
\caption{Modeled-to-total ratios, $f_\phi$, used in the RT PANS computations at $0.00<f_k<1.00$.}
\label{tab:3_1}    
\begin{tabular}{C{1.1cm}C{1.2cm}C{1.7cm}C{1.7cm}C{1.7cm}}
\hline
 	& $f_k$				&	$0.25$ & $0.35$ & $0.50$	 \\ [3pt] \hline
		&$f_\varepsilon$	&	$1.00$ &  $1.00$ &  $1.00$ \\[3pt] 
$S_1$ 	&$f_a$			&	$1.00$ &  $1.00$ &  $1.00$ \\[3pt] 
		&$f_b$			&	$1.00$ &  $1.00$ &  $1.00$ \\[3pt] \hline
		&$f_\varepsilon$	&	$1.00$ &  $1.00$ &  $1.00$ \\[3pt] 
$S_2$ 	&$f_a$			&	$0.50$ &  $0.59$ &  $0.71$ \\[3pt] 
		&$f_b$			&	$1.00$ &  $1.00$ &  $1.00$ \\[3pt] \hline
		&$f_\varepsilon$	&	$1.00$ &  $1.00$ &  $1.00$ \\[3pt] 
$S_3$ 	&$f_a$			&	$0.15$ &  $0.20$ &  $0.25$ \\[3pt] 
		&$f_b$			&	$0.25$ &  $0.35$ &  $0.50$ \\[3pt] \hline
\end{tabular}
\end{table}
%
%
\section{Flows and Simulations Details}
\label{sec:4}
%
%
\subsection{Taylor-Green vortex}
\label{sec:4.1}
\begin{figure}[t!]
\centering
\includegraphics[scale=0.15,trim=0 0 0 0,clip]{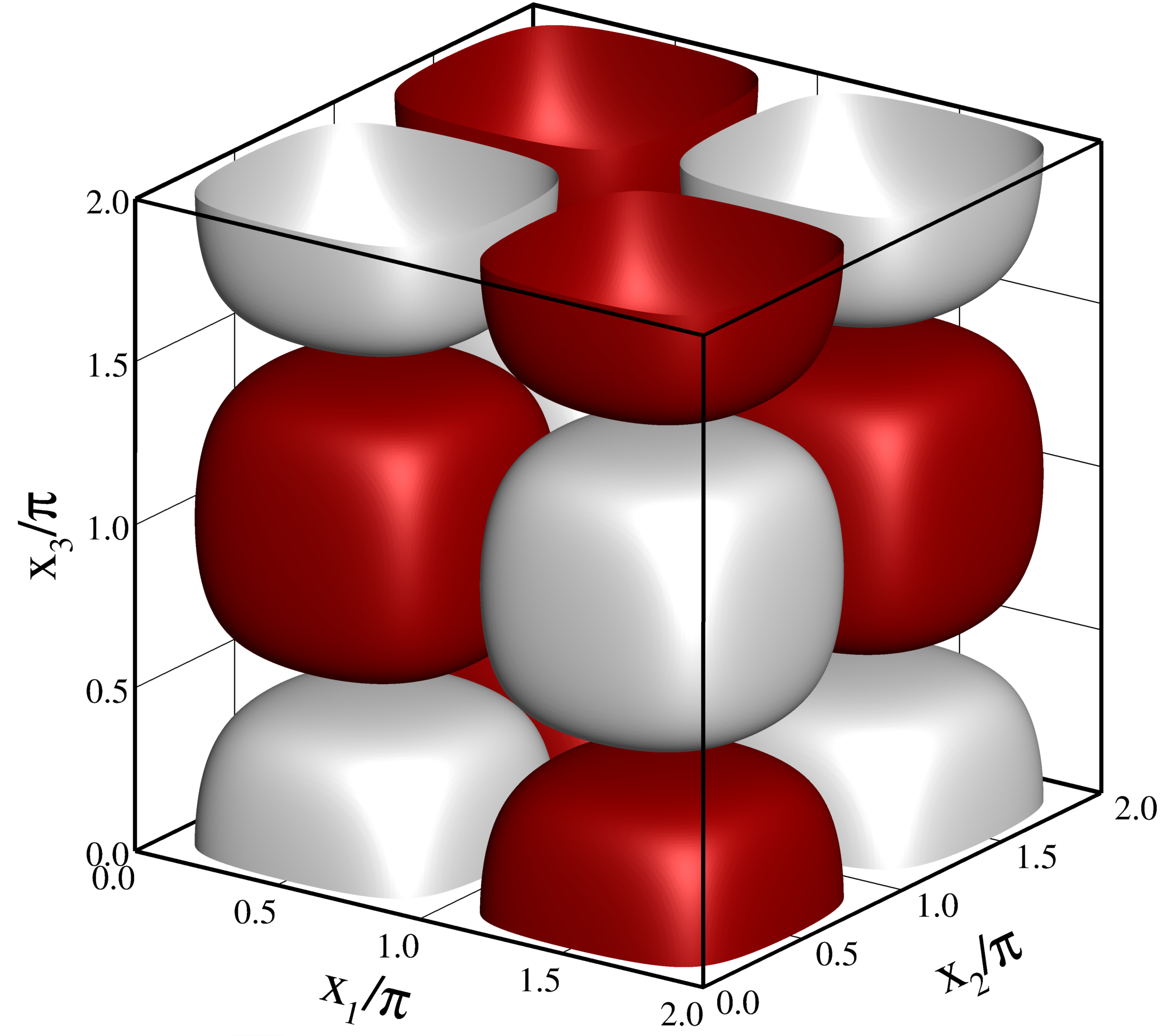}
\caption{Vortical structures present in the Taylor-Green vortex flow at $t=0$. Structures defined by the isosurfaces of the vorticity $x_3$ component.}
\label{fig:4.1_1}
\end{figure}

The Taylor-Green vortex flow \cite{TAYLOR_RSA_1937} is a canonical test case used to investigate the modeling and simulation of onset, development, and decay of turbulence \cite{YANG_JFM_2011,BRACHET_JFM_1983,SHU_JSC_2005,DRIKAKIS_JOT_2007,YANG_JFM_2010,CHAPELIER_AIAA42_2012,DEBONIS_NASAREP_2013,SHIROKOV_JOT_2014,BULL_AIAA_2015,DAIRAY_JCP_2017,MOURA_JCP_2017,PENG_PRF_2018,SHARMA_POF_2019,GRINSTEIN_CMA_2019,PEREIRA_PRE_2021,PEREIRA_PRF_2021}. The flow is initially characterized by multiple laminar, well-characterized, and single-mode vortices. These are illustrated in figure \ref{fig:4.1_1}, and defined by \cite{TAYLOR_RSA_1937,BRACHET_JFM_1983},
\begin{equation}
\label{4.1_1}
V_1(\mathbf{x},t_o)=V_o \sin(x_1)\cos(x_2)\cos(x_3) \; ,
\end{equation}
\begin{equation}
\label{4.1_2}
V_2(\mathbf{x},t_o)=-V_o \cos(x_1)\sin(x_2)\cos(x_3) \; ,
\end{equation}
\begin{equation}
\label{4.1_3}
V_3(\mathbf{x},t_o)=0 \; ,
\end{equation}
where $V_o$ is the initial velocity magnitude. The corresponding pressure field is obtained from solving the Poisson equation, 
\begin{equation}
\label{4.1_4}
P(\mathbf{x},t_o)=  P_o + \frac{\rho_o V_o^2}{16} \left[ 2 +\cos \left( 2x_3 \right)\right] \left[ \cos \left( 2x_1\right) +\cos \left( 2x_2 \right)\right]\; ,
\end{equation}
where $P_o$ and $\rho_o$ are the pressure and density at $t=0$. The vortical structures of figure \ref{fig:4.1_1} interact and evolve in time, and vortex stretching processes generate vortex-sheets that gradually get closer. Afterward, these vortex-sheets roll-up and reconnect \cite{KIDA_ARFM_1994,YANG_JFM_2011}, leading to the onset of turbulence and subsequent intensification of vorticity. 
The coherent structures breakdown and high-intensity turbulence appears. Finally, the turbulence kinetic energy dissipates rapidly by the action of viscous effects.

The analyzed TGV flow is characterized by a Reynolds number $\mathrm{Re} \equiv \rho L_o V_o/\mu=3000$ \cite{BRACHET_JFM_1983,DRIKAKIS_JOT_2007}, and an initial Mach number $\mathrm{Ma}_o=0.28$. Such a $\mathrm{Ma}_o$ leads to maximum instantaneous and averaged ($L_1$ norm) variations of $\rho$ smaller than $11.0\%$ and $1.4\%$ of $\rho_o$ for $f_k=0.00$, respectively. The computational domain of this problem is a cube with length equal to $L=2\pi L_o$. Periodic boundary conditions are applied on all boundaries. The initial thermodynamic and flow properties are the following: $V_o=10^4$cm/s, $L_o=1.00$cm, $\rho_o=1.178\times 10^{-3}$g/cm$^3$, $P_o=10^5$Pa, $\mu= 3.927 \times 10^{-3}$g/(cm.s), heat capacity ratio $\gamma=1.40$, $k_o=10^{-7}$cm$^2$/s$^2$, and $S_o=6.136\times 10^{-3}$cm.
%
%
\subsection{Raleigh-Taylor flow}
\label{sec:4.2}

The RT flow \cite{RAYLEIGHT_PLMS_1882,TAYLOR_PRSA_1950} is a benchmark problem of variable-density turbulent mixing, which has been intensely studied through numerous numerical experiments \cite{COOK_JFM_2001,DIMONTE_POF_2004,RISTORCELLI_JFM_2004,CABOT_N_2006,BANERJEE_IJHMT_2009,LIVESCU_JOT_2009,VLADIMIROVA_POF_2009,LIVESCU_PTRSA_2013,YOUNGS_PS_2017,KONNIKAKIS_PRE_2019}. 
Its importance to the variable-density flow community motivated diverse validation initiatives such as the Alpha-Group collaboration \cite{DIMONTE_POF_2004}.

\begin{figure}[t!]
\centering
\includegraphics[scale=0.10,trim=0 0 200 0,clip]{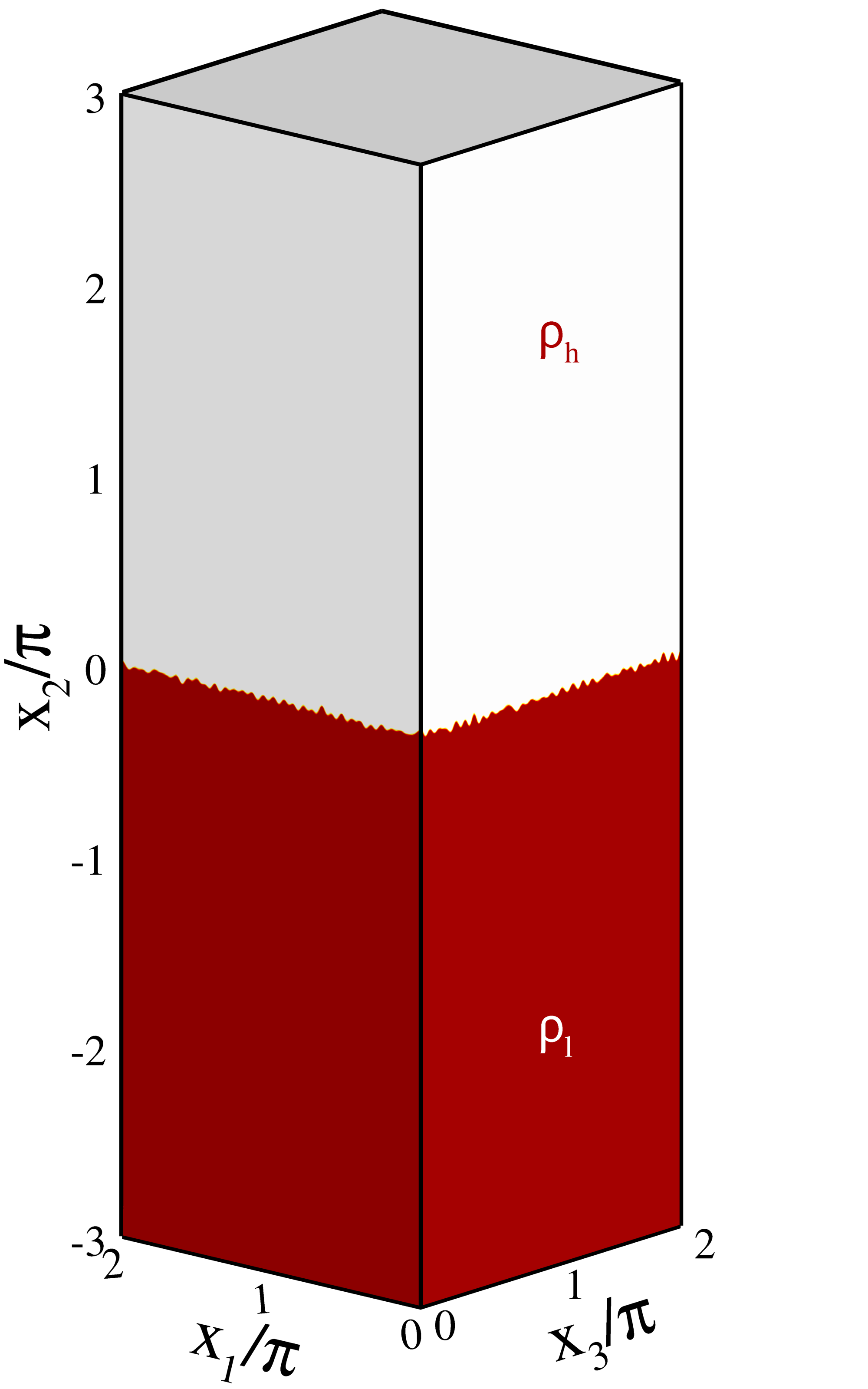}
\caption{Density field of the Rayleigh-Taylor flow at $t=0$.}
\label{fig:4.2_1}
\end{figure}

The flow is initially characterized by a perturbed interface separating two fluids of different densities, figure \ref{fig:4.2_1}. These materials are at rest, and the dense fluid, $\rho_h$, is on top of the light medium, $\rho_l$. The Atwood number of the flow is defined as $\mathrm{At} \equiv (\rho_h-\rho_l)/(\rho_h+\rho_l)$. After this instant, the heavy fluid starts accelerating downwards by the action of gravity, whereas the light material moves upwards. The interface perturbations create a misalignment between the density gradient and the pressure, which induces the RT instability. The resulting upward moving structures, named bubbles, are the penetration of heavy fluid into the light medium and, conversely, downward moving spikes are the penetration of light fluid into the heavy medium. The shearing motion on the edges of these coherent structures triggers a Kelvin-Helmholtz instability, leading to the onset and development of turbulence. As a result, the mixing rate of the two materials and the mixing-layer width increase. The temporal evolution of the RT flow comprises a linear (laminar flow) and a non-linear (laminar, transitional, and turbulent flow) regimes. A comprehensive description of the flow is given by \citeauthor{SHARP_PD_1984} \cite{SHARP_PD_1984}, \citeauthor{ZHOU_PR1_2017} \cite{ZHOU_PR1_2017,ZHOU_PR2_2017}, and \citeauthor{BOFFETA_ARFM_2017} \cite{BOFFETA_ARFM_2017}.

\begin{figure}[t!]
\centering
\subfloat[Wave-number space]{\label{fig:4.2_2a}
\includegraphics[scale=0.13,trim=0 0 0 0,clip]{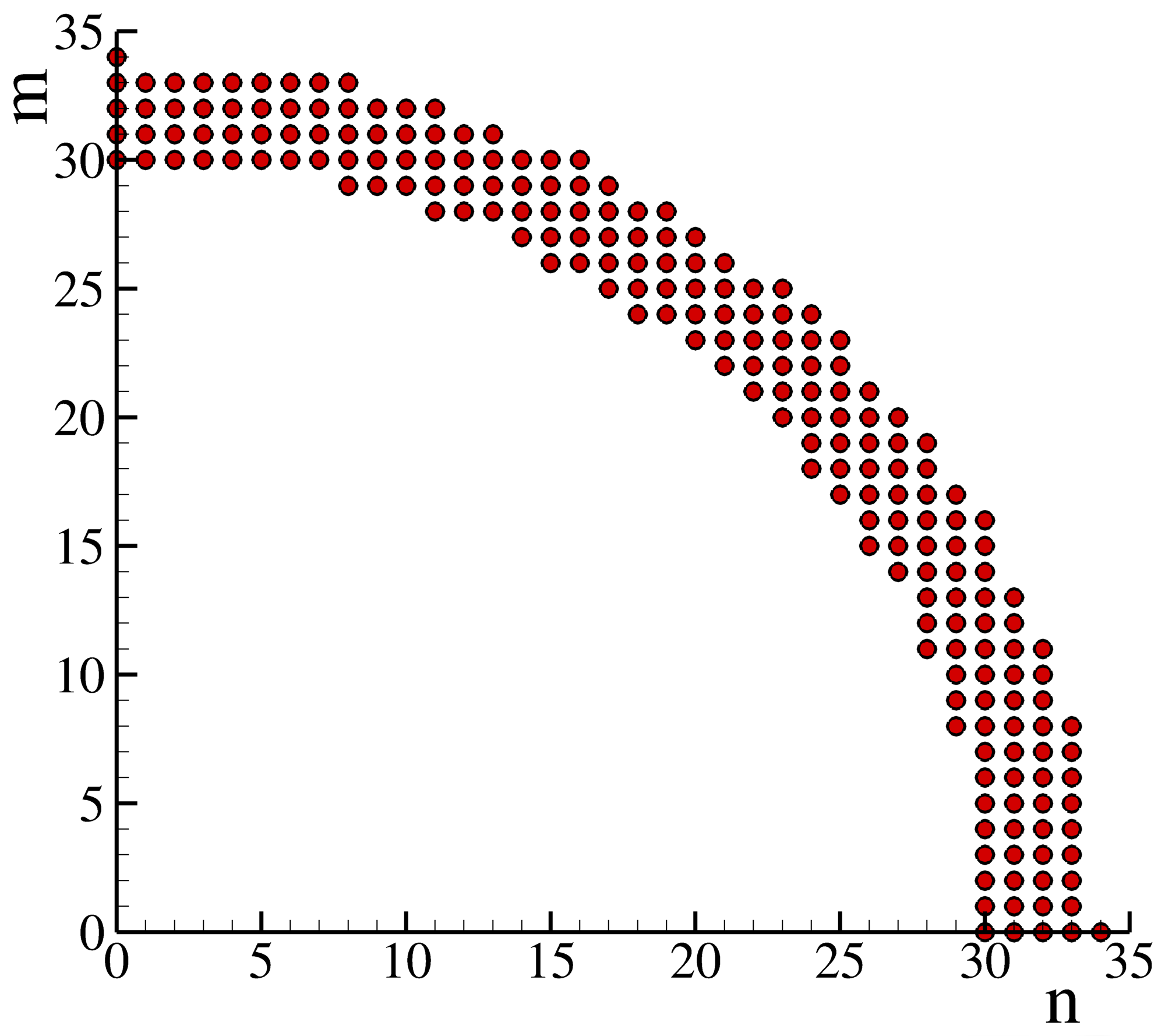}}
\\
\subfloat[Physical space]{\label{fig:4.2_2b}
\includegraphics[scale=0.15,trim=0 0 0 0,clip]{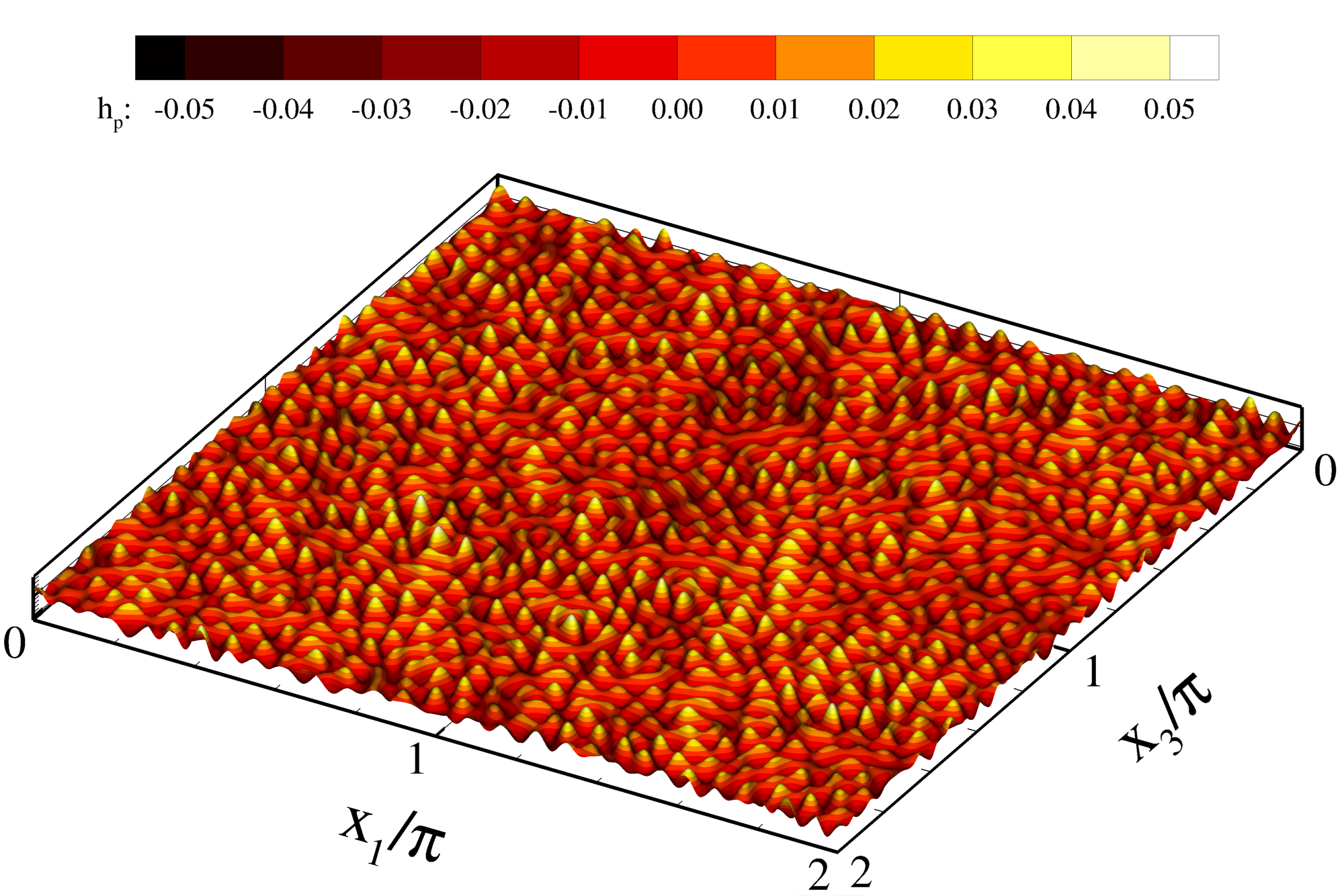}}
\caption{Initial perturbations at the interface ($x_2=0$) in wave-number (modes) and physical space.}
\label{fig:4.2_2}
\end{figure}

The flow configuration analyzed here is based on the DNS of \citeauthor{LIVESCU_PD_2020} \cite{LIVESCU_PD_2020} at $\mathrm{At}=0.5$. The computational domain is a rectangular prism defined in a Cartesian coordinate system $(x_1,x_2,x_3)$, figure \ref{fig:4.2_1}. Its cross-section is $L=2\pi$cm wide, and the height is $3L$ to ensure a negligible influence of the vertical boundaries on the simulations during the simulated time $T=25$ time-units (time normalized by $t^*=\sqrt(L/(32g\mathrm{At}))$ \cite{LIVESCU_PD_2020}). The bulk Reynolds number defined as $\mathrm{Re} \equiv h\dot{h}/\nu$ can reach $\mathrm{Re} \approx 500$ for the current settings ($h$ and $\dot{h}$ are the mixing-layer height and its temporal derivative). Periodic boundary conditions are applied on the lateral walls and reflective conditions on the vertical boundaries, $x_2=\pm 1.5L$. The domain height and simulation time guarantee that the simulations are not disturbed by the latter boundary condition.

The location of the interface between the two fluids is perturbed  by,
\begin{equation}
\label{4.2_1}
\begin{split}
h_p(x_1,x_3)= \sum_{n,m} 	& \cos \left[ 2 \pi \left( n \frac{x_1}{L} + r_1 \right)\right]  \\
							& \cos \left[ 2 \pi \left( m\frac{x_3}{L} + r_3 \right)\right]
\end{split}
\; .
\end{equation}
These perturbations possess wavelengths ranging from modes 30 to 34 {\color{black}($30\leq \sqrt{n^2 + m^2} \leq 34$)}, and amplitudes with standard-deviation not exceeding $0.04L$ \cite{BANERJEE_IJHMT_2009,GRINSTEIN_CMA_2019}, figure \ref{fig:4.2_2}. Note that $m$ and $n$ are selected to include the most unstable mode of the linearized problem \cite{DUFF_POF_1962,LIVESCU_PD_2020}. In equation \ref{4.2_1}, $r_1$ and $r_3$ are random numbers between 0 and 1. The numerical experiments rely on the ideal gas equation of state, and the initial temperature is set to maintain the flow $\mathrm{Ma}<0.10$ and guarantee incompressible flow. The initial thermodynamic and flow properties are defined as follows: $\mu_l=0.002$g/(cm.s), $\mu_h=0.006$g/(cm.s), $\rho_l=1.0$g/cm$^3$, $\rho_h=3.0$g/cm$^3$, $\gamma_l=\gamma_h=1.40$, $g=-980$cm/s$^2$, $k_o=10^{-6} \mathrm{cm^2/s^2}$, $S_o=10^{-6} \mathrm{cm}$, and Schmidt and Prandtl numbers are set equal to one.
%
%
%
\subsection{Numerical settings}
\label{sec:4.3}

All calculations are conducted with the flow solver xRAGE \cite{GITTINGS_CSD_2008}. This code utilizes a finite volume approach to solve the compressible and multi-material conservation equations for mass, momentum, energy, and species concentration. The resulting system of governing equations is resolved \cite{PEREIRA_CAF_2020} through the Harten-Lax-van Leer-Contact \cite{TORO_SW_1994} Riemann solver using a directionally unsplit strategy, direct remap, parabolic reconstruction \cite{COLLELA_JCP_1987}, and the low Mach number correction proposed by \citeauthor{THORNBER_JCP_2008} \cite{THORNBER_JCP_2008}. The equations are discretized with second-order accurate methods: the spatial discretization is based on a Godunov scheme, whereas the temporal discretization relies on an explicit Runge-Kutta scheme known as Heun's method. The time-step, $\Delta t$, is defined by prescribing the maximum instantaneous CFL number,
\begin{equation}
\label{eq:4.3_1}
\Delta t = \frac{ \Delta x \times \text{CFL}}{3(|V| + c)} \; ,
\end{equation}
where $c$ is the speed of sound, and $\Delta x$ is the grid cell size. The CFL is set equal to $0.45$ for the TGV and $0.50$ for the RT. The code can utilize an Adaptive Mesh Refinement (AMR) algorithm for following waves, especially shock-waves and contact discontinuities. This option is not used in the work to prevent hanging-nodes \cite{PEREIRA_MASTER_2012} and the simulations use orthogonal uniform hexahedral grids. For the TGV, these have $512^3$ elements for simulations at $f_k \ge 0.25$, and $1024^3$ cells for computations at $f_k=0.00$. This option keeps the numerical accuracy of computations at different $f_k$ uniform \cite{PEREIRA_PRF_2021}. On the other hand, the RT uses a mesh with $256^2\times 768$ cells  \cite{PEREIRA_POF_2021}. 

\begin{figure*}[t]
\centering
\subfloat[$t=3.0$]{\label{fig:5.1_1a}
\includegraphics[scale=0.073,trim=0 0 0 0,clip]{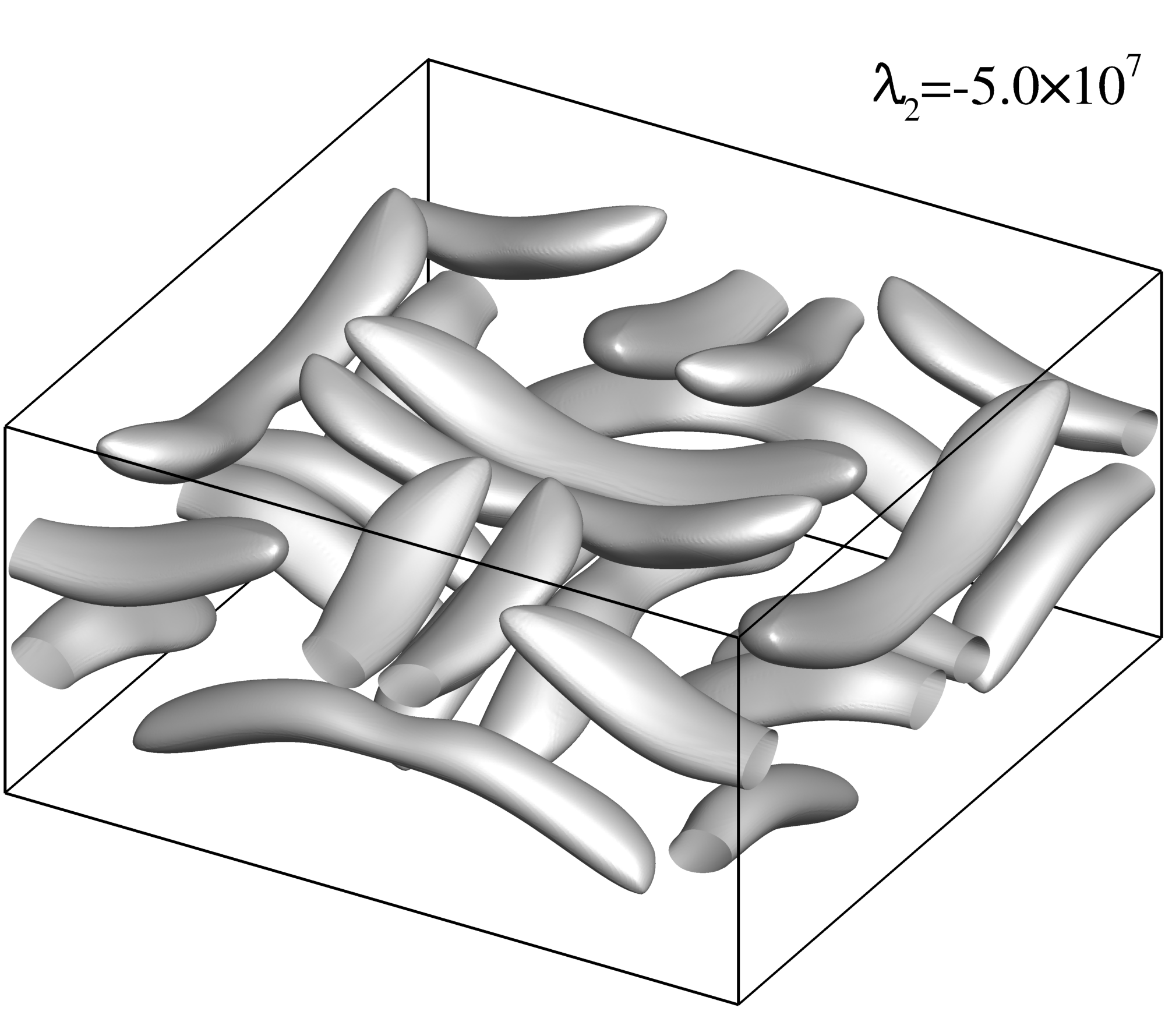}}
~
\subfloat[$t=5.0$]{\label{fig:5.1_1b}
\includegraphics[scale=0.073,trim=0 0 0 0,clip]{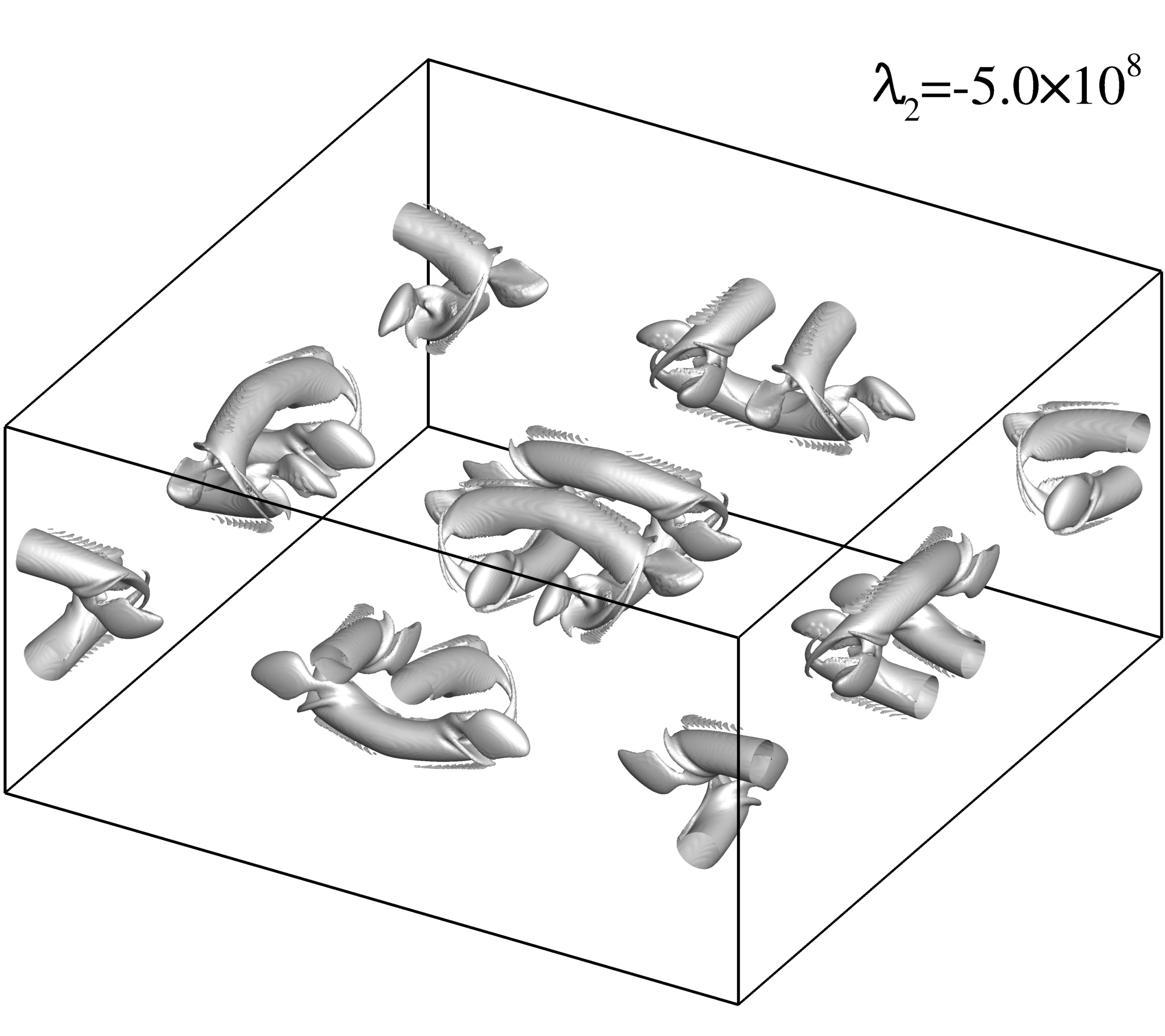}}
~
\subfloat[$t=7.0$]{\label{fig:5.1_1c}
\includegraphics[scale=0.073,trim=0 0 0 0,clip]{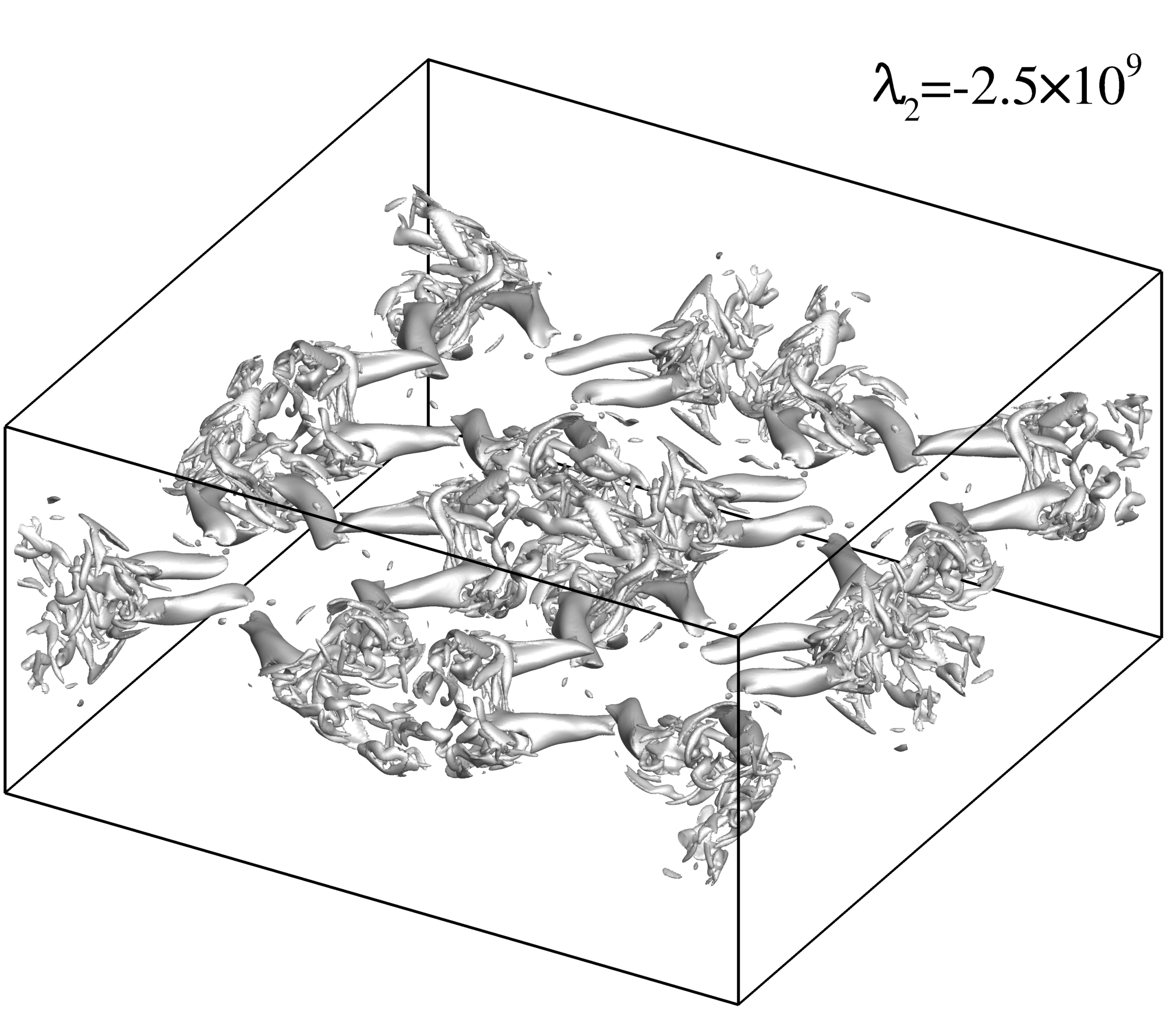}}
\\
\subfloat[$t=9.0$]{\label{fig:5.1_1d}
\includegraphics[scale=0.073,trim=0 0 0 0,clip]{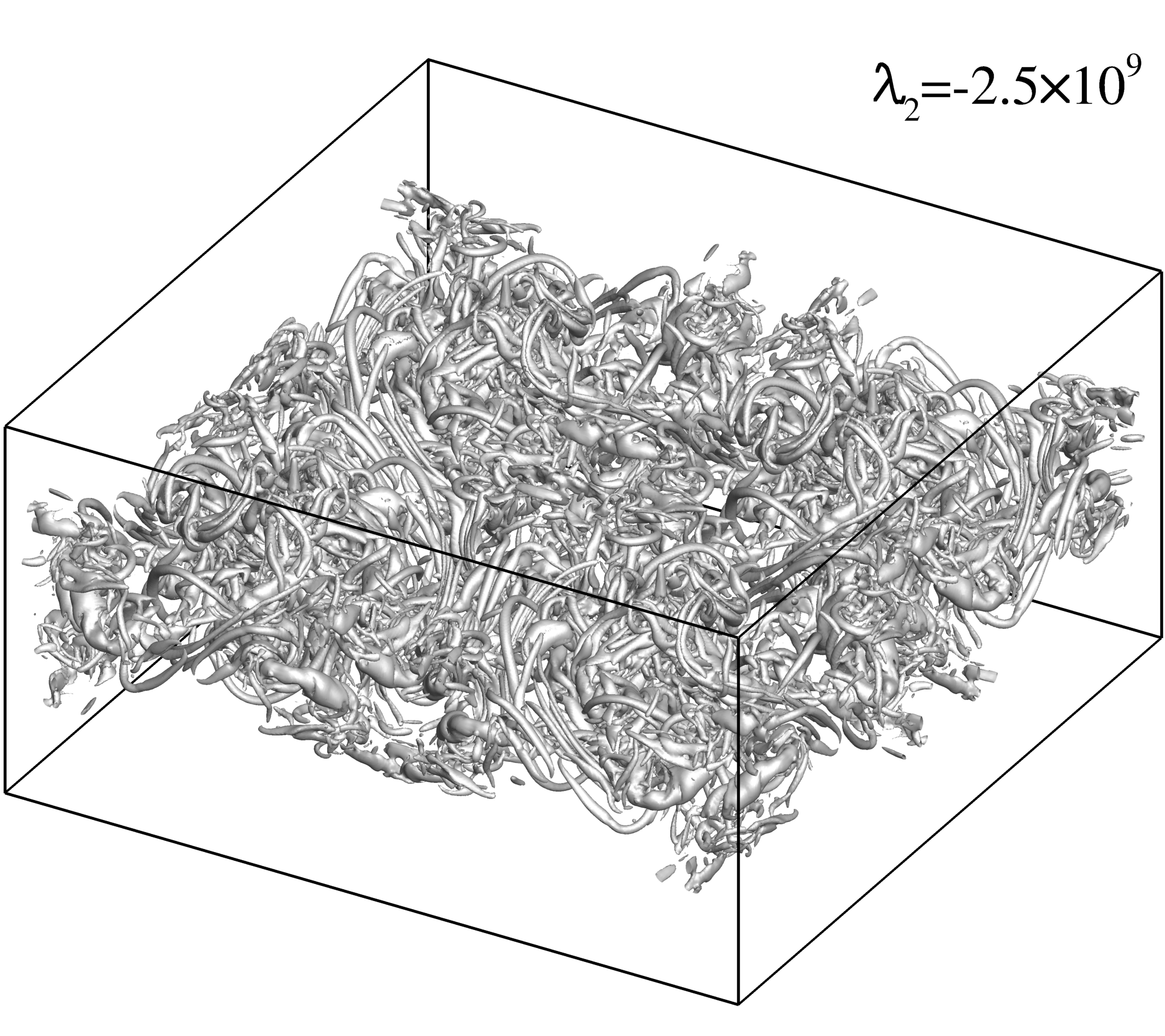}}
~
\subfloat[$t=12.0$]{\label{fig:5.1_1e}
\includegraphics[scale=0.073,trim=0 0 0 0,clip]{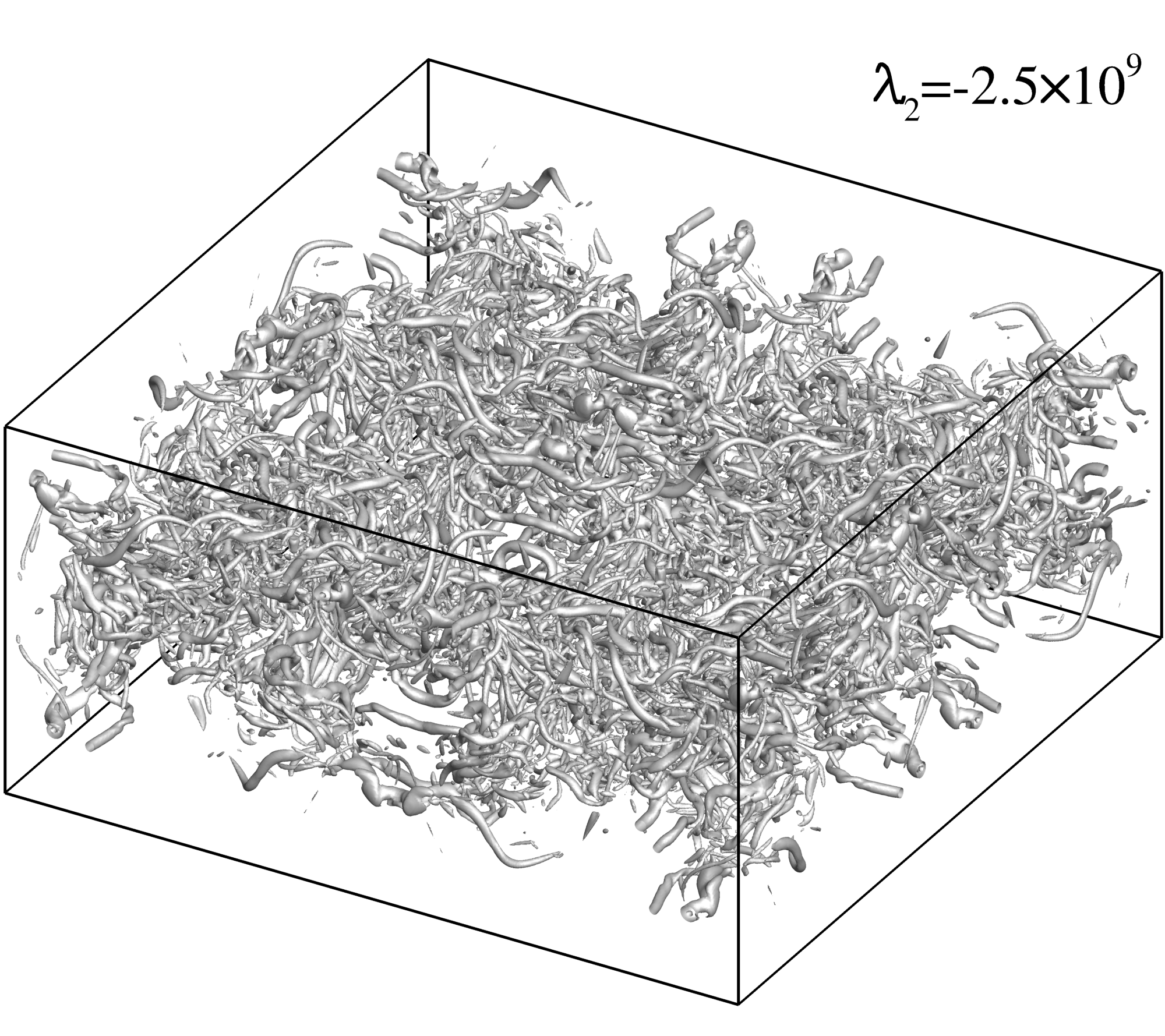}}
~
\subfloat[$t=20.0$]{\label{fig:5.1_1f}
\includegraphics[scale=0.073,trim=0 0 0 0,clip]{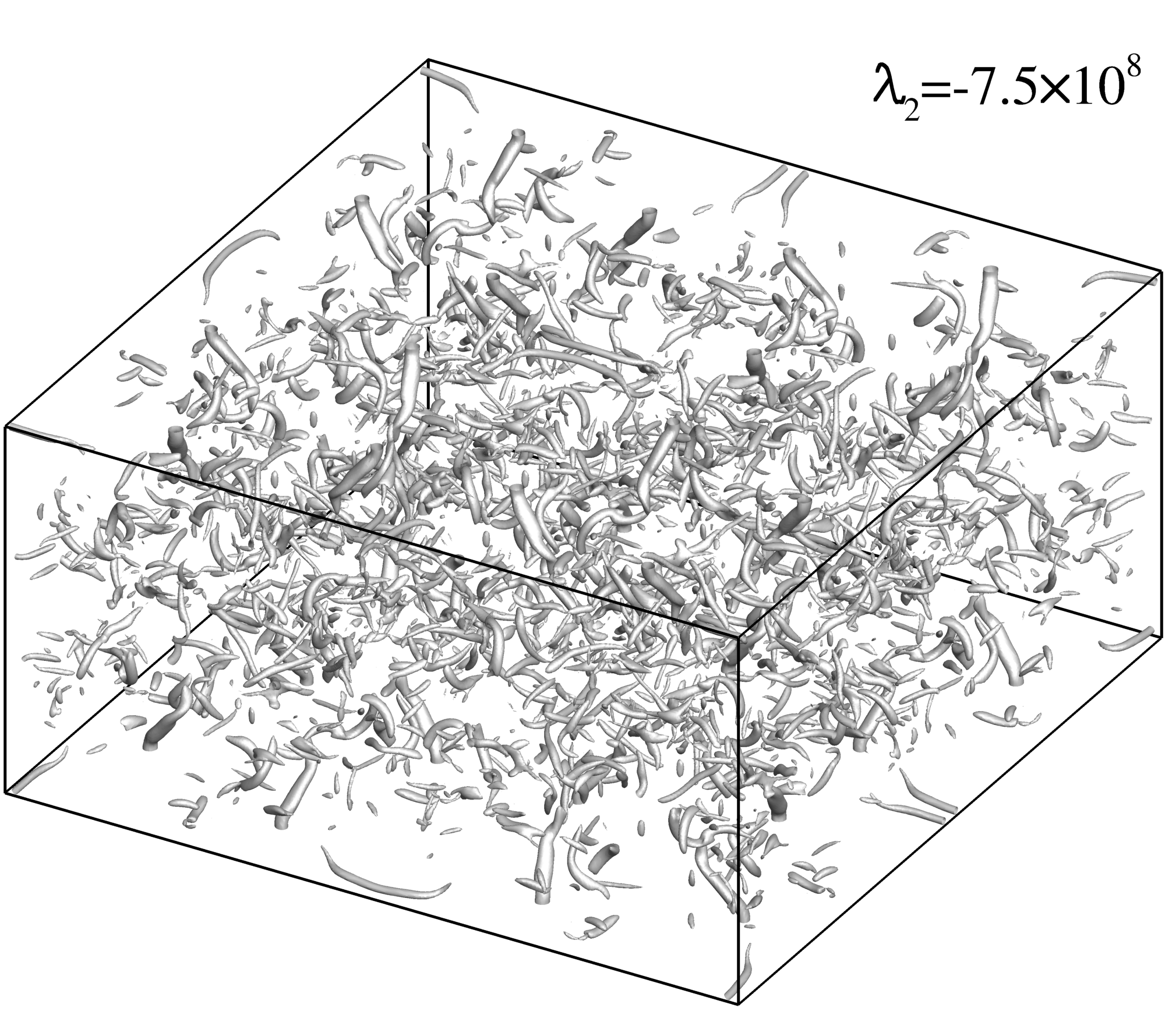}}
\caption{Temporal evolution of the coherent and turbulent structures of the TGV predicted with $f_k=0.00$ \cite{PEREIRA_PRF_2021}. Vortical structures identified with the $\lambda_2$ criterion \cite{JEONG_JFM_1995}.}
\label{fig:5.1_1}
\end{figure*}

xRAGE models miscible material interfaces and high convection-driven flows with a van-Leer limiter \cite{LEER_JCP_1997}, without artificial viscosity, and no material interface treatments \cite{GRINSTEIN_PF_2011,HAINES_PRE_2014}. The solver uses the assumption that cells containing more than one material are in pressure and temperature equilibrium as a mixed cell closure. The effective kinematic viscosity in multi-material problems \cite{PEREIRA_PRE_2021} is defined as
\begin{equation}
\label{eq:4.3_2}
\nu=\sum_{n=1}^{n_t} \nu_n f_n \; ,
\end{equation}
where $n$ is the material index, $n_t$ is the number of materials, and $f_n$ is the volume fraction of material $n$. For the RT flow, the diffusivity ${\cal{D}}$ and thermal conductivity $\kappa$ are defined by imposing Schmidt ($\mathrm{Sc}\equiv \nu/\cal{D}$) and Prandtl ($\mathrm{Pr}\equiv c_p\mu/\kappa$) numbers equal to one.

The RT computations test all three strategies for setting the parameters $f_\phi$ given is Section \ref{sec:3.1.3}. For the TGV simulations, where there are no $a_{i_u}$ and $b_u$ equations, the three strategies are equivalent.
%
%
\section{Results and Discussion}
\label{sec:5}

This section summarizes results for the TGV and RT results to illustrate the accuracy and potential of the proposed variable-density PANS formulation. Additional details of the TGV results are given in \citeauthor{PEREIRA_PRF_2021} \cite{PEREIRA_PRF_2021}. More detailed analysis of the RT will be the subject of a subsequent manuscript \cite{PEREIRA_POF_2021}. Apart from two- and three-dimensional field plots, all results have been spatially averaged.
%
%
\subsection{Taylor-Green vortex}
\label{sec:5.1}

As previously mentioned, the TGV initially features the laminar, single-mode, and well-defined vortical structures depicted in figure \ref{fig:4.1_1}. Immediately after $t=0$, these coherent structures start interacting and deforming, leading to vortex-stretching processes that generate the pairs of long sheet-like vortices observed in figure \ref{fig:5.1_1} at $t=3.0$. Between $t=3.0$ and $7.0$, these structures get closer and undergo a complex vortex-reconnection mechanism \cite{BRACHET_JFM_1983,KIDA_ARFM_1994,PEREIRA_PRF_2021} between pairs of counter-rotating vortices, figure \ref{fig:5.1_1b}. This triggers the onset of the turbulence at $t\approx 7.0$. Figure \ref{fig:5.1_1c} shows the bursts of small turbulence scales at this instant. Afterward, turbulence further develops and eventually decays. This is illustrated in figures \ref{fig:5.1_1d} to \ref{fig:5.1_1f}. Considering the flow evolution, the physics of the first nine time-units is expected to pose the greatest challenges to modeling and simulation of the TGV flow.

\begin{figure}[t!]
\centering
\includegraphics[scale=0.11,trim=0 0 0 0,clip]{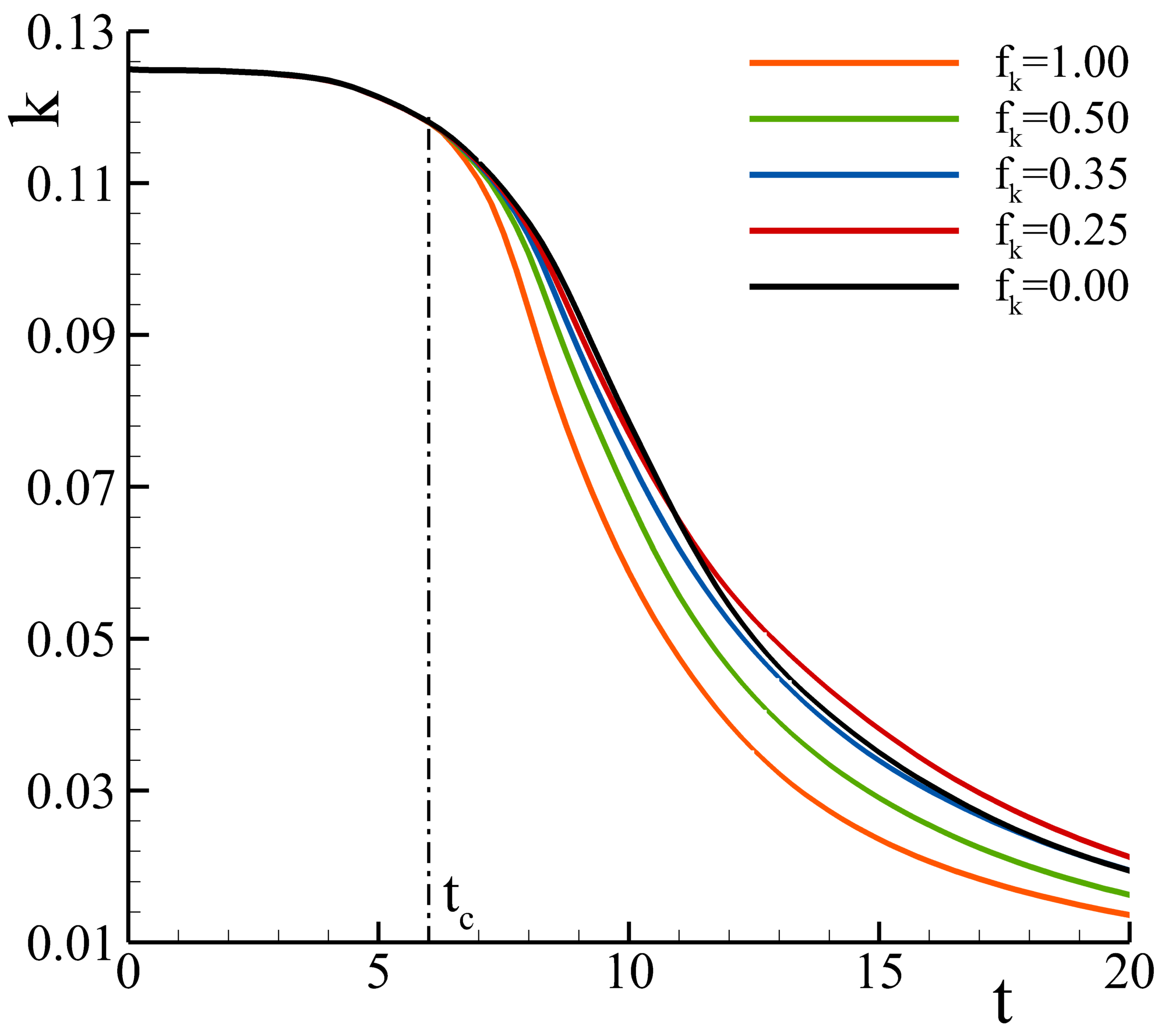}
\caption{Temporal evolution of the total kinetic energy, $k$, for predictions at different $f_k$.}
\label{fig:5.1_2}
\end{figure}

Figure \ref{fig:5.1_2} presents the temporal evolution of the total kinetic energy, $k$, predicted by PANS BHR-LEVM at different degrees of physical resolution, $f_k$. Note that $k$ comprises a resolved, $k_r$, and unresolved, $k_u$, component,
\begin{equation}
\label{eq:5.1_1}
k=k_r+k_u \; ,
\end{equation}
which are obtained from the resolved velocity field and the turbulence closure, respectively. It is important to stress that the resolved component of $k$ comprises a non-turbulent \cite{PALKIN_FTC_2016} and turbulent component, while the unresolved part entails the turbulent fraction of $k$ being modeled. The results indicate that $k$ is initially nearly constant, and independent of $f_k$ until $t=t_c\approx 6.0$. At this instant, in which the flow undergoes vortex-reconnection processes, the simulations become strongly dependent on $f_k$. This result allows us to categorize the simulations into high- (HPR, $f_k<0.50$) and low- (LPR, $f_k\ge 0.50$) physical resolution. The data show that LPR simulations lead to a pronounced non-physical decay of $k$. As discussed later, this is caused by a rapid increase and overprediction of the modeled turbulent stresses. In contrast, HPR computations exhibit smaller energy decay rates and, as such, larger values of $k$ at late times. Yet, the most significant result of figure \ref{fig:5.1_2} is that the solutions convergence upon physical resolution refinement ($f_k\rightarrow 0$). It also shows that all HPR solutions are in good agreement. This behavior is particularly evident until $t=10$.

\begin{figure}[t!]
\centering
\includegraphics[scale=0.11,trim=0 0 0 0,clip]{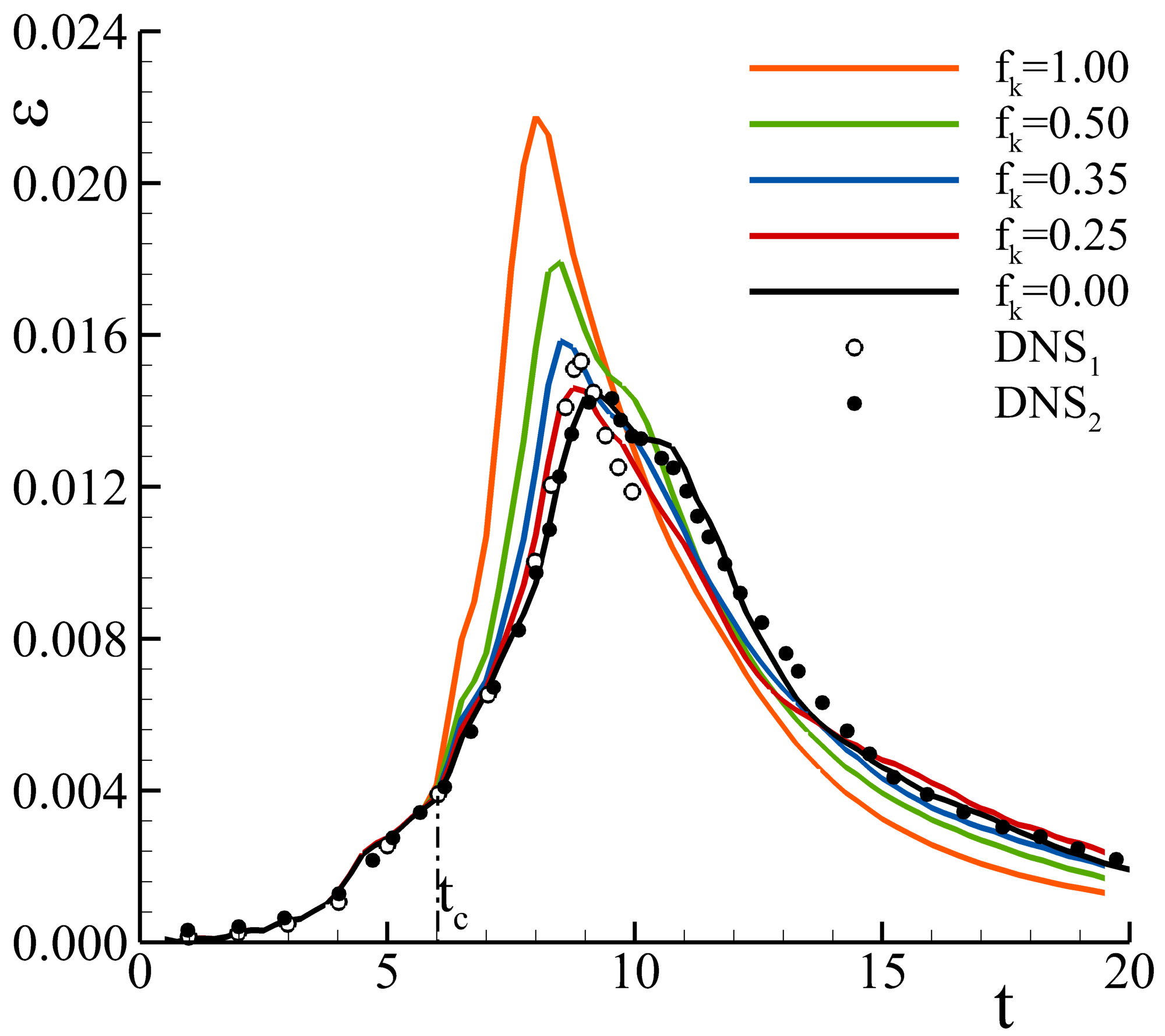}
\caption{Temporal evolution of the total kinetic energy dissipation, $\varepsilon$ $(\mathrm{s^{-1}})$, for predictions at different $f_k$.}
\label{fig:5.1_3}
\end{figure}

Next, figure \ref{fig:5.1_3} depicts the temporal evolution of the dissipation of total kinetic energy, $\varepsilon$,
\begin{equation}
\label{eq:5.1_2}
\varepsilon=-\frac{d k}{d t}\; ,
\end{equation}
and compares the results against the DNS of \citeauthor{BRACHET_JFM_1983} \cite{BRACHET_JFM_1983} at $\mathrm{Ma}=0$ ($\mathrm{DNS_1}$) and \citeauthor{DRIKAKIS_JOT_2007} \cite{DRIKAKIS_JOT_2007} at $\mathrm{Ma}=0.28$ ($\mathrm{DNS_2}$). The results exhibit similar tendencies to those of $k$. Until $t=t_c$, all simulations are independent of the physical resolution and in excellent agreement with the reference DNS solutions. After this instant, the computations become closely dependent on $f_k$, but their solutions converge toward the reference DNS data upon physical resolution refinement, $f_k \rightarrow 0$. Most notably, it is once again possible to distinguish between HPR and LPR computations. The first show a diminishing dependence on $f_k$, and a good agreement with the reference numerical experiments \cite{BRACHET_JFM_1983,DRIKAKIS_JOT_2007}. Considering the cases at $f_k \leq 0.25$, the maximum values of $\varepsilon$ at $8.8 \leq t \leq 9.3$ range from $0.145$ to $0.146$, whereas the DNS studies report values between $0.143$ and $0.153$. The largest discrepancies between HPR and DNS computations occur at late times. These are likely caused by numerical uncertainty and compressibility effects \cite{VIRK_JFM_1995,PEREIRA_PRF_2021}.

On the other hand, LPR computations lead to large discrepancies compared to the reference DNS studies. The peak of dissipation occurs prematurely, and its magnitude is clearly overpredicted. As for $k$, the differences grow as the physical resolution coarsens, $f_k\rightarrow 1.0$. For instance, the magnitude of the dissipation peak can reach $0.218$ ($f_k=1.00$), exceeding the value reported by the reference DNS studies in more than $50\%$. Also, the dissipation peak occurs at $t\approx 8$ instead of between $t=9$ and $9.3$ \cite{BRACHET_JFM_1983,DRIKAKIS_JOT_2007}. 

\begin{figure}[t!]
\centering
\includegraphics[scale=0.11,trim=0 0 0 0,clip]{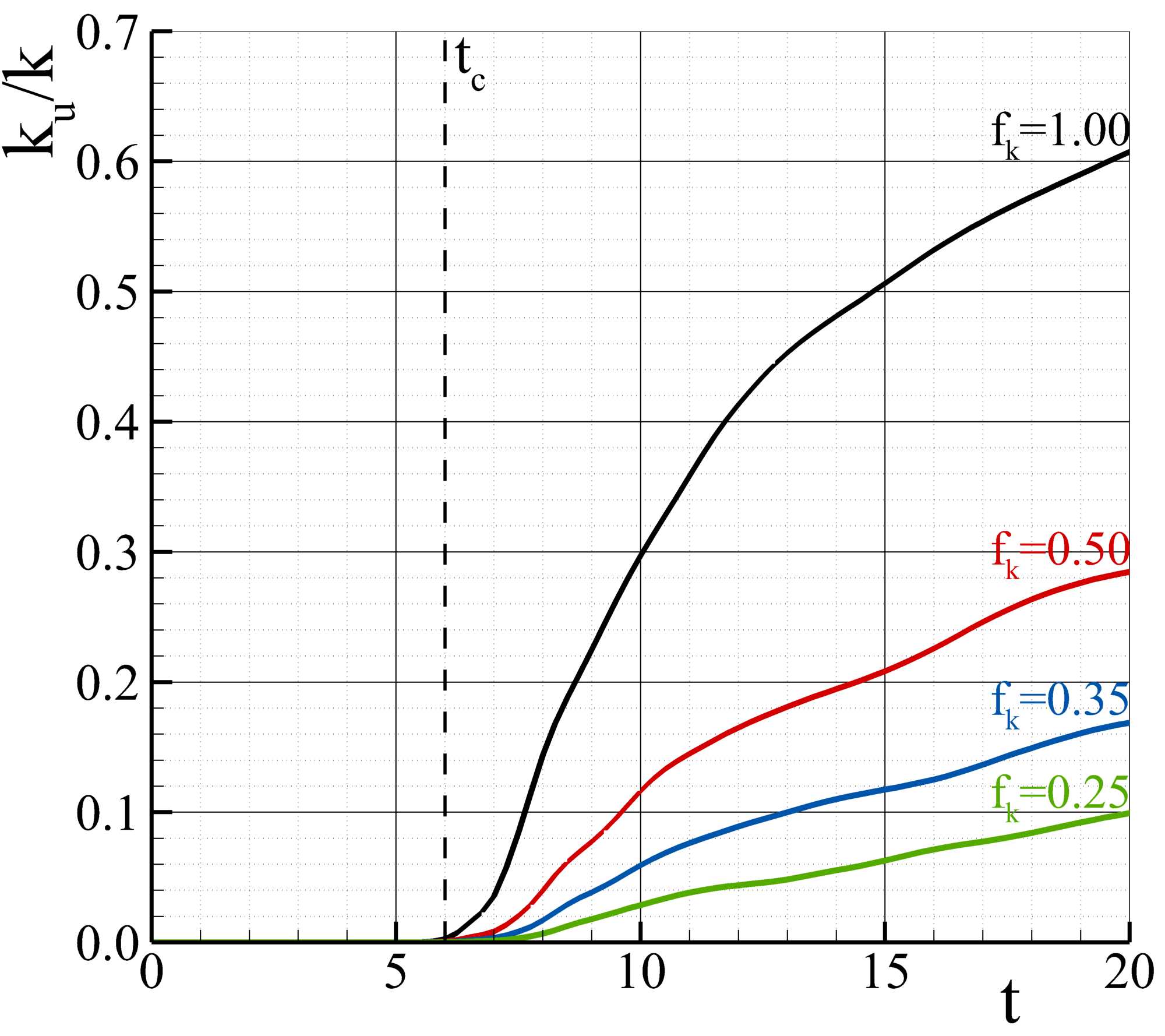}
\caption{Temporal evolution of the ratio modeled-to-total kinetic energy, $k_u/k$, for predictions at different $f_k$.}
\label{fig:5.1_4}
\end{figure}

\begin{figure}[t]
\centering
\subfloat[$f_k=0.00$]{\label{fig:5.1_5a}
\includegraphics[scale=0.105,trim=0 0 0 0,clip]{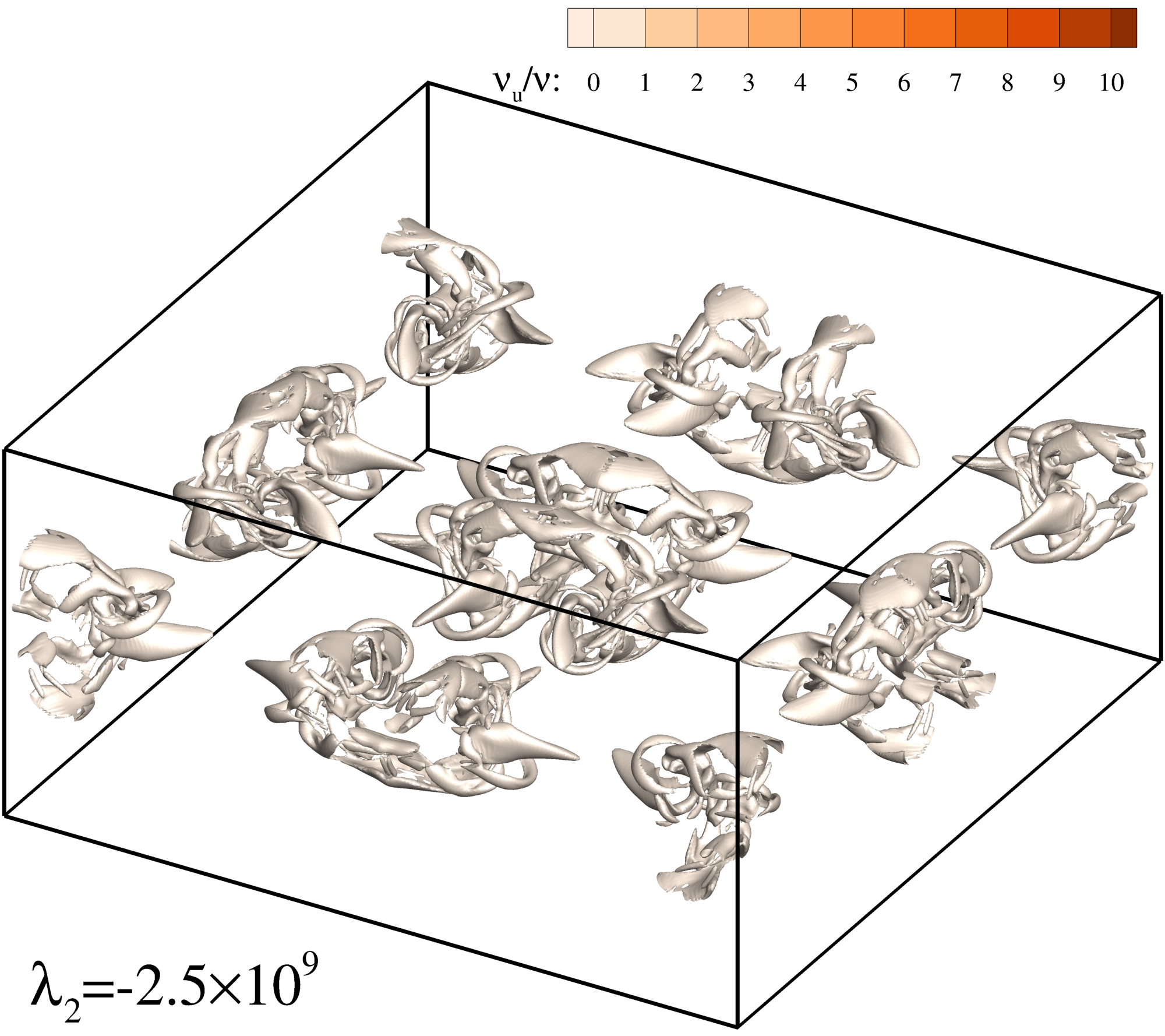}}
\\
\subfloat[$f_k=0.35$]{\label{fig:5.1_5b}
\includegraphics[scale=0.105,trim=0 0 0 0,clip]{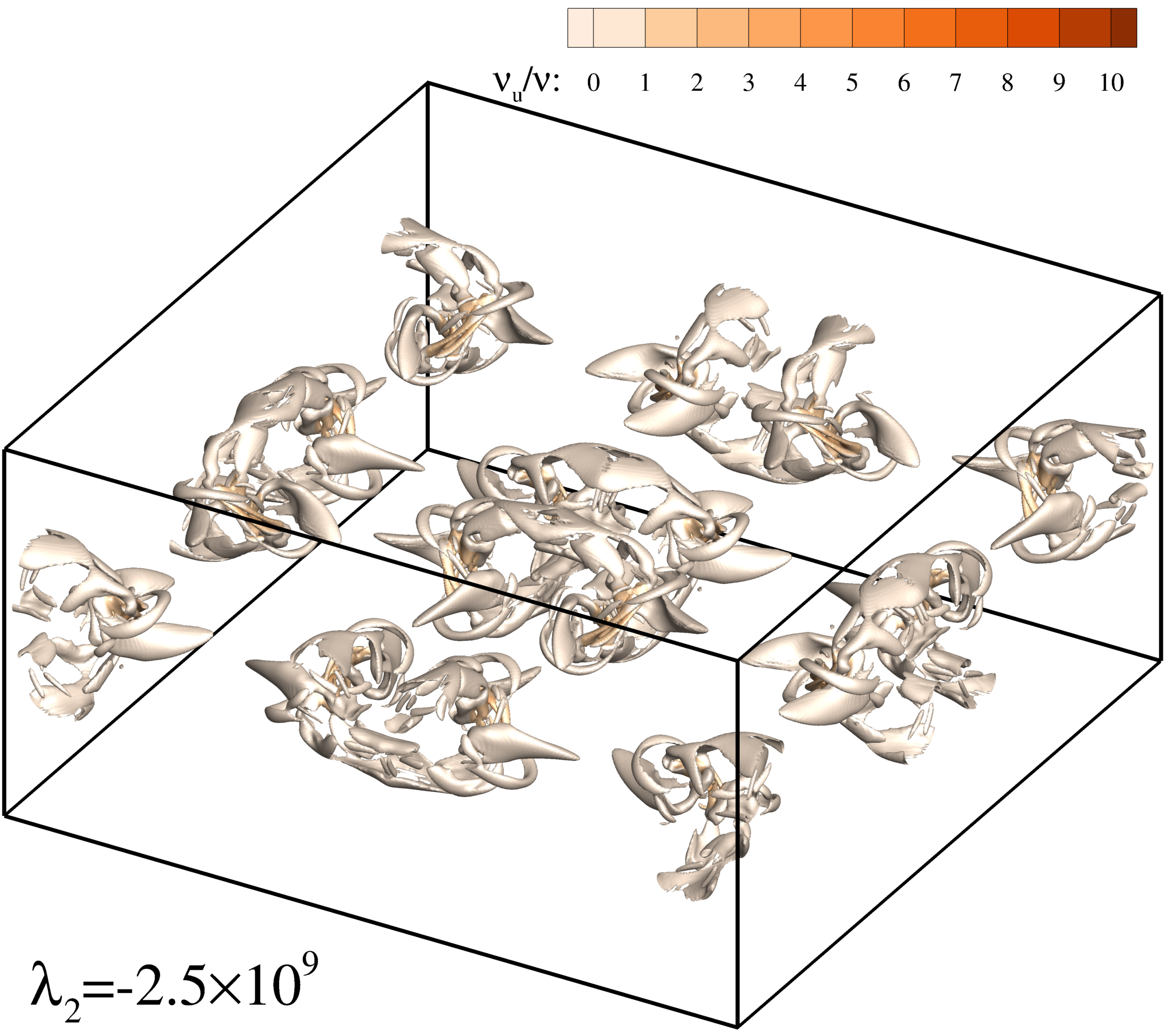}}
\\
\subfloat[$f_k=1.00$]{\label{fig:5.1_5c}
\includegraphics[scale=0.105,trim=0 0 0 0,clip]{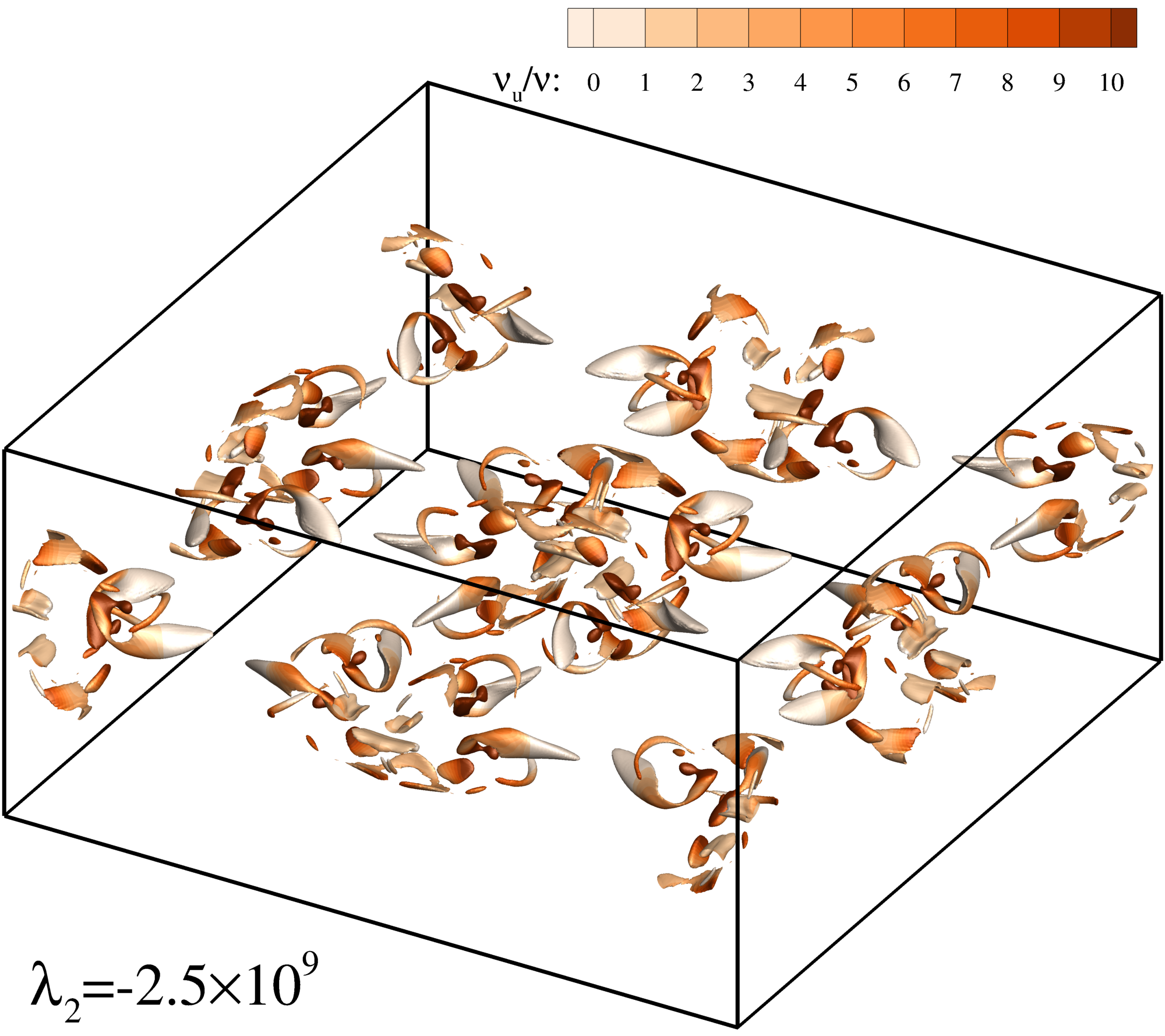}}
\caption{Temporal evolution of the coherent and turbulent structures of the TGV at $t=6.5$ predicted with different $f_k$ \cite{PEREIRA_PRF_2021}. Vortical structures identified with the $\lambda_2$ criterion \cite{JEONG_JFM_1995}.}
\label{fig:5.1_5}
\end{figure}

The results of figures \ref{fig:5.1_2} and \ref{fig:5.1_3} suggest that LPR simulations prematurely predict the onset of turbulence, overpredicting the unresolved turbulent stress tensor. This would explain the rapid decay of $k$ observed in figure \ref{fig:5.1_2}. These ideas are supported by the ratio of unresolved-to-total kinetic energy, $k_u/k$, depicted in figure \ref{fig:5.1_4}. The data indicate that $k_u/k$ is negligible and independent of $f_k$ until $t=t_c$. After this instant, $k_u/k$ grows considerably, and its magnitude becomes closely dependent on $f_k$. Also, it is visible that the growth of $k_u/k$ starts earlier and it is more rapid for simulations at $f_k=1.00$ than at $f_k=0.25$. For example, $k_u/k$ predicted at $t=20$ and $f_k=1.00$ is six times larger than that at $f_k=0.25$.  Considering the results for $\varepsilon$, this shows that simulations at $f_k=1.00$ overpredict the turbulent stresses. We emphasize that transient flows are highly sensitive to history effects, and high-resolution PANS simulations showing $f_k\approx (f_k)_e=k_u/k_{u(f_k=1.00)}$ would indicate that the RANS closure can accurately represent the mean-flow field of the selected problem. This is often observed in statistically steady flows.

It is also interesting to note that $k_u/k$ only starts growing rapidly at $t \ge 7.0$ for HPR simulations. As shown in figure \ref{fig:5.1_1}, this corresponds to the instant when the onset of turbulence is expected to occur \cite{BRACHET_JFM_1983}. Hence, we can infer that LPR simulations misrepresent the onset of turbulence due to the overprediction of the unresolved turbulent stress tensor, leading to a poor prediction of the vortex-reconnection process and consequent premature onset of turbulence. This is illustrated in figure \ref{fig:5.1_5} which depicts the coherent and turbulent flow structures predicted at $t=6.5$ at representative values of $f_k$. The plots show that the LPR simulation ($f_k=1.00$) dissipates the laminar coherent structures involved in the vortex-reconnection processes. This is caused by the overprediction of turbulence since $\nu_t$ (see equation \ref{eq:2_16}) can exceed its laminar counterpart by a factor of $30$. This does not occur at $f_k=0.35$ (highest HPR $f_k)$ nor at $f_k=0.00$. A comprehensive assessment of this flow is given in \cite{PEREIRA_PRF_2021}.

In summary, the results indicate that the proposed PANS BHR-LEVM model can accurately predict the shear driven TGV flow using $f_k<0.50$. At such values of $f_k$, the computations exhibit a relatively small dependence on the physical resolution and are able to resolve the phenomena not amenable to straightforward closure modeling. Next, we evaluate the performance of the model predicting the buoyancy driven RT flow.
%
%
%
%
\subsection{Rayleigh-Taylor}
\label{sec:5.2}

\begin{figure*}[t]
\centering
\subfloat[$t=0.0$]{\label{fig:5.2_1a}
\includegraphics[scale=0.08,trim=0 0 0 0,clip]{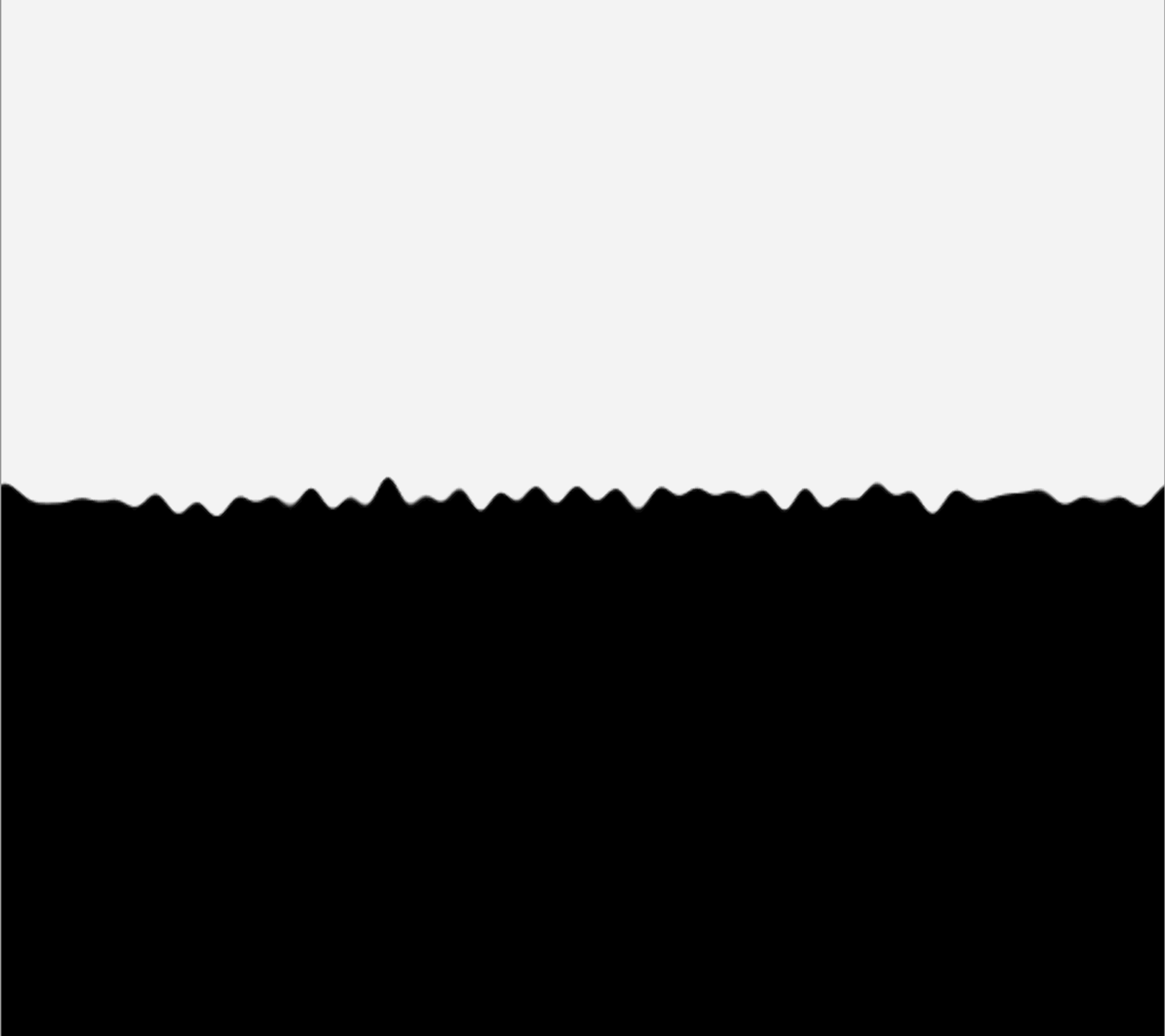}}
~
\subfloat[$t=1.0$]{\label{fig:5.2_1b}
\includegraphics[scale=0.08,trim=0 0 0 0,clip]{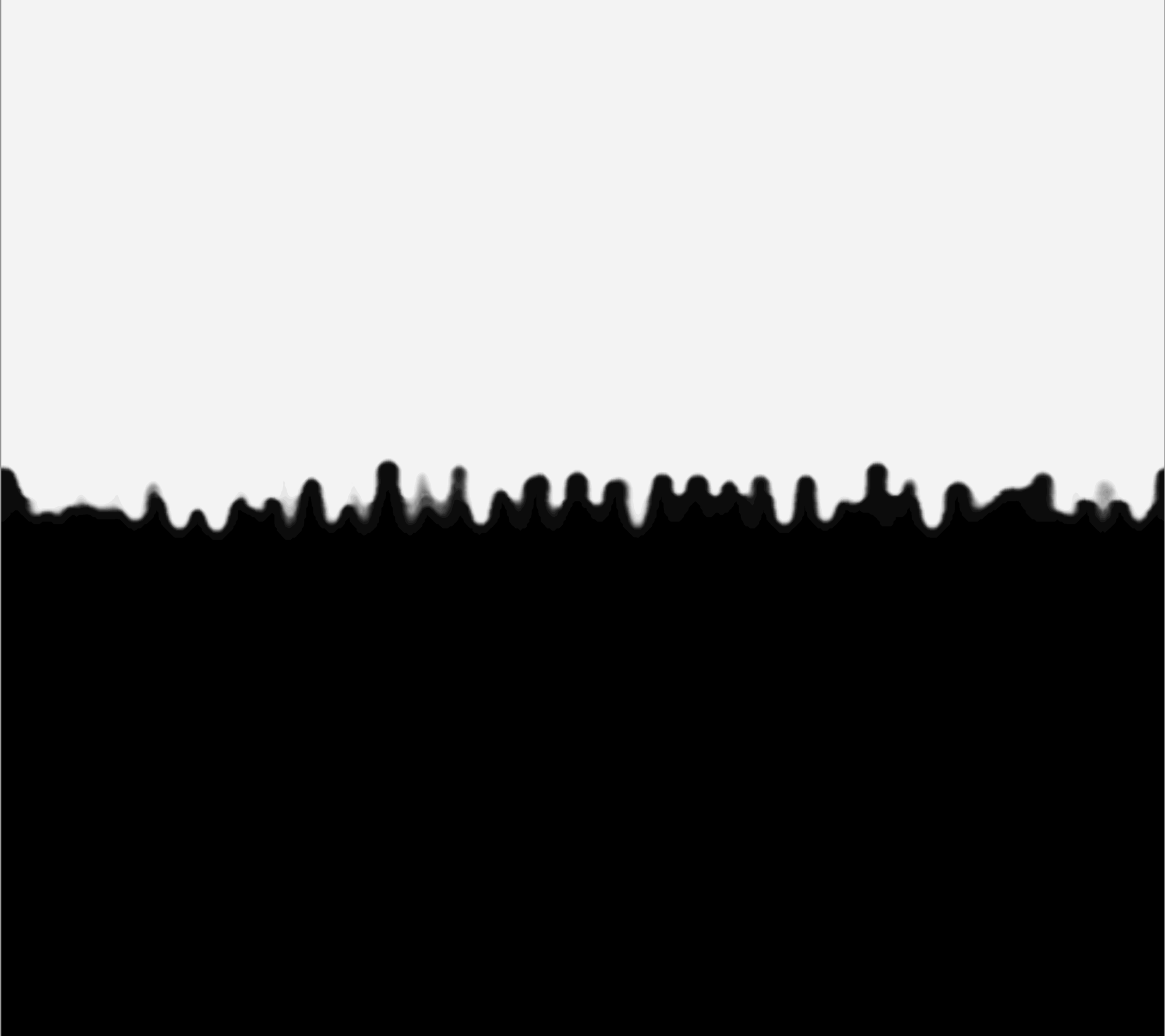}}
~
\subfloat[$t=1.5$]{\label{fig:5.2_1c}
\includegraphics[scale=0.08,trim=0 0 0 0,clip]{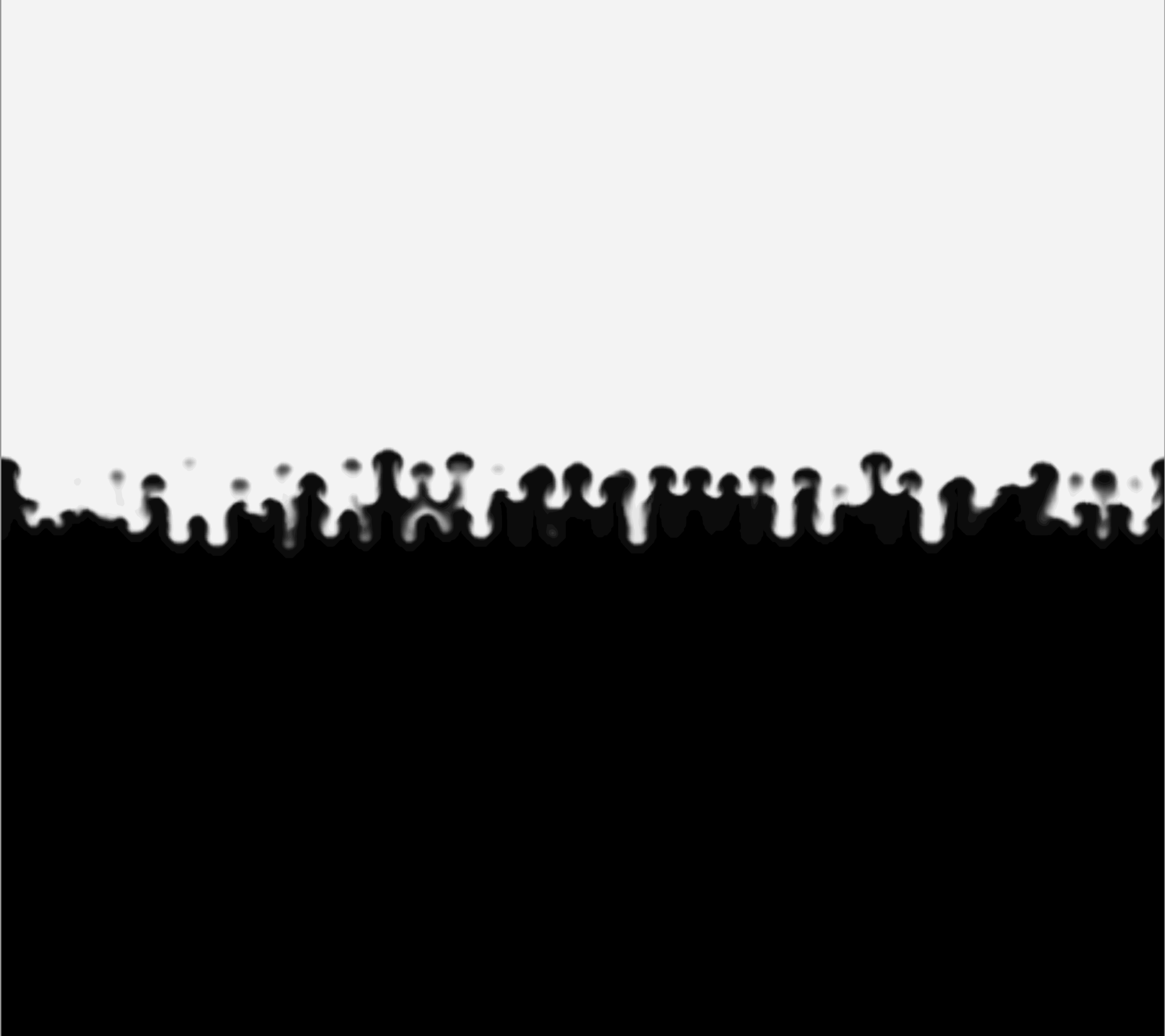}}
\\
\subfloat[$t=2.0$]{\label{fig:5.2_1d}
\includegraphics[scale=0.08,trim=0 0 0 0,clip]{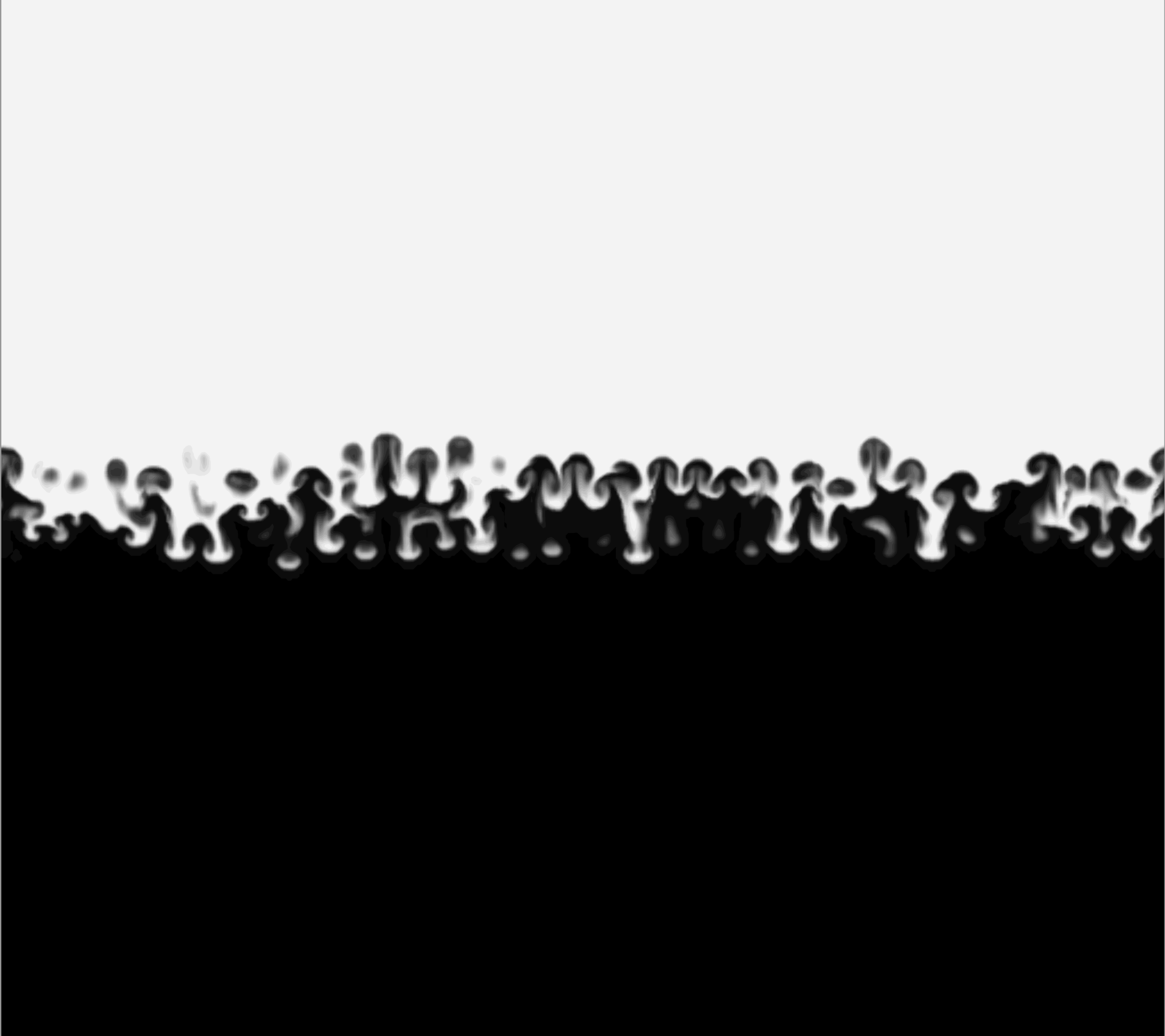}}
~
\subfloat[$t=2.5$]{\label{fig:5.2_1e}
\includegraphics[scale=0.08,trim=0 0 0 0,clip]{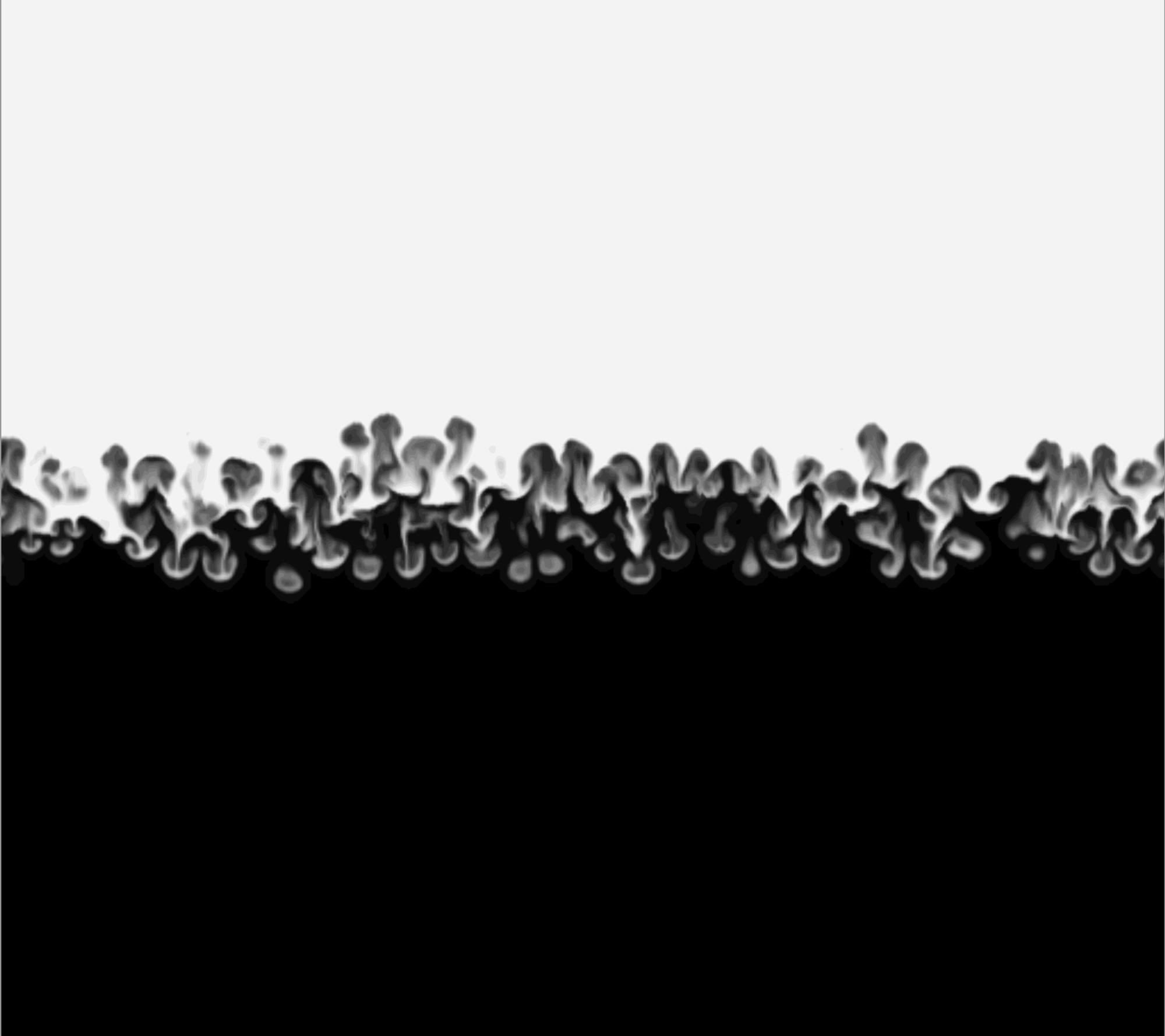}}
~
\subfloat[$t=3.0$]{\label{fig:5.2_1f}
\includegraphics[scale=0.08,trim=0 0 0 0,clip]{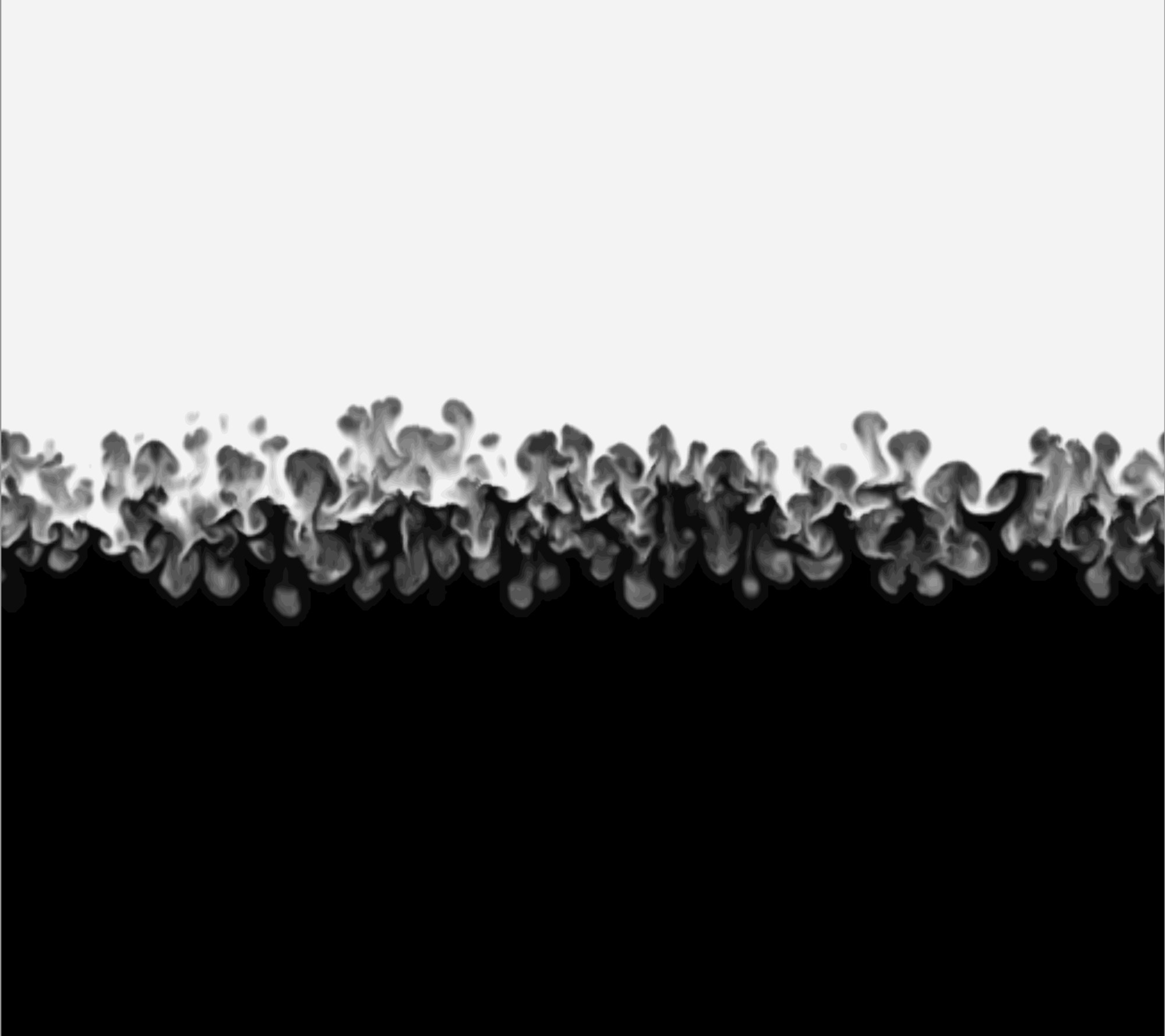}}
\\
\subfloat[$t=3.5$]{\label{fig:5.2_1g}
\includegraphics[scale=0.08,trim=0 0 0 0,clip]{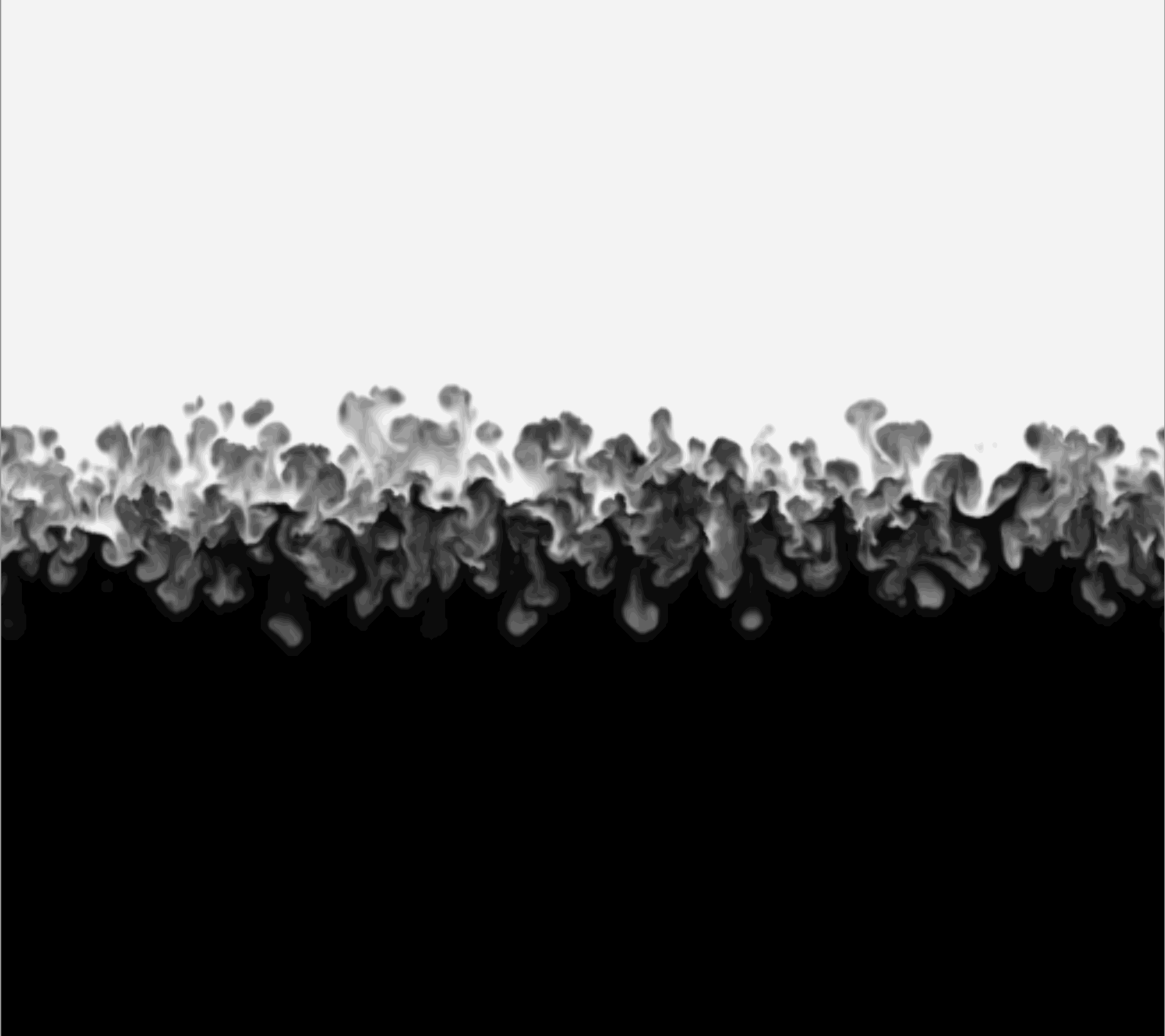}}
~
\subfloat[$t=4.0$]{\label{fig:5.2_1h}
\includegraphics[scale=0.08,trim=0 0 0 0,clip]{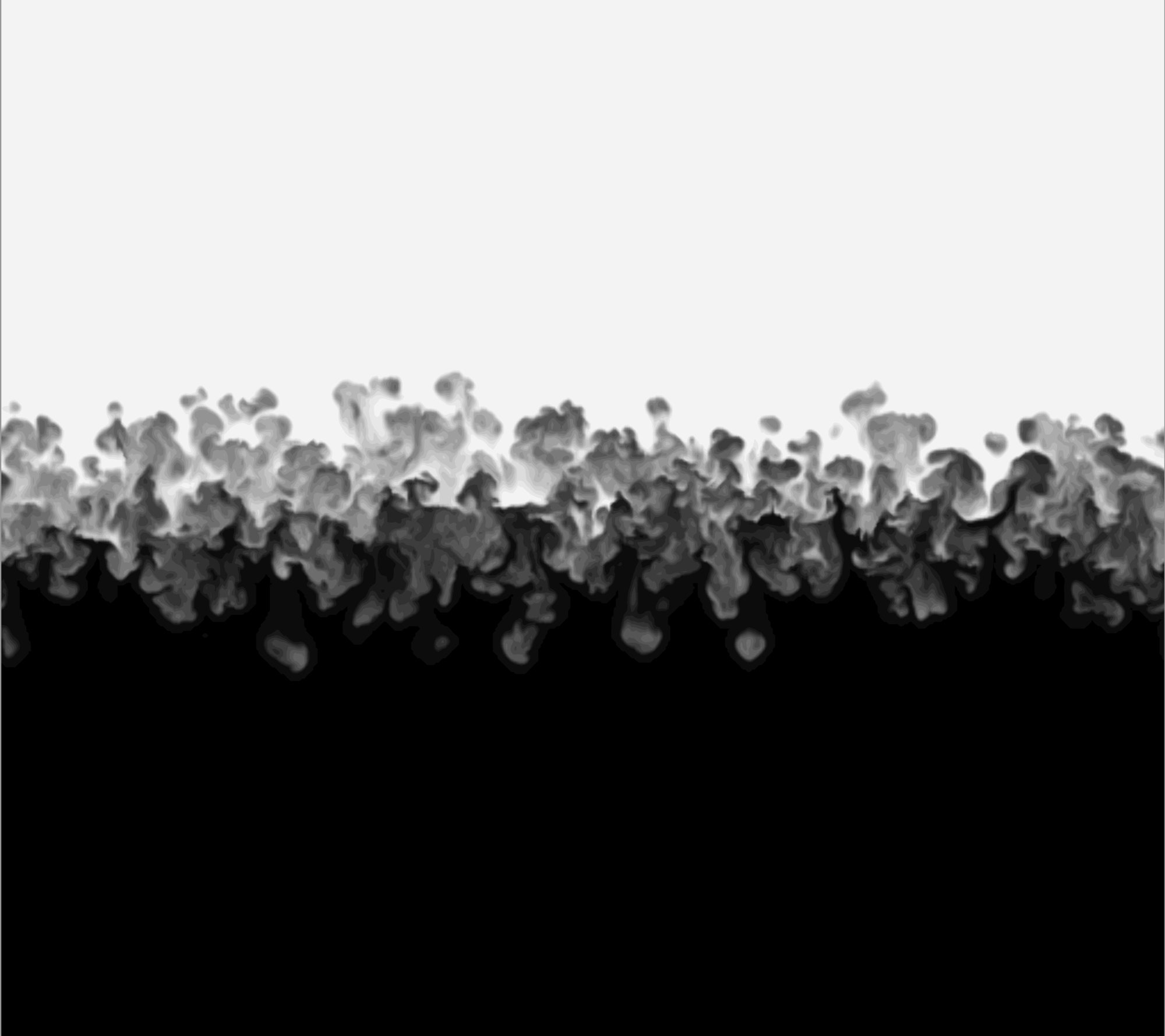}}
~
\subfloat[$t=5.0$]{\label{fig:5.2_1i}
\includegraphics[scale=0.08,trim=0 0 0 0,clip]{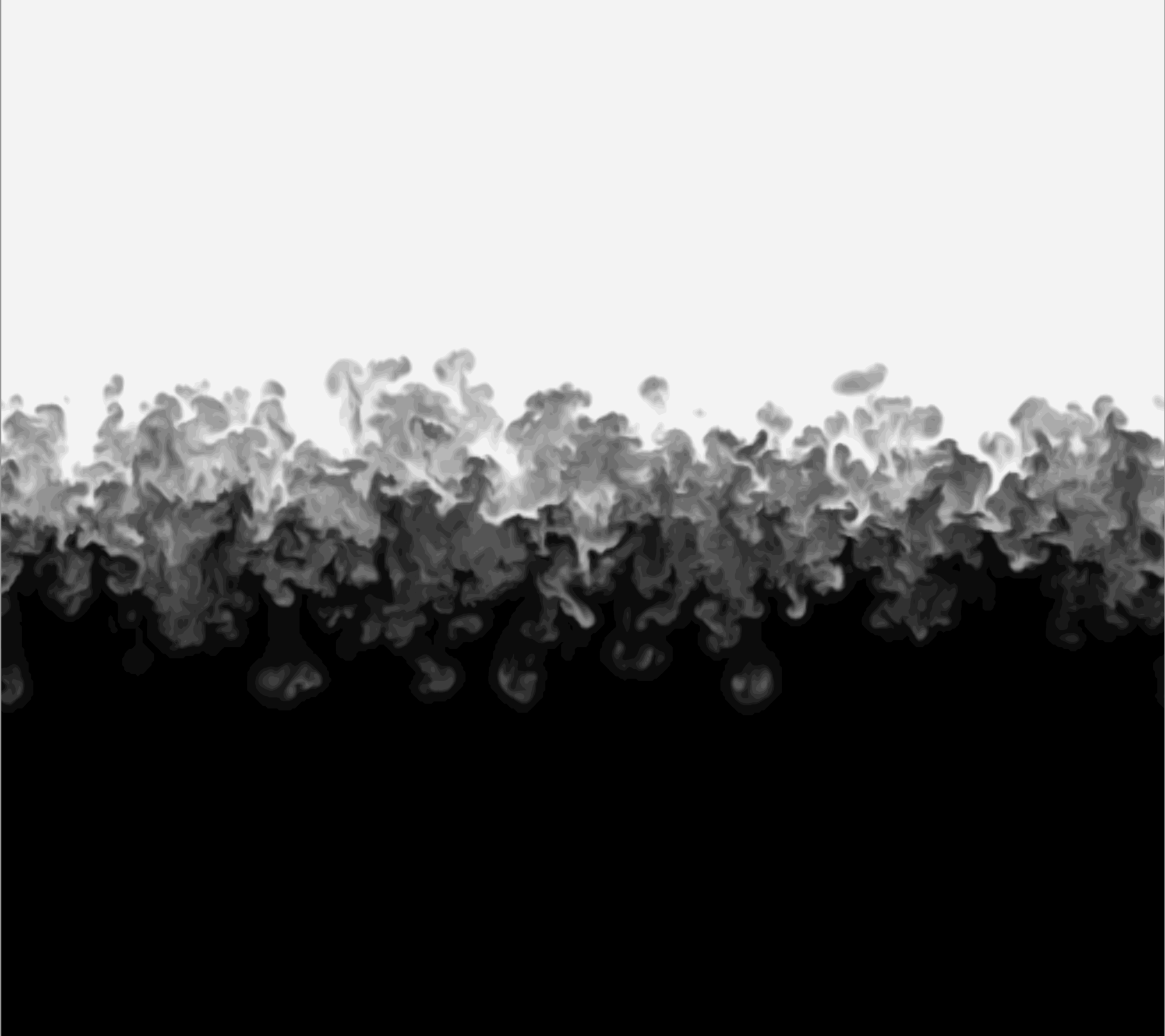}}
\caption{Temporal evolution of the RT density field predicted at early flow stages using $f_k=0.00$.}
\label{fig:5.2_1}
\end{figure*}

\begin{figure*}[t]
\centering
\subfloat[$t=2.5$]{\label{fig:5.2_2a}
\includegraphics[scale=0.095,trim=0 0 370 0,clip]{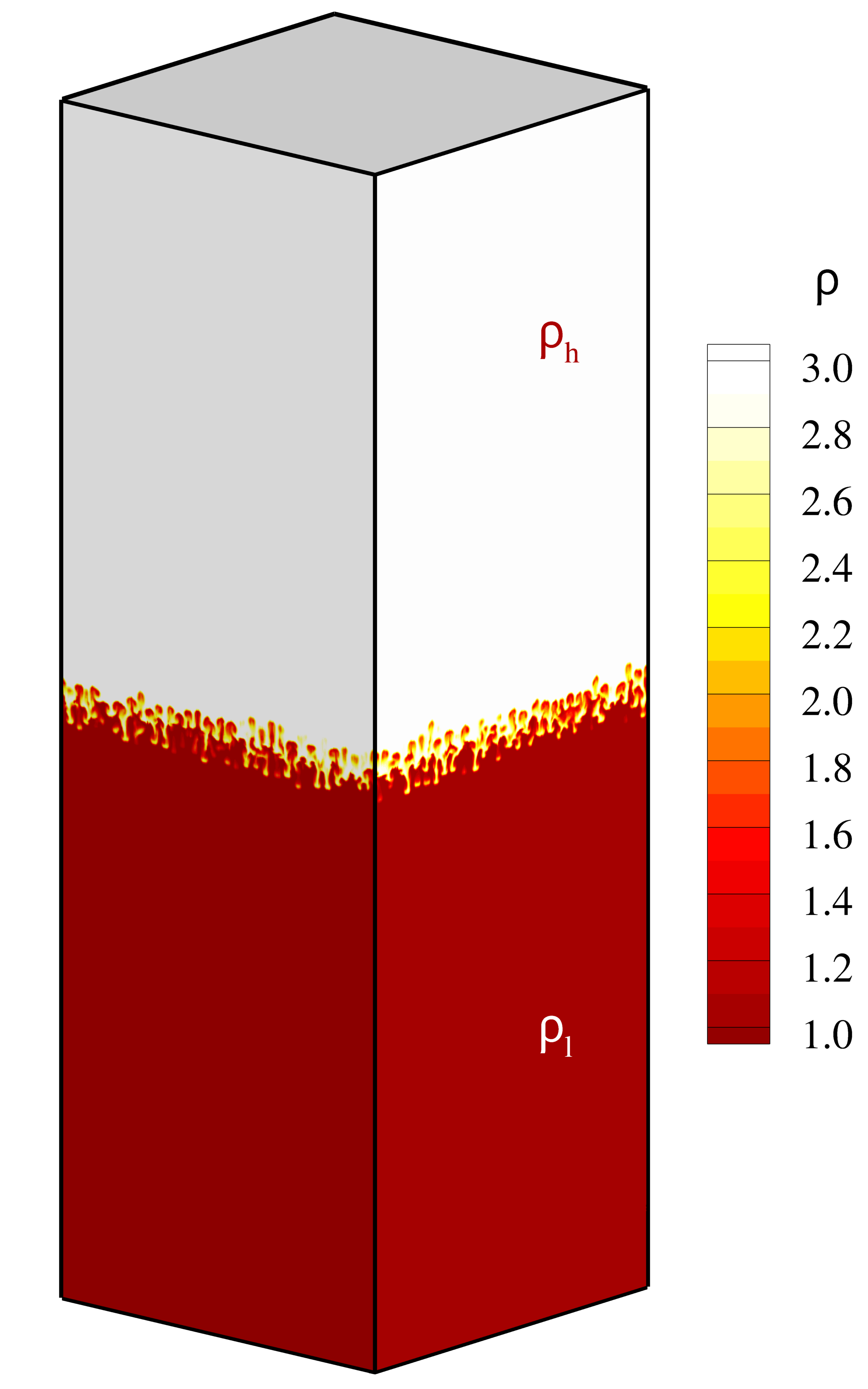}}
~
\subfloat[$t=5.0$]{\label{fig:5.2_2b}
\includegraphics[scale=0.095,trim=0 0 370 0,clip]{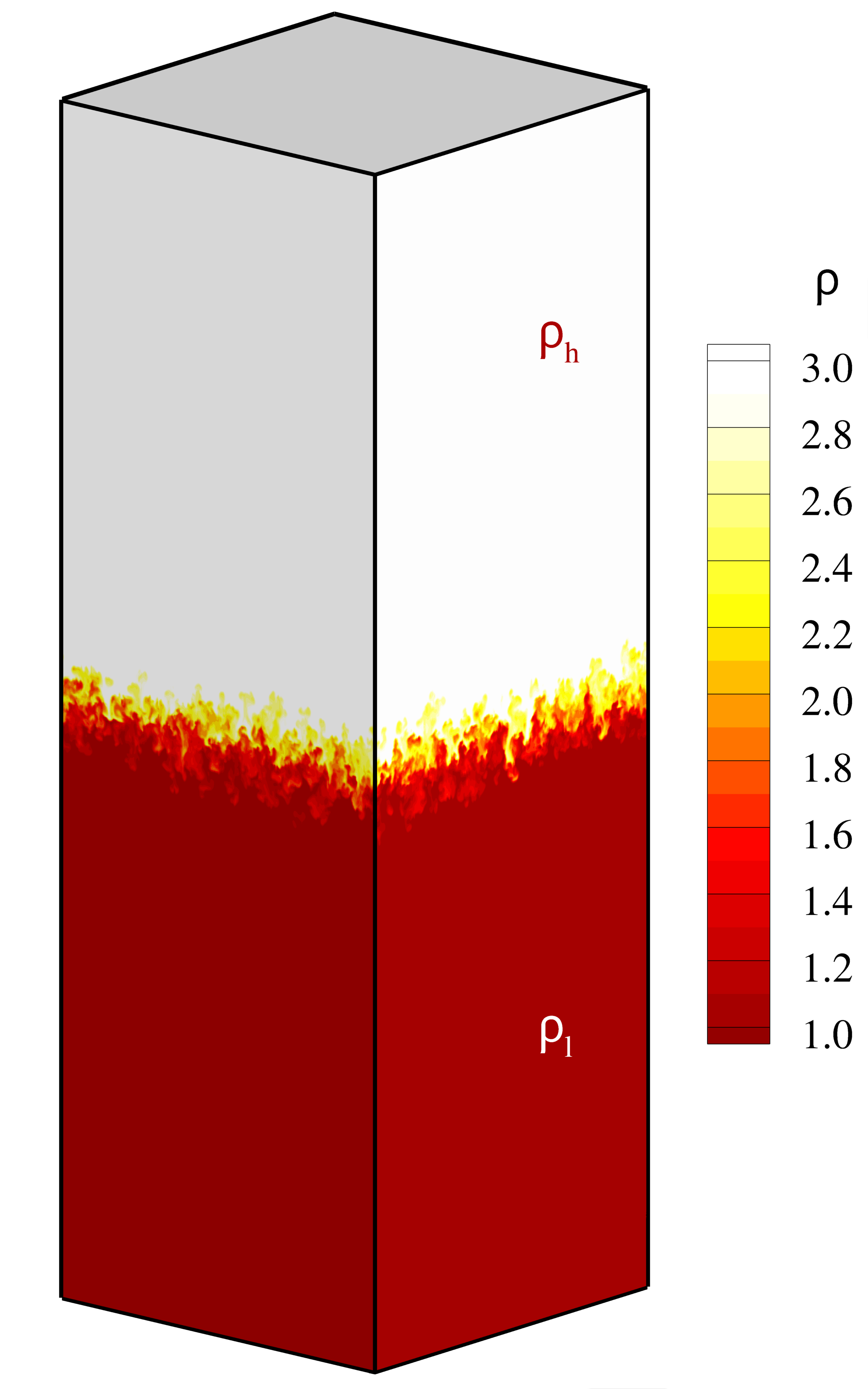}}
~
\subfloat[$t=10.0$]{\label{fig:5.2_2c}
\includegraphics[scale=0.095,trim=0 0 370 0,clip]{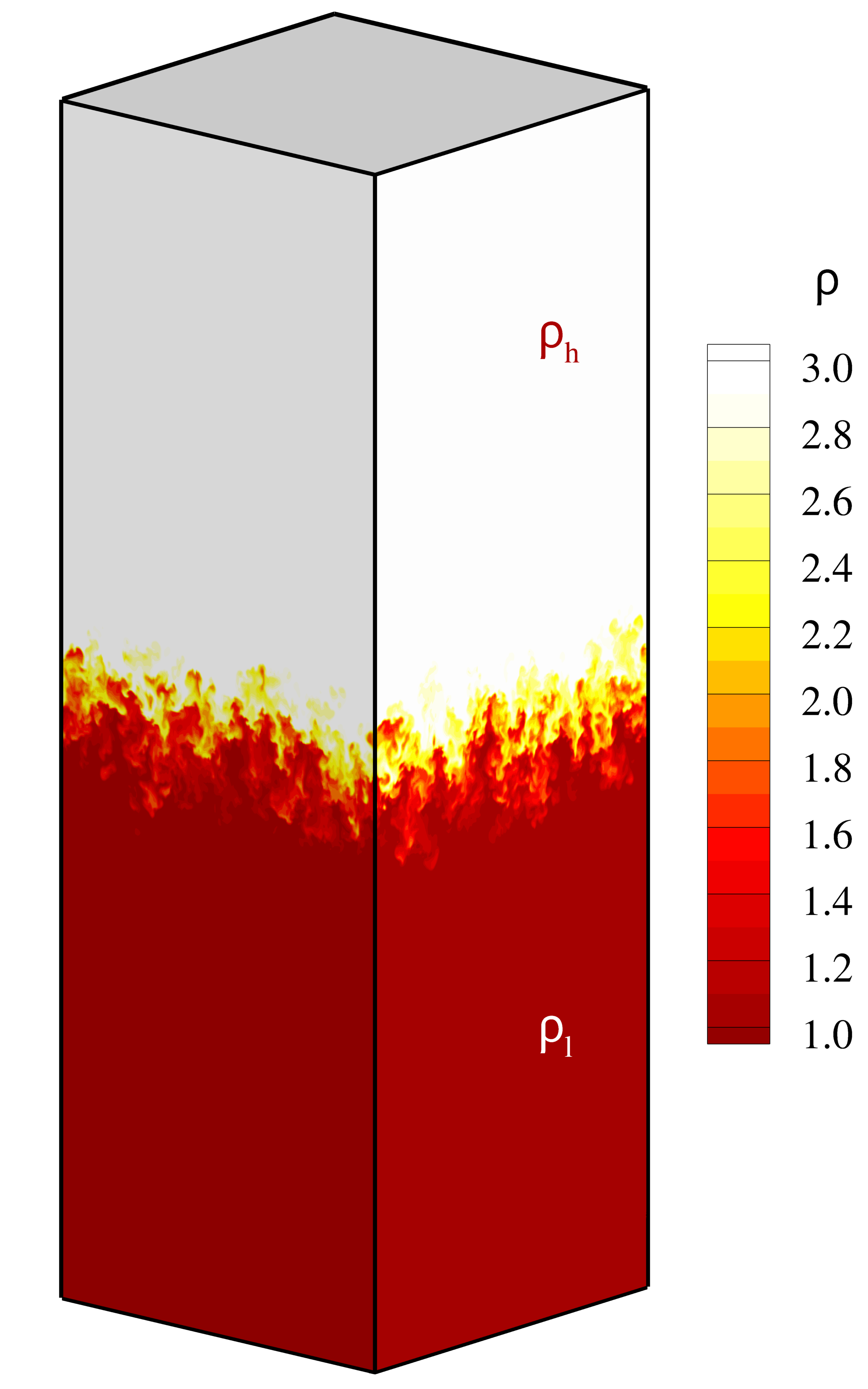}}
~
\subfloat[$t=20.0$]{\label{fig:5.2_2d}
\includegraphics[scale=0.095,trim=0 0 370 0,clip]{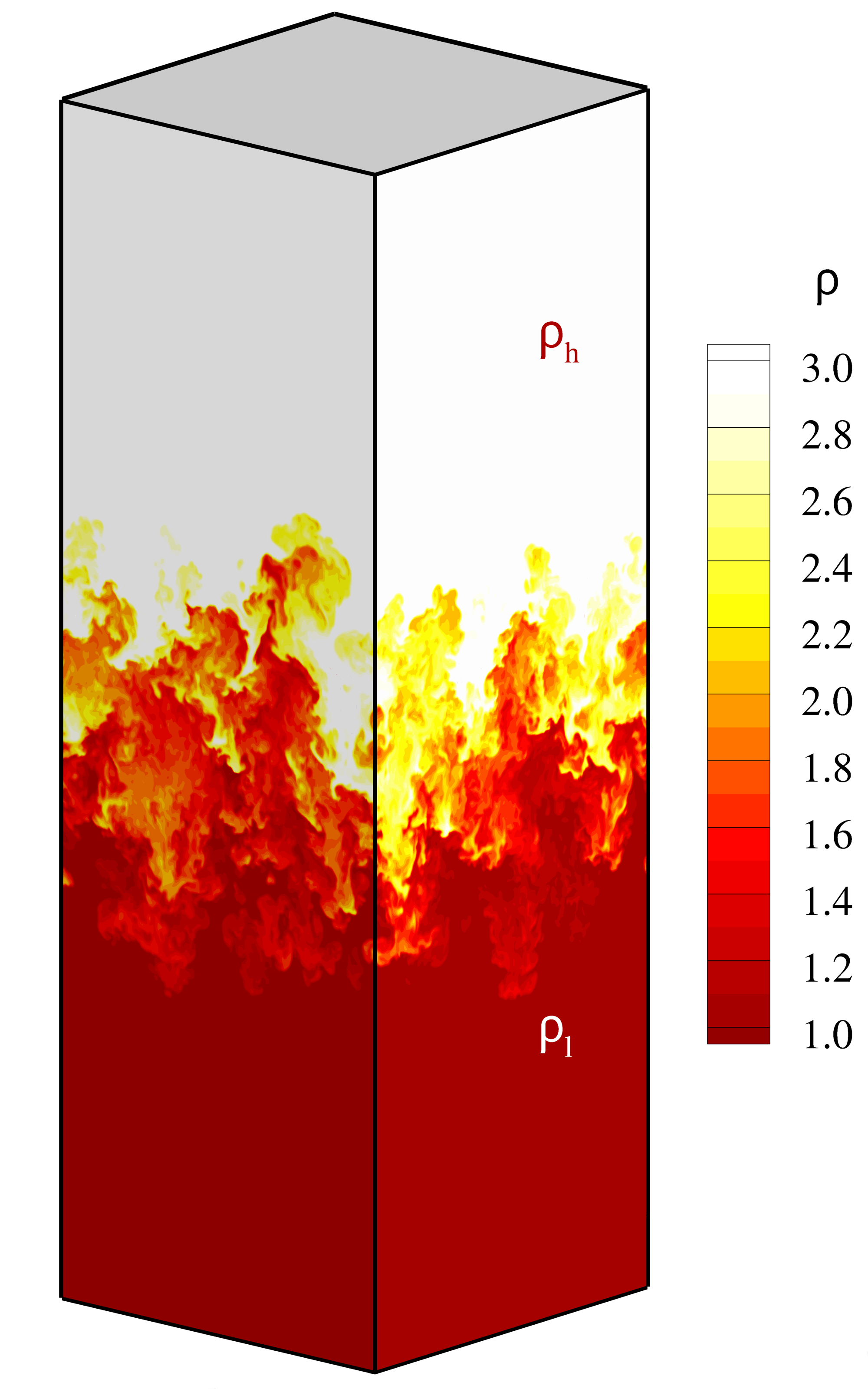}}
\caption{Temporal evolution of the RT density field predicted with $f_k=0.00$.}
\label{fig:5.2_2}
\end{figure*}

The present RT flow is initialized with the perturbed interface shown in figures \ref{fig:4.2_2} and \ref{fig:5.2_1a}. Immediately after this instant, the two fluids accelerate, and the interface perturbations create a misalignment between the density gradient and the pressure. This leads to the generation of coherent structures called spikes and bubbles \cite{ZHOU_PR1_2017,ZHOU_PR2_2017} with the mushroom-like shape illustrated in figures \ref{fig:5.2_1c}-\ref{fig:5.2_1e} and \ref{fig:5.2_2a} at $t \leq 2.5$. During this period, the flow is laminar, and it is in the so-called linear regime \cite{ZHOU_PR1_2017,ZHOU_PR2_2017}. In the following instants, figures \ref{fig:5.2_1f}-\ref{fig:5.2_1i} and \ref{fig:5.2_2b}-\ref{fig:5.2_2d}, the mixing-layer continues growing, and the initially linear structure and the Kelvin-Helmholtz secondary instability will eventually trigger the onset and development of turbulence. This phenomenon increases the mixing rate and enhances the mixture homogeneity, occurring in the non-linear regime \cite{ZHOU_PR1_2017,ZHOU_PR2_2017}. It is particularly pronounced at $t=20.0$ (figure \ref{fig:5.2_2d}). As in the TGV case, the onset and development of turbulence is expected to pose major challenges to modeling and simulation the RT problem.

\begin{figure}[t!]
\centering
\subfloat[$S_1$.]{\label{fig:5.2_3a}
\includegraphics[scale=0.105,trim=0 0 0 0,clip]{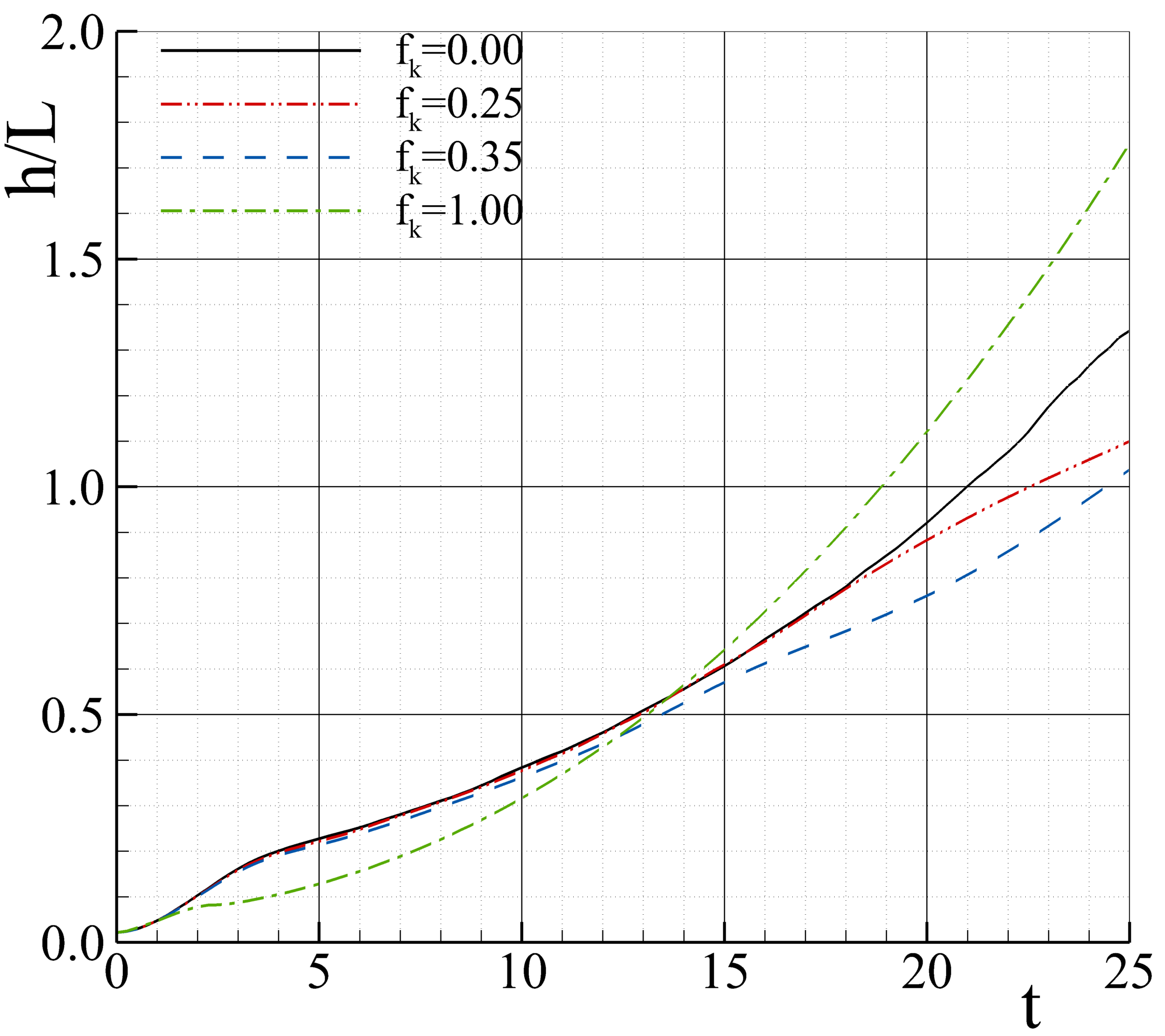}}
\\
\subfloat[$S_2$.]{\label{fig:5.2_3b}
\includegraphics[scale=0.105,trim=0 0 0 0,clip]{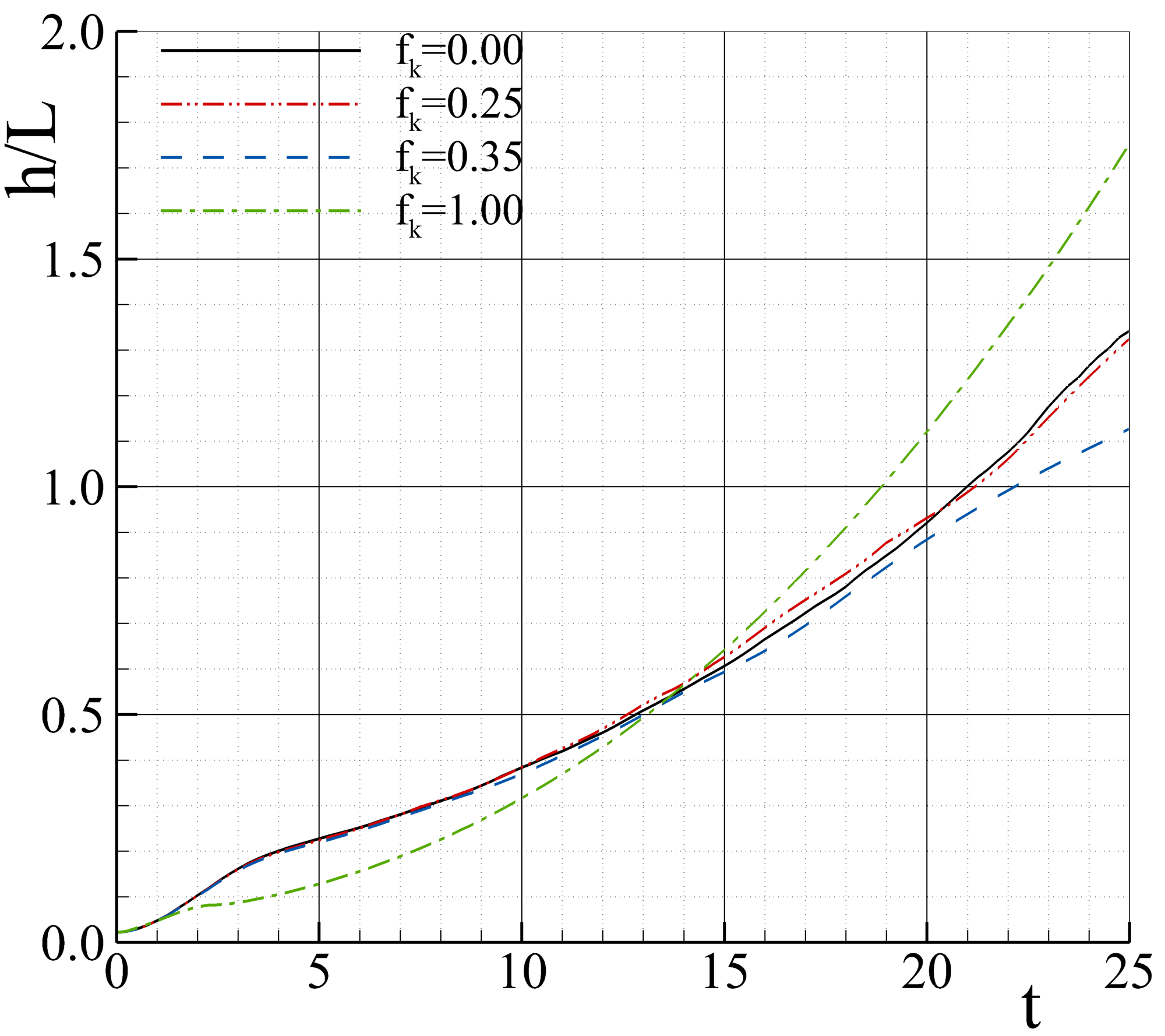}}
\\
\subfloat[$S_3$.]{\label{fig:5.2_3c}
\includegraphics[scale=0.105,trim=0 0 0 0,clip]{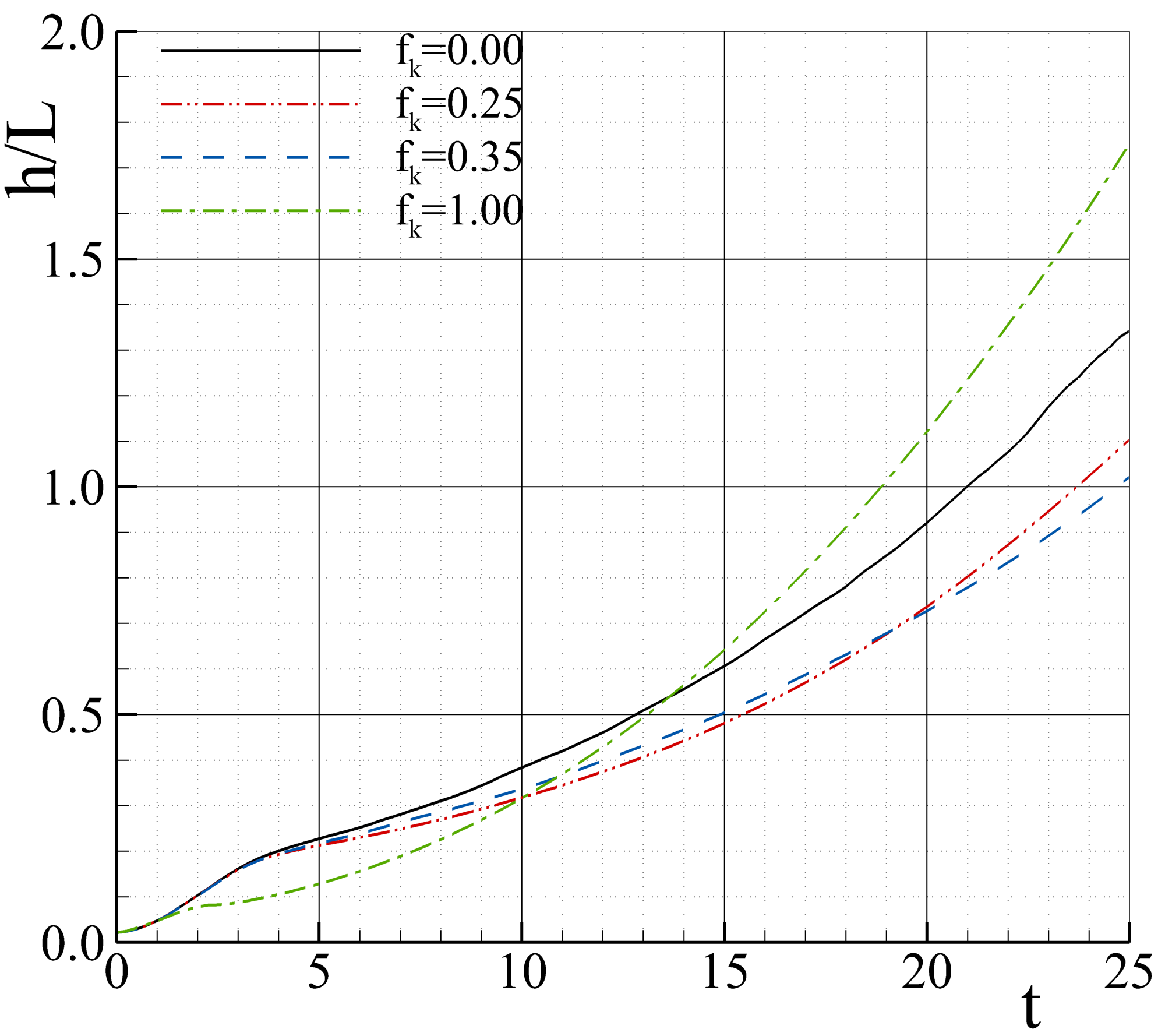}}
\caption{Temporal evolution of the mixing-layer height, $h$, predicted with different $f_k$ and $S_i$.}
\label{fig:5.2_3}
\end{figure}

\begin{figure}[th!]
\centering
\subfloat[$f_k=0.00$.]{\label{fig:5.2_4a}
\includegraphics[scale=0.105,trim=0 0 0 0,clip]{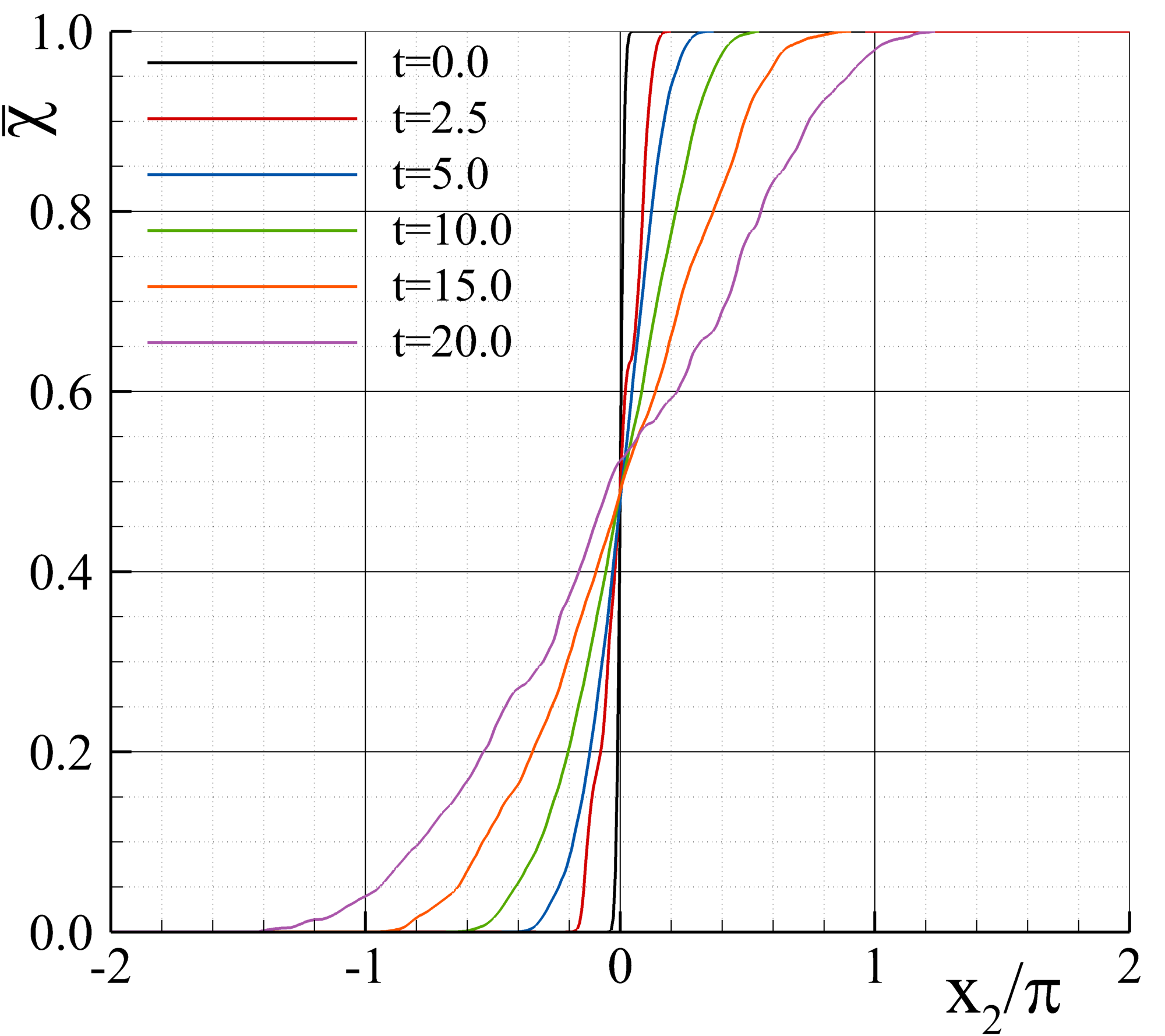}}
\\
\subfloat[$f_k=0.25$ using $S_1$.]{\label{fig:5.2_4b}
\includegraphics[scale=0.105,trim=0 0 0 0,clip]{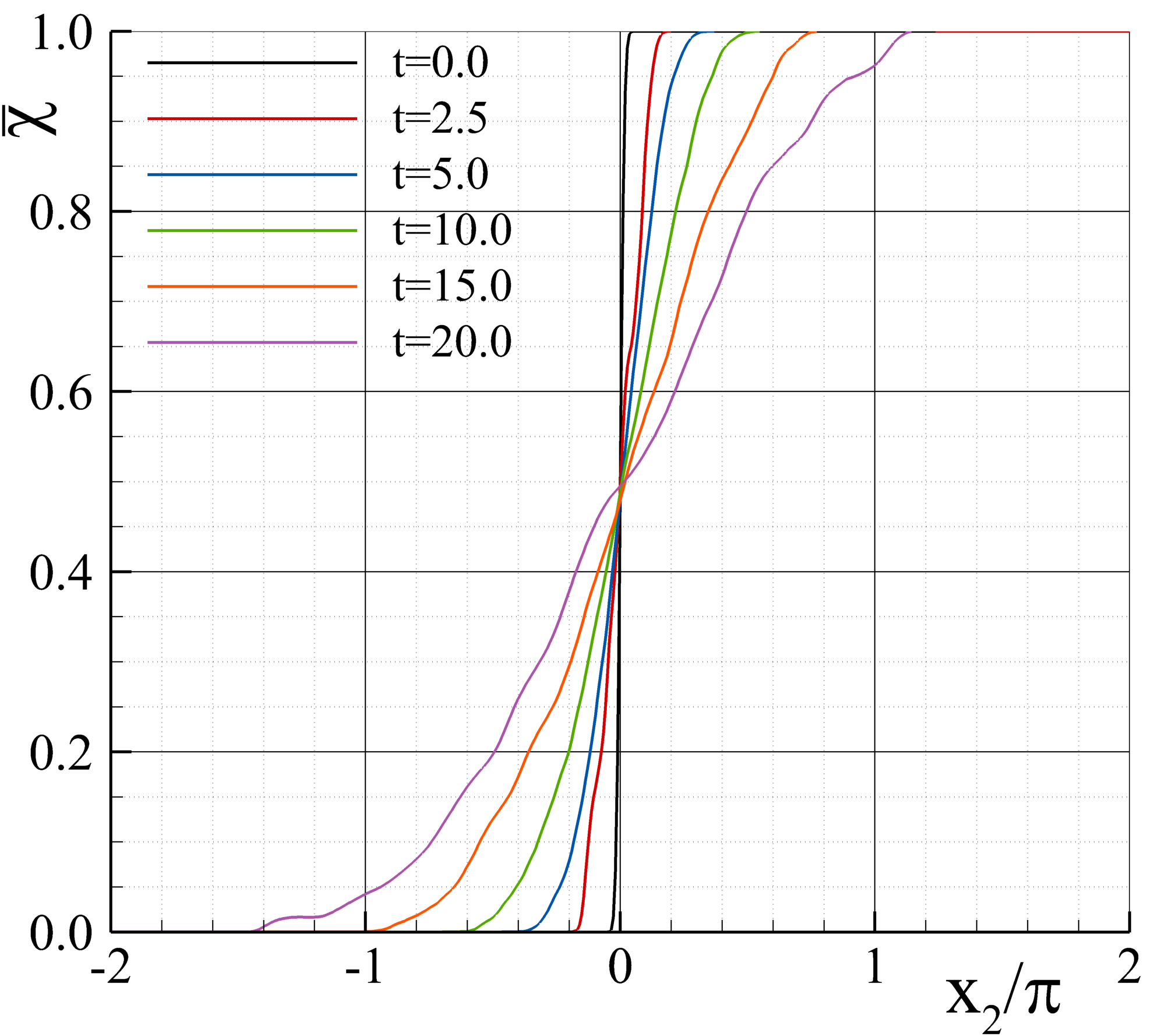}}
\\
\subfloat[$f_k=1.00$.]{\label{fig:5.2_4e}
\includegraphics[scale=0.105,trim=0 0 0 0,clip]{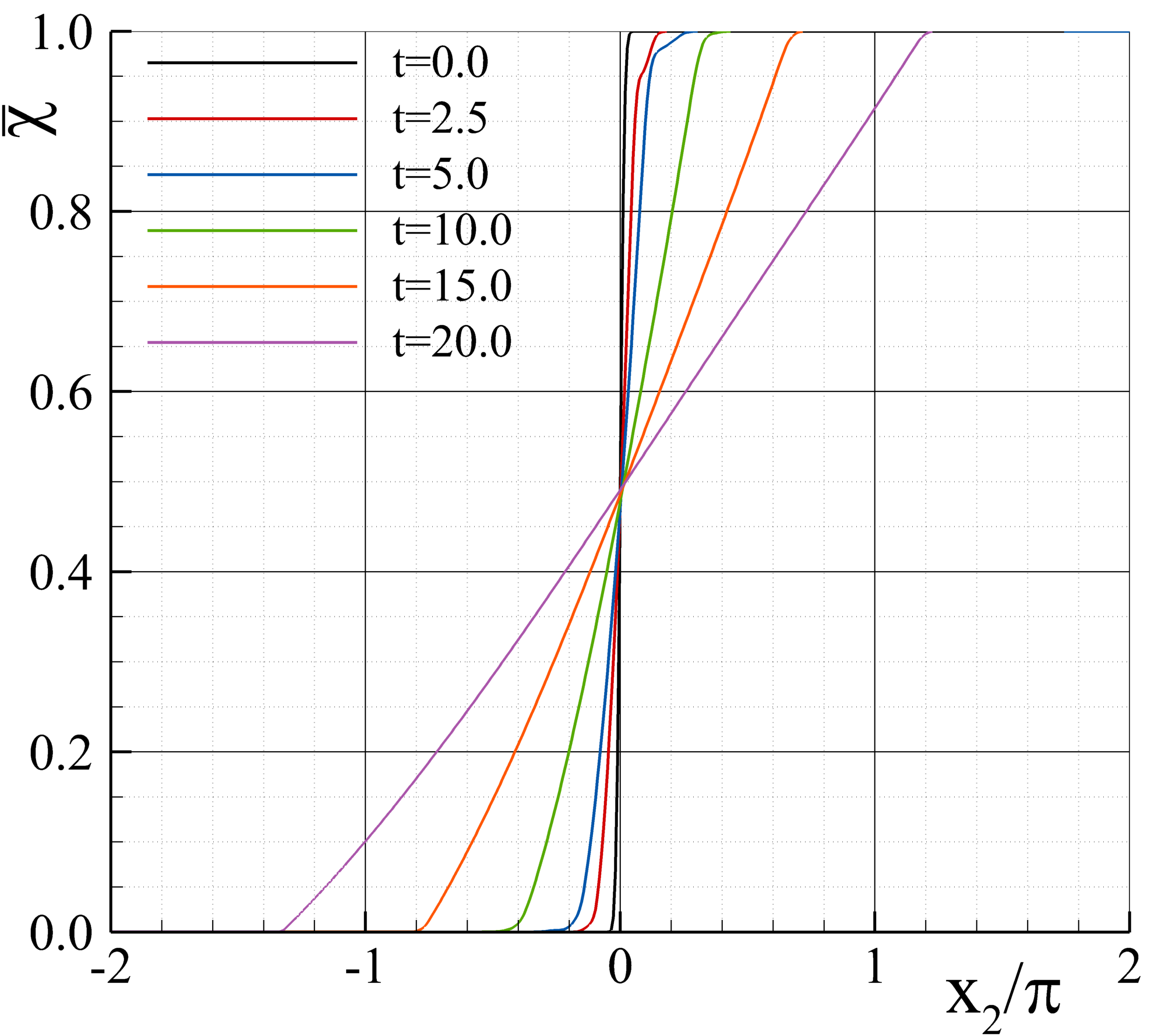}}
\caption{Temporal evolution of the density field, $\overline{\chi}$, predicted with different $f_k$.}
\label{fig:5.2_4}
\end{figure}

To evaluate the accuracy of the PANS BHR-LEVM model, figures \ref{fig:5.2_3} and \ref{fig:5.2_4} depict the evolution of the mixing-layer height, $h$, and density field, $\chi$, predicted at different $f_k$. Also, the simulations test the three strategies proposed in Section \ref{sec:3.1.3} to define the relationship between the parameters $f_\phi$ ($f_k$, $f_\varepsilon$, $f_a$, and $f_b$). Here, the quantity $\overline{\chi}$ used to analyze the density field is defined as
\begin{equation}
\label{eq:5.2_1}
\overline{\chi} = \frac{\overline{\rho} - \rho_l }{\rho_h - \rho_l}\; .
\end{equation}
The mixing-layer width $h$ is defined as the distance between the locations $\bar{\chi}=0.05$ and $0.95$, where here the bar indicates a planar average normal to gravity. Both $h$ and $x_2$ are normalized by $L$. Since the physics and statistics of RT flow are highly dependent on initial conditions and settings, we use the solutions obtained at $f_k=0.00$ as reference.

Figure \ref{fig:5.2_3} indicates that the simulations are closely dependent on the value of $f_k$, and converge upon this parameter's refinement (except $S_3$), $f_k\rightarrow 0.00$. Comparing the solution at $f_k=0.00$ against that at $f_k=1.00$ shows that the second case leads to a significantly thicker mixing-layer at late times, and shorter linear region ($t < 2$). Once again, this result suggests that the simulation at $f_k=1.00$ prematurely predicts the onset of turbulence by overpredicting the total turbulent stresses. The refinement of $f_k$ improves the simulations by reducing the discrepancies against the reference solution ($f_k=0.00$). Also, figure \ref{fig:5.2_3} illustrates that all solutions at $f_k \leq 0.25$ are in good agreement.

The exceptions are the simulations using $S_3$. The data indicate that this approach leads to poorer results than the remaining strategies, which do not improve for $f_k\leq 0.35$ (recall that $f_k=0.00$ does not use a turbulence closure). This outcome stems from the fully developed turbulence assumption embedded in this strategy, which is not verified at early flow stages. As comprehensively discussed in \cite{PEREIRA_POF_2021}, $S_3$ leads to an inconsistent definition of $f_\phi$ at early flow stages and, consequently, to large modeling errors and numerical robustness issues. The latter increase upon grid resolution refinement. This result illustrates the importance of a precise and robust selection of $f_\phi$. Regarding the remaining strategies, $S_2$ leads to the smallest comparison errors between simulations ($f_k>0$) and the reference solution.

Figure \ref{fig:5.2_4} shows how the density field evolves in time and space with $f_k$. For conciseness, we only show three representative cases: $f_k=0.00$, $f_k=0.25$ ($S_2$), and $f_k=1.00$. The remaining cases are in line with the results of figure \ref{fig:5.2_3}. As for the quantity $h$, the results show distributions of $\overline{\chi}$ quite similar between solutions obtained at $f_k\leq 0.25$. In contrast, those obtained from simulations at $f_k=1.00$ exhibit linear and smoother profiles. This behavior stems from the shortcomings of one-point closures to fully model ($f_k=1.00$) transient turbulence and the fact that this modeling strategy does not resolve turbulence (lower numerical requirements).

\begin{figure}[th!]
\centering
\subfloat[$S_1$.]{\label{fig:5.2_5a}
\includegraphics[scale=0.105,trim=0 0 0 0,clip]{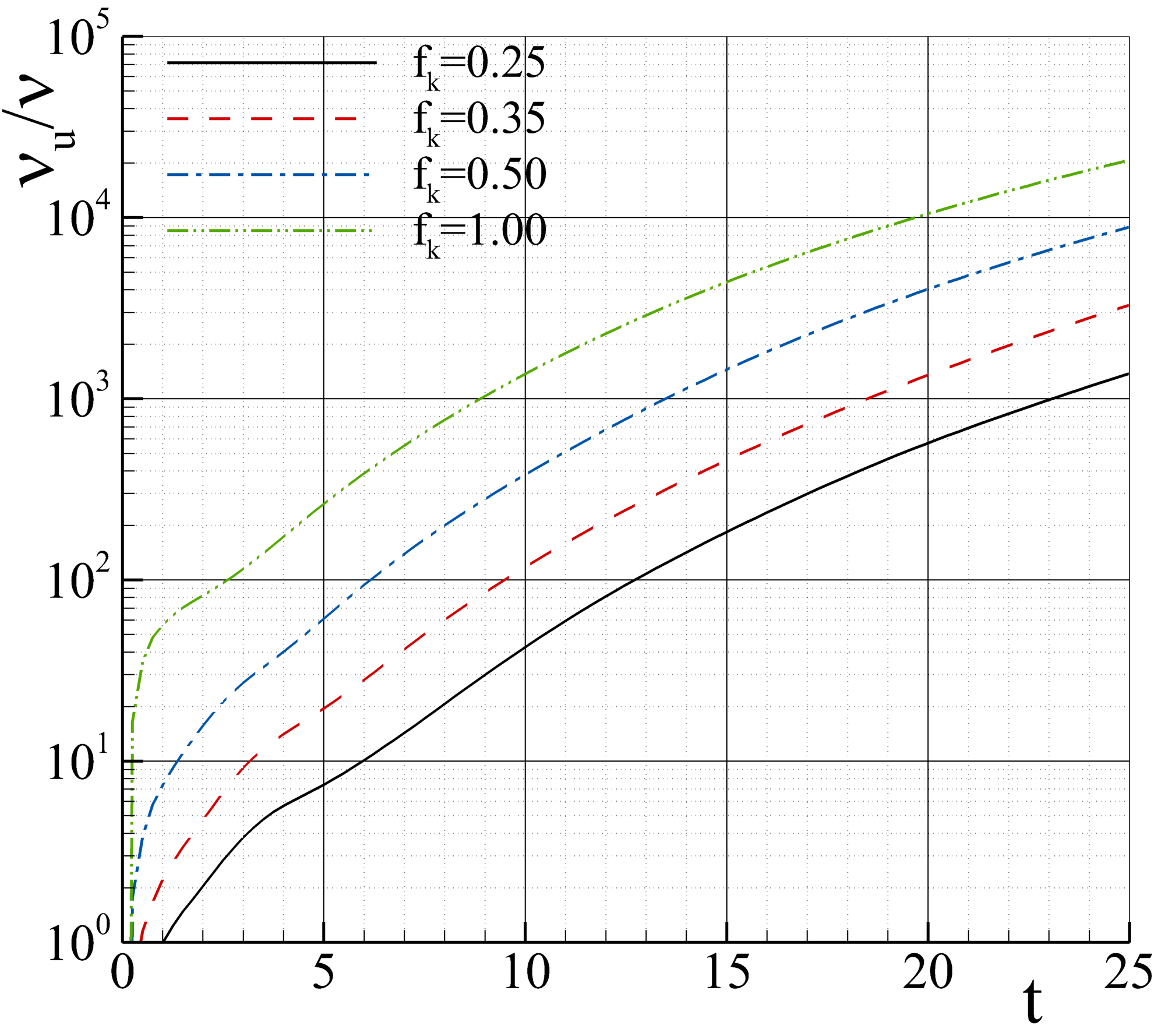}}
\\
\subfloat[$S_2$.]{\label{fig:5.2_5b}
\includegraphics[scale=0.105,trim=0 0 0 0,clip]{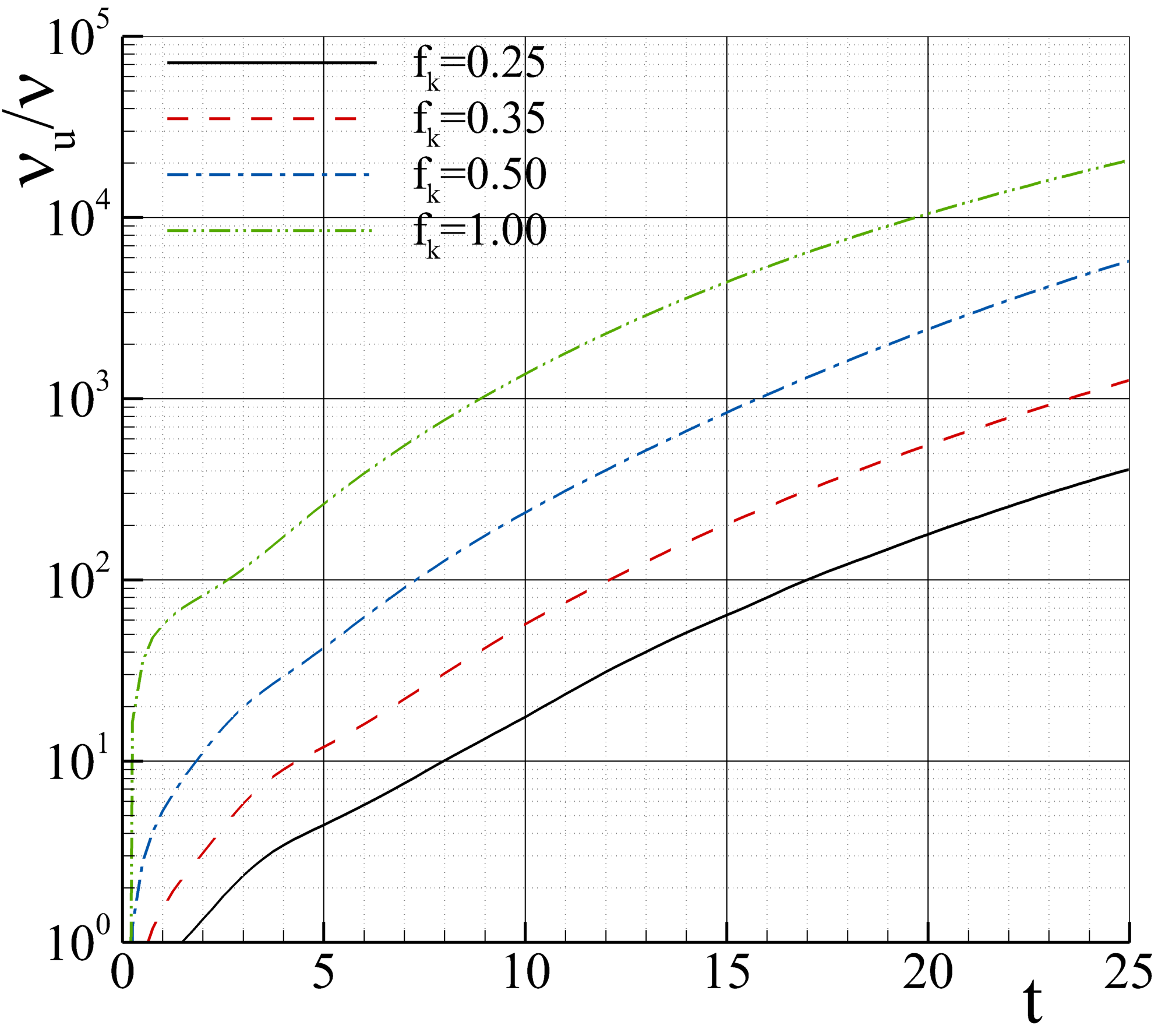}}
\\
\subfloat[$S_3$.]{\label{fig:5.2_5c}
\includegraphics[scale=0.105,trim=0 0 0 0,clip]{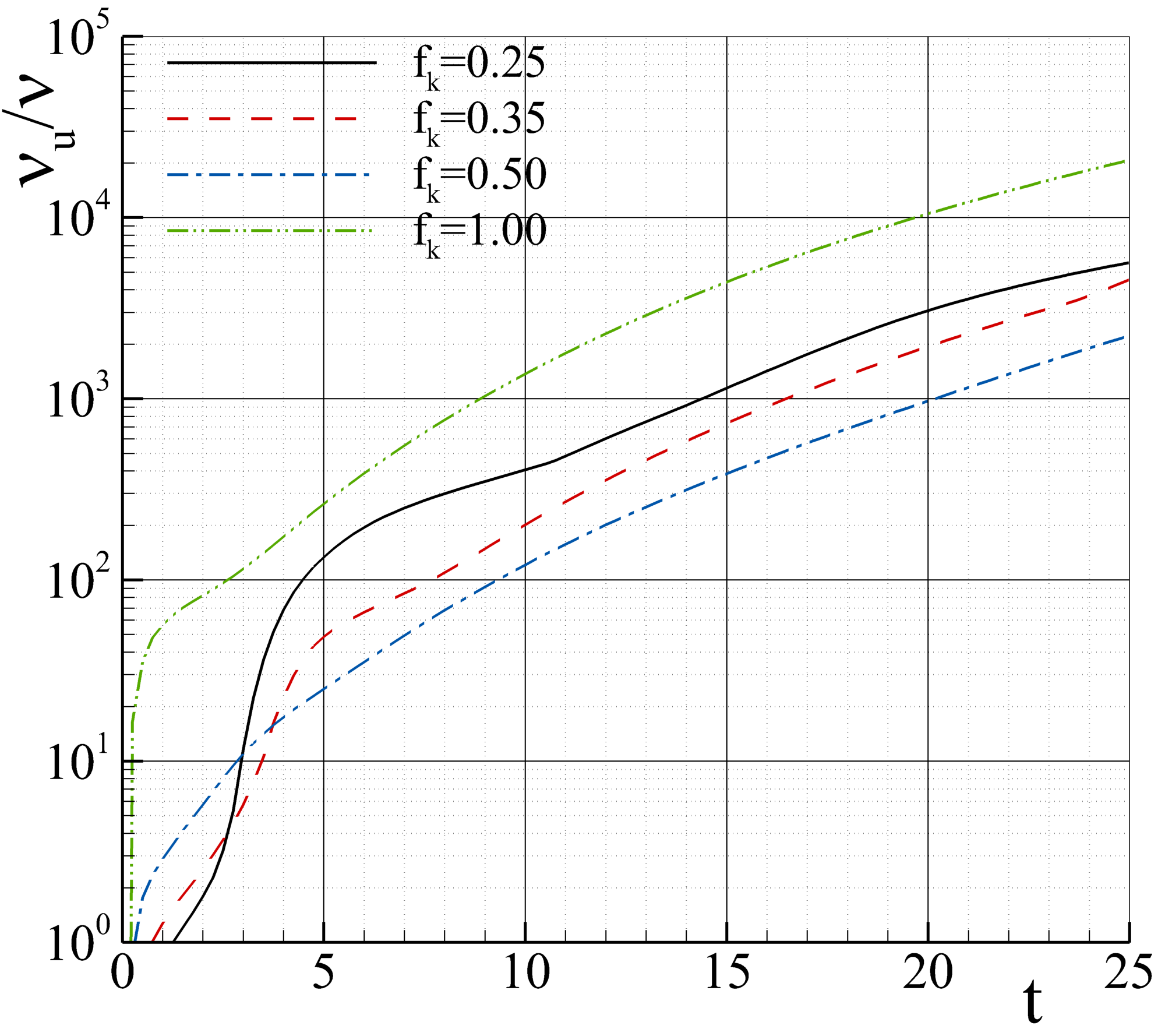}}
\caption{Temporal evolution of the ratio $\nu_u/\nu$ predicted with different $f_k$ and $S_i$.}
\label{fig:5.2_5}
\end{figure}

\begin{figure}[th!]
\centering
\subfloat[$f_k=0.25$ and $S_1$.]{\label{fig:5.2_6a}
\includegraphics[scale=0.08,trim=0 0 0 0,clip]{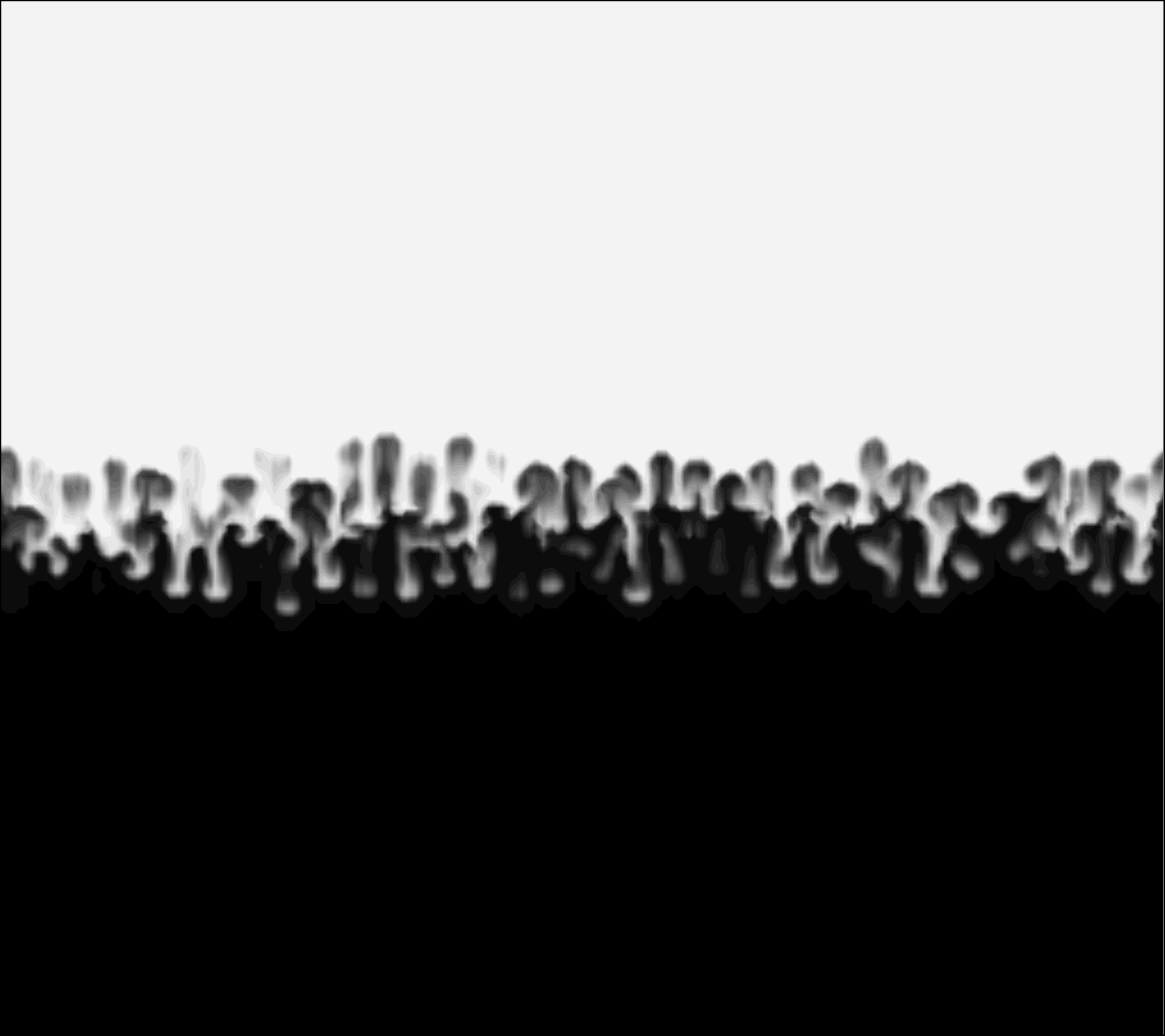}}
\\
\subfloat[$f_k=0.25$ and $S_2$.]{\label{fig:5.2_6b}
\includegraphics[scale=0.08,trim=0 0 0 0,clip]{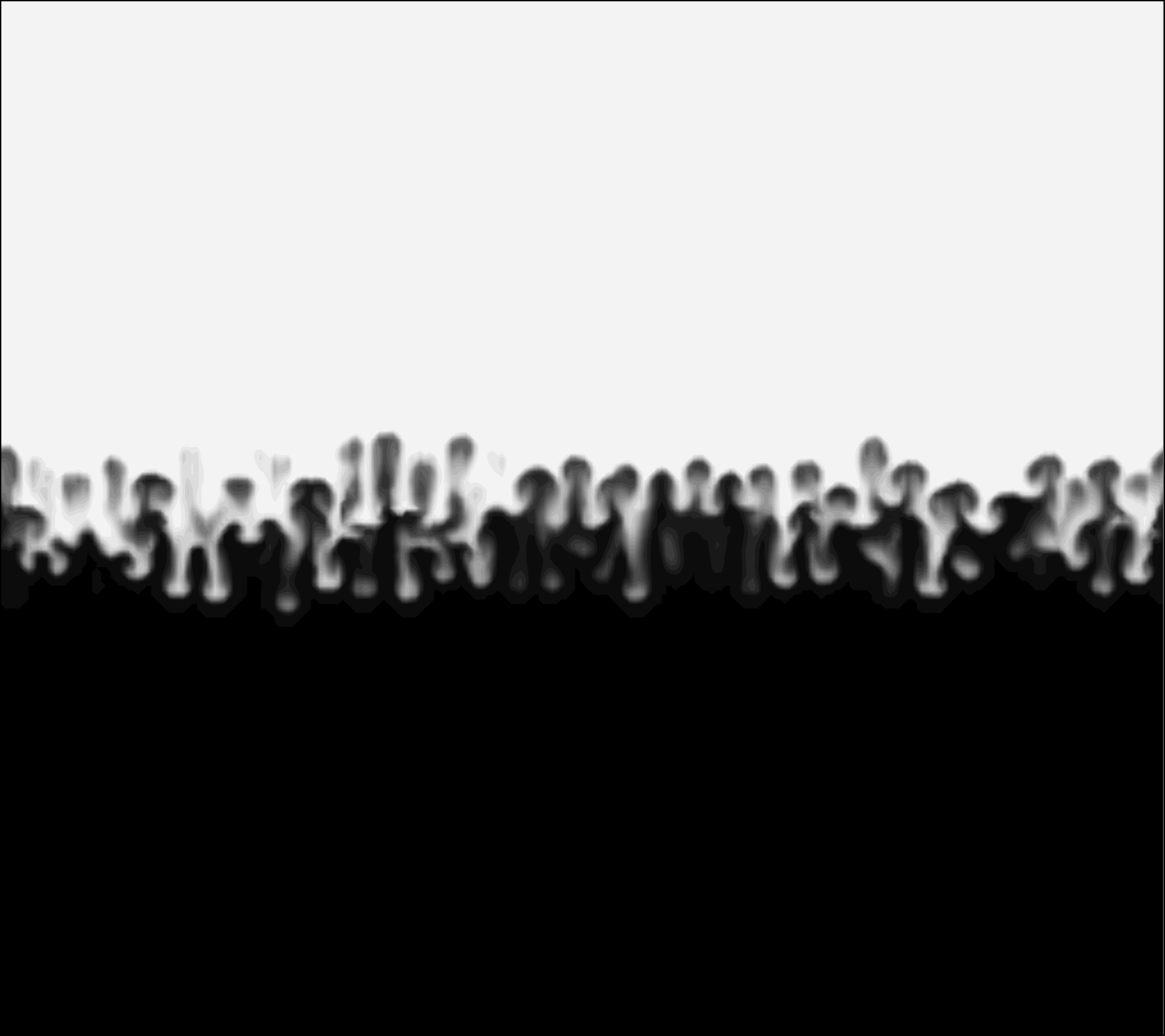}}
\\
\subfloat[$f_k=1.00$.]{\label{fig:5.2_6c}
\includegraphics[scale=0.08,trim=0 0 0 0,clip]{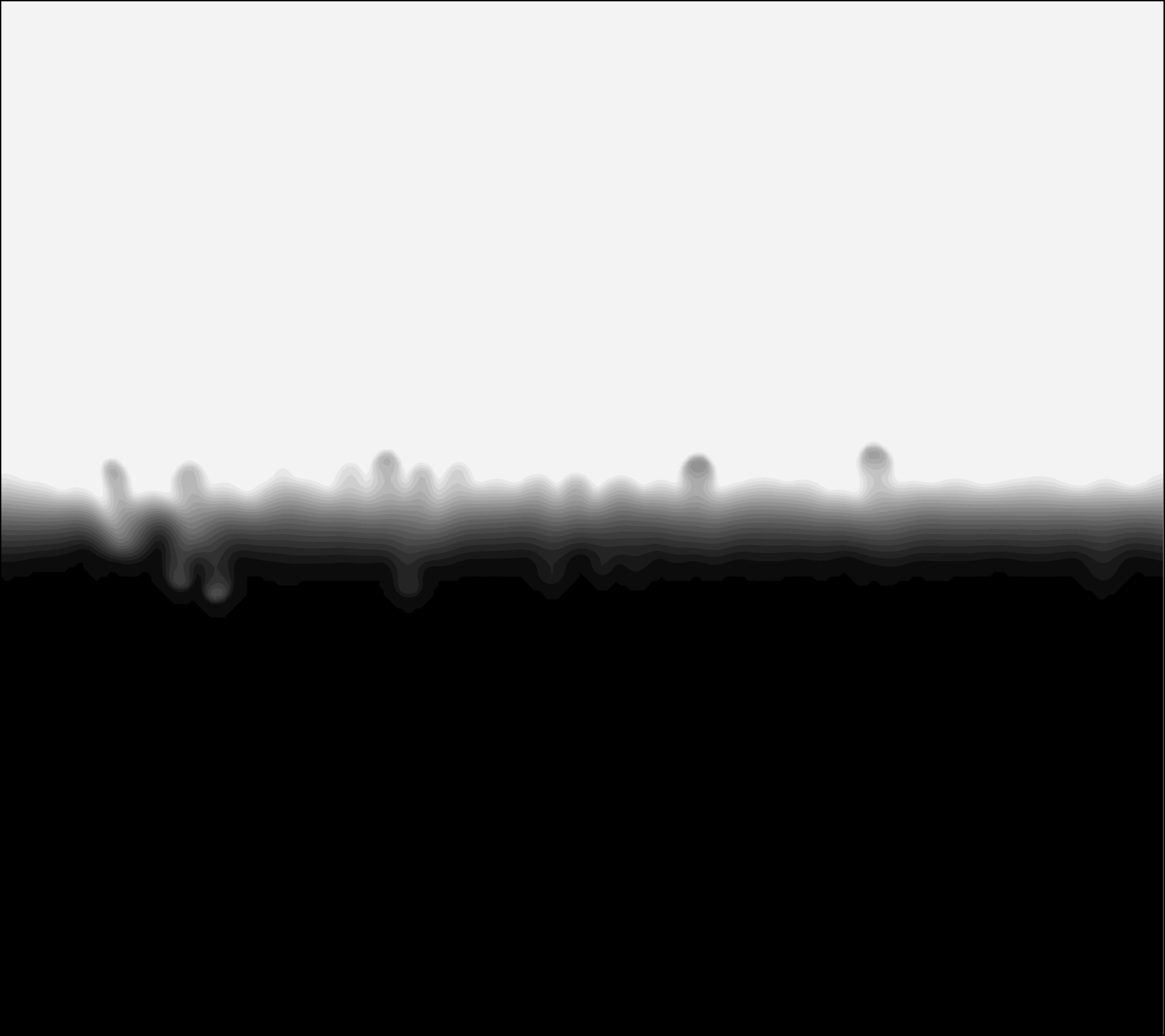}}
\caption{RT structures at $t=2.5$ predicted with $f_k=0.25$ ($S_1$ and $S_2$) and $1.00$. Vortical structures identified through the density field.}
\label{fig:5.2_6}
\end{figure}

Next, figure \ref{fig:5.2_5} presents the evolution of the maximum planar ($x_1-x_3$) averaged value of $\nu_u$ obtained at different physical resolutions. Note that $\nu_u$ is here utilized to evaluate how the unresolved turbulent stresses evolve with $f_k$ (see equation \ref{eq:2_16}). Also, it important to emphasize that ratios $\nu_u/\nu$ exceeding $\mathcal{O}(1)$ are usually attributed to turbulence effects. As expected, the results show that $\nu_u/\nu$ decreases with $f_k$. Considering $t=25$, $\nu_u/\nu$ reduces from $20736.6$ at $f_k=1.00$ to $1376.0$ ($S_1$), $407.8$ ($S_2$), and $5634.4$ ($S_3$) at $f_k=0.25$. The large values obtained with $S_3$ at the smallest $f_k$ are caused by the aforementioned consistency issues selecting $f_\phi$ with this strategy. {\color{black}These results also show} that the strategy used to define the relationship between the different $f_\phi$ has an important impact on the magnitude of $\nu_u$ and, consequently, $\tau^1(V_i,V_j)$. 

However, the most significant result in figure \ref{fig:5.2_5} is the fact that $\nu_u/\nu$ does not exceed $5.7$ for $f_k=0.25$ ($S_1$ and $S_2$) and $t\leq 4$ (linear regime and laminar flow), whereas this quantity exceeds $173.7$ for $f_k=1.00$. Such a result indicates that the simulation with $f_k=1.00$ leads to a premature onset of turbulence. On the other hand, PANS at $f_k=0.25$ can capture the coherent structures that later are involved on the onset and development of turbulence. This outcome explains the results of figures \ref{fig:5.2_3} and \ref{fig:5.2_4} and can be seen in figure \ref{fig:5.2_6}, in which the flow coherent structures are depicted for $f_k=0.25$ ($S_1$ and $S_2$) and $1.00$ at $t=2.5$. Compared to figure \ref{fig:5.2_1}, the results show that simulations at $f_k=1.00$ dissipate the laminar coherent structures. This is caused by the overprediction of turbulence, i.e., the magnitude of $\nu_t$ or $\tau^1(V_i,V_j)$. In clear contrast, simulations at $f_k=0.25$ can accurately predict these coherent structures, this being the reason for the good agreement between simulations at $f_k \leq 0.25$ ($S_1$ and $S_2$). Hence, the RT computations reinforce the importance of resolving the flow phenomena not amenable to modeling to obtain efficient high-fidelity simulations \cite{PEREIRA_JCP_2018}. 

Overall, the results have shown that PANS BHR-LEVM model can predict the current RT flow accurately using sufficiently small values of $f_k$. Regarding the strategies to prescribe $f_\phi$, $S_2$ leads to the lowest values of $\nu_u$, allowing the utilization of larger $f_k$ than $S_1$. $S_3$ causes consistency issues between the different $f_\phi$ due to the fully developed turbulence assumption. A comprehensive analysis of this problem is given in a subsequent paper.
%
%
\section{Conclusions}
\label{sec:6}

We extended the framework of the PANS model to variable-density flow, i.e., multi-material and/or compressible mixing problems including density fluctuations and production of turbulence kinetic energy by shear and buoyancy mechanisms. The framework was utilized to derive the PANS version of the $k-S-a_i-b$ equation BHR-LEVM closure. The parameters defining the physical resolution of the model ($f_k$, $f_\varepsilon$, $f_a$, and $f_b$) have been studied through \textit{a-priori} testing. Three strategies are proposed to set these parameters as a function of $f_k$: \textit{i)} define $f_\varepsilon=f_{a_i}=f_b=1.0$; \textit{ii)} prescribe $f_{a_i}=\sqrt{f_k}$ and $f_\varepsilon=f_b=1.0$; and \textit{iii)} set $f_\varepsilon=1.0$, $f_a=0.5 f_k$, and $f_b=f_k$. The first two strategies lead to high-fidelity simulations, whereas the third leads to consistency issues between different $f_\phi$. Thus, $S_1$ and, in particular, $S_2$ seem better approaches to select $f_\phi$. Future studies will further investigate these strategies. The initial validation space of the PANS BHR-LEVM model comprises the TGV at $\mathrm{Re}=3000$ and the RT at $\mathrm{At}=0.5$ and $(\mathrm{Re})_{\max}\approx 500$ flows. The results are promising and confirm the ability of the model to calculate these representative flows accurately. Hence, this initial validation space and the theoretical justification demonstrate the new methodology's potential to predict complex problems of variable-density flow. Nevertheless, subsequent studies will be needed to study and extend the validation space of the model further. Finally, all simulations indicate the importance of resolving the phenomena not amenable to modeling by the closure. This dictates the required physical resolution to obtain high-fidelity simulations.
%
%
%
%
%
\section*{Acknowledgments}

We would like to thank C .B. da Silva, D. Aslangil, and D. Livescu for sharing their DNS data sets. Also, we would like to thank the two reviewers for their suggestions that improved our paper. Los Alamos National Laboratory (LANL) is operated by TRIAD National Security, LLC for the US DOE NNSA. This research was funded by LANL Mix and Burn project under the DOE ASC, Physics and Engineering Models program.

\bibliography{references}
\end{document}